\documentclass[a4paper,11pt]{article}
\usepackage{jheppub} 
\usepackage{rotating}
\usepackage[T1]{fontenc}

\title{\boldmath The current status of fine-tuning in supersymmetry}

\author[a,b]{Melissa van Beekveld}
\author[a,b]{, Sascha Caron}
\author[c]{, Roberto Ruiz de Austri}

\affiliation[a]{Theoretical High Energy Physics, Radboud University
Nijmegen, Heyendaalseweg 135, 6525 AJ Nijmegen, the Netherlands}
\affiliation[b]{Nikhef, Science Park 105, 1098 XG Amsterdam, the Netherlands}
\affiliation[c]{Instituto de F\'isica Corpuscular, IFIC-UV/CSIC, Valencia, Spain}

\abstract{
In this paper, we minimize and compare two different fine-tuning measures in four high-scale supersymmetric models that are embedded in the MSSM. In addition, we determine the impact of current and future dark matter direct detection and collider experiments on the fine-tuning. We then compare the low-scale electroweak measure with  the high-scale Barbieri-Giudice measure, which generally do not agree. However, we find that they do reduce to the same value when the higgsino parameter drives the degree of fine-tuning. 
Depending on the high-scale model and fine-tuning definition, we find a minimal fine-tuning of $3-38$ (corresponding to  $\mathcal{O}(10-1)\%$) for the low-scale measure, and $63-571$ (corresponding to $\mathcal{O}(1-0.1)\%$) for the high-scale measure. In addition, minimally fine-tuned spectra give rise to a dark matter relic density that is between $10^{-3} < \Omega h^2 < 1$, when $\mu$ determines the minimum of the fine-tuning. We stress that it is too early to conclude on the fate of supersymmetry, based only on the fine-tuning paradigm. 
}

\begin{document} 

\begin{flushright}
Nikhef/2019-023
\end{flushright}

\maketitle
\flushbottom

\section{Introduction}
\label{S:0}

The instability of the mass of a scalar particle under quantum corrections was firstly discussed in the pioneering work of Susskind \cite{Susskind:1978ms},
Veltman \cite{Veltman:1980mj} and 't Hooft \cite{tHooft:1979rat}. They showed that the existence of both a scalar particle, like the Higgs boson, and a much heavier particle that couples to this scalar particle implies that the mass of the scalar particle will receive radiative corrections that are quadratic in the mass of the heavy particle. They pointed out that if the Higgs boson indeed has a mass around the weak scale, and if the Standard Model (SM) is to be embedded in a high-scale (HS) theory, this HS theory would generally require extreme forms of fine-tuning (FT) to prevent the Higgs mass from becoming of the same order as that of the HS. This would result in an \emph{unnatural} theory unless one constructs the HS theory in such a way that it is free from FT. Many years later, the Higgs boson with a mass of 125 GeV has been found \cite{Aad:2012tfa, Chatrchyan:2012ufa}, a value that is certainly not far from the weak scale. The existence of a HS theory could, therefore, introduce a FT problem. \\
There are many reasons to believe in a theory that complements the Standard Model (SM) at higher mass scales: there is no explanation/solution for the origin of neutrino masses or the hierarchies among fermion masses, the nature of dark matter, the origin of electroweak symmetry breaking (EWSB) or the possible instability of the Higgs vacuum. In addition, as gravity becomes strongly coupled around the Planck scale, the SM needs to be complemented by a theory of quantum gravity at this high scale. And beyond the Planck scale,  at energies around $\mathcal{O}(10^{40})$~GeV, we encounter the Landau Pole of $U(1)$, signaling that the perturbative nature of the SM is bound to break down eventually. If the HS theory is to solve some or all of the aforementioned problems, it is likely that it somehow couples to the SM particles. As mentioned before, this HS theory would then have to be constructed in a careful way in order to avoid the Higgs mass FT problem. 
\\
The FT problem is closely related to how we perform physics and set up experiments: we always assume that physics on small length scales is not affected by physics on large length scales. For example, we do not expect that gravity influences the physics that we are probing at colliders and consequently do not take its effects into account when making predictions for the LHC. Therefore, the apparent break down of this assumption in the Higgs sector \emph{does} require an explanation, as it is the only example we know of where low-scale physics is \emph{extremely} sensitive to high-scale physics. 
\\

\noindent Supersymmetry (SUSY) is a theory that is able to solve the Higgs FT problem by imposing a space-time symmetry between bosons and fermions \cite{Martin:1997ns,WITTEN1981513,KAUL198219}. Realized in its minimal form, SUSY can explain the observed Higgs mass, predict EWSB from SUSY breaking and allow for a dark matter candidate if R-parity conservation is assumed. In addition, any (softly broken) supersymmetric extension of the SM ensures that quadratic quantum corrections to masses of scalar particles are absent, which remedies the FT problem. \\
The FT problem in SUSY is well-studied \cite{Sakai:1981gr, PhysRevLett.109.161802, PhysRevD.73.095004, Hall2012, Baer:2012cf, Barbieri:1987fn, Feng:1999zg,Kitano:2005wc,Cabrera:2012vu,Fichet:2012sn,Baer:2012mv,Boehm:2013gst,Balazs:2013qva,Kim:2013uxa,Boehm:2013gst,Casas:2014eca,PhysRevD.91.075005,Drees:2015aeo,Baer:2015rja,Kim:2016rsd,Casas:2016xnl,vanBeekveld:2016hug,Cici:2016oqr,Cabrera:2016wwr,Buckley:2016kvr,Baer:2017pba,Abdughani:2017dqs,Fundira:2017vip, Baer:2018rhs}. Since many years it is claimed that in order to have a {\it natural} version of the minimal supersymmetric extension of the SM (the MSSM), one expects SUSY particles with masses that lie well below the TeV scale \cite{Ellis:1986yg, Kane:1993td,Anderson:1994dz, Anderson:1994tr, Dimopoulos:1995mi,Chankowski:1997zh, Chankowski:1998xv,Casas:2003jx, Kitano:2005wc, Papucci:2011wy, Liu:2013ula, Kowalska:2013ica,Arvanitaki:2013yja,Lykken:2014bca}. However, to date, there has not been any evidence for the existence of SUSY particles at these scales. This has motivated some to go beyond the MSSM (see e.g. Refs. \cite{Cassel:2009ps,Ross:2011xv,Gogoladze:2012yf,Gogoladze:2013wva,Kaminska:2013mya,Cao:2016nix,Cao:2016cnv,Li:2017fbg,Zhu:2017moa,Athron:2017fxj,Alvarado:2018rfl,Du:2018pko,Badziak:2018nnf,Kobakhidze:2018vuy,Yanagida:2018arr,Badziak:2018ijy,Cao:2018rix,Wang:2018vxp}). \\
In this paper, we reanalyze the question: does the non-observation of SUSY particles make any minimal realization of a supersymmetric SM necessarily an unnatural theory? We show that the answer to this question is both a measure dependent and a model dependent one. A natural follow up question, which is also one we will address here, is: what sort of experiment is needed to exclude the natural MSSM?
To answer both questions, one needs a measure of FT to quantify the degree of naturalness. We will explore two widely used FT measures: the electroweak (EW) measure \cite{Baer:2013gva} and the Barbieri-Giudice (BG) measure \cite{Ellis:1986yg, Barbieri:1987fn}, whose definitions will be given in Sec. \ref{S:1}. Low values of these FT measures mean low degrees of fine-tuning: an FT value of 100 corresponds to $\mathcal{O}(1\%)$ fine-tuning\footnote{We note in passing that the inverse of the FT measure can also be interpreted either as a p-value \cite{Casas:2014eca} or in a Bayesian context \cite{Fichet:2012sn,Cabrera:2008tj} to get the correct value of the Higgs boson mass.}. Note that `how much' FT one allows in the theory depends on the level of cancellation between different theory parameters one is willing to accept. There is no general consensus on this, which brings a certain degree of subjectivity to the discussion. But even if one would agree on an absolute maximum amount of FT that one is willing to tolerate, there is, in fact, a second complication. The sort of experiment that will exclude a natural version of the MSSM is, as mentioned above, a measure {\it dependent} one, which we will clarify in Sec. \ref{S:1}. \\
To asses the impact of worldwide data on the FT, we will \emph{minimize}, like was done for the first time in Ref. \cite{vanBeekveld:2016hug}, the FT for both measures in four different HS SUSY models (described in Sec. \ref{S:1}). All of these HS models are embedded in the low-scale (LS) phenomenological Minimal Supersymmetric Standard Model (pMSSM) \cite{Djouadi:1998di}. 
We then show how past and current experiments have constrained the FT. In addition, we study the ability of future ton-scale dark matter direct detection experiments and future colliders to constrain the FT in the four HS realizations. The set-up for this analysis can be found in Sec. \ref{sec:analysis} and the results are reported in Sec. \ref{S:3}. We conclude in Sec. \ref{sec:conclusion}.

\section{Supersymmetric models and their fine-tunings}
\label{S:1}

Whether SUSY is realized in nature is unknown and as such, we do not know which and how many fundamental parameters exist for the actual HS theory.  A generic broken SUSY theory has two relevant energy scales: a HS one ($M_{\rm GUT}$) at which SUSY breaking takes place, and a LS one ($M_{\rm SUSY}$) where the resulting SUSY particle spectrum is situated and the EWSB conditions must be satisfied. The breaking conditions link the $Z$-boson mass ($M_Z$) to the input parameters via the minimization of the one-loop scalar potential of the Higgs fields. The resulting equation is \cite{Coleman:1973jx,Baer:2012cf}:
\begin{equation}
\label{eq:mz}
\frac{M_Z^2}{2} = \frac{m_{H_d}^2 + \Sigma_d^d - (m_{H_u}^2 + \Sigma_u^u)\tan^2\beta}{\tan^2\beta-1} - \mu^2,
\end{equation}
where $m_{H_u}$ and $m_{H_d}$ are the soft SUSY breaking Higgs masses, $\mu$ is the SUSY version of the SM Higgs mass parameter and $\tan\beta$ is the ratio of the vacuum expectation values of the two Higgs doublets.  The two effective potential terms $\Sigma_u^u$ and $\Sigma_d^d$ denote the one-loop corrections to the soft SUSY breaking Higgs masses (explicit expressions are shown in the appendix of Ref. \cite{Baer:2012cf}). All terms in Eq. (\ref{eq:mz}) are evaluated at $M_{\rm SUSY}$. \\
In order to obtain the observed value of $M_Z = 91.2$ GeV, one needs some degree of cancellation between the SUSY parameters appearing in Eq.~(\ref{eq:mz}). In the general case: if the needed cancellation is large, small changes in the SUSY parameters will result in a widely different value of $M_Z$, in which case the considered spectrum is {\it fine-tuned}. FT measures aim to quantify the sensitivity of $M_Z$ to the SUSY input parameters. In the literature one can find two main classes of SUSY FT measures: one that does take underlying model assumptions into account, such as the BG measure \cite{Ellis:1986yg, Barbieri:1987fn}, and one that does not, such as the EW measure \cite{Baer:2013gva}. 
To assess the differences between these two measures, we will look at four different HS SUSY models that can all be embedded in the (LS) pMSSM. After defining these HS models we will review the two FT measures in more detail. 

\subsection{SUSY models}
\label{sec:susymodels}
All of the HS SUSY models we will consider in this paper are embedded in the pMSSM \cite{Djouadi:1998di}, which is constructed as follows:
\begin{itemize}
\item The first and second generation squark and slepton masses are degenerate. 
\item All trilinear couplings of the first and second generation sfermions are set to zero.
\item There are no new sources of CP violation.
\item All sfermion mass matrices are assumed to be diagonal to ensure minimal flavor violation.
\end{itemize}
After applying these conditions one ends up with a 19-dimensional model that can be parametrized  as follows: the sfermion soft-masses are described by the first and second generation squark masses $m_{\tilde{Q}_1}$, $m_{\tilde{u}_{R}}$ and $m_{\tilde{d}_{R}}$, the third generation squark masses $m_{\tilde{Q}_3}$, $m_{\tilde{t}_R}$ and $m_{\tilde{b}_R}$, the first and second generation of slepton masses $m_{\tilde{L}_1}$, $m_{\tilde{e}_{R}}$, and the third generation of slepton masses $m_{\tilde{L}_3}$, $m_{\tilde{\tau}_R}$. Only the trilinear couplings of the
third generation of sfermions $A_{\tilde t}$, $A_{\tilde b}$ and $A_{\tilde \tau}$ are assumed to be non-zero. The Higgs sector is described by the ratio of the Higgs vacuum expectation values tan $\beta$ and the soft Higgs masses $m_{H_{u,d}}$. Instead of these Higgs masses, it is custom to use the higgsino mass parameter $\mu$ and the mass of the pseudoscalar Higgs $m_A$ as free(input) parameters, which set the values for the soft Higgs masses via the requirement of EWSB (Eq.\eqref{eq:mz}). Finally, the gaugino sector is described by $M_1$, $M_2$ and $M_3$. All of these parameters are defined at the SUSY breaking scale $M_{\rm SUSY}$, which is taken to be the geometric average of the two stop pole masses ($\sqrt{m_{\tilde{t}_1}m_{\tilde{t}_2}}$).  \\

\noindent The HS models that we will consider are:
\begin{itemize}
\item {\bf mSUGRA} \cite{Chamseddine:1982jx,Barbieri:1982eh,Ohta:1982wn,Hall:1983iz}, defined by a global scalar mass $m_0$ giving mass to all scalar particles, a gaugino mass $M_{1/2}$,  a trilinear soft term $A_0$, $\tan \beta$, the sign of $\mu$  and gauge coupling unification at the high scale $M_{\rm GUT}$. Apart from $\tan \beta$, all of these parameters are defined at the high scale $M_{\rm GUT}$. This model has 4 free parameters and the undefined sign of $\mu$. We will probe both signs of $\mu$.
\item {\bf mSUGRA-var}, defined mostly in the same way as mSUGRA except for one modification: we allow for free ratios of the gaugino masses such that $f_1 M_1 = f_2 M_2 = M_3 = M_{1/2}$. This model has 6 free parameters and the sign of $\mu$ that one can choose. The number of parameters that are assumed to be independent for the computation of the BG FT measure (see Sec.~\ref{sec:bgft}) is the same as for the mSUGRA model. This model allows us to study the impact of HS model dependence.  
\item {\bf NUHGM} \cite{Cabrera:2013jya}, where we use two independent mass parameters for the slepton and squark sector. We use $m_{0,L}$ as a soft-breaking SUSY mass parameter for all sparticles of the left-handed SM particles and $m_{0,R}$ for all sparticles of the right-handed SM particles. The gaugino masses are not required to unify and are given by three independent parameters: $M_1$, $M_2$ and $M_3$. Furthermore, there is one trilinear soft term $A_0$, the supersymmetric higgsino mass term $\mu$, $\tan\beta$ and finally the pseudo-scalar Higgs boson pole mass $m_A$. We demand gauge coupling unification at the high scale $M_{\rm GUT}$. All SUSY parameters, except $m_A$, $\mu$ and $\tan\beta$, are defined at the high scale $M_{\rm GUT}$. This model has 9 free parameters. 
\item {\bf pMSSM-GUT} \cite{Peiro:2016ykr}, defined by the gaugino masses $M_1$, $M_2$, $M_3$, first/second generation scalar masses $m_{\tilde{Q}_1}, m_{\tilde{u}_R}$, $m_{\tilde{d}_R}$, $m_{\tilde{L}_1}$, $m_{\tilde{e}_R}$, third generation scalar masses $m_{\tilde{Q}_3}$, $m_{\tilde{t}_R}$, $m_{\tilde{b}_R}$, $m_{\tilde{L}_3}$, $m_{\tilde{\tau}_R}$, trilinear soft terms $A_t$, $A_b$, $A_{\tau}$, the pseudo-scalar Higgs boson pole mass $m_A$, the higgsino mass term $\mu$ and the ratio of weak scale Higgs vevs $\tan\beta$. All parameters, except $m_A$, $\mu$ and $\tan\beta$,  are defined at the high scale $M_{\rm GUT}$, the scale where the coupling constants unify. This model has 19 free parameters. The model is very closely related to the pMSSM as defined above, and indeed the only difference is that in the pMSSM the parameters are defined at the low-energy SUSY breaking scale instead of the GUT scale. Since the number of LS parameters is the same as the number of HS ones, we can study the impact of the RGE running on the FT by defining the matching conditions at the HS. 
\end{itemize}
In what follows, we will compute the minimal possible amount of FT that each of these models has after imposing all experimental constraints. To compute the amount of FT, we will use the EW FT measure and the BG FT measure, which are explained in more detail below. 

\subsection{The electroweak fine-tuning measure}
\label{sec:ewft}
The EW FT measure ($\Delta_{\rm EW}$) was first proposed in Ref. \cite{PhysRevLett.109.161802}. It parameterizes how sensitive $M_Z$ (Eq.~(\ref{eq:mz})) is to variations in each of the coefficients $C_i$ (as defined below). The measure is defined as
\begin{equation}
\label{eq:FT}
\Delta_{\rm EW} \equiv \max_i \left\lvert\frac{C_i}{M_Z^2/2}\right\rvert,
\end{equation}
where the $C_i$ are
\begin{align*}
C_{m_{H_d}} &= \frac{m_{H_d}^2}{\tan^2\beta-1},\hspace{1em} C_{m_{H_u}} =  \frac{-m_{H_u}^2\tan^2\beta}{\tan^2\beta-1},\hspace{1em}  C_{\mu} = -\mu^2, \\
C_{\Sigma_d^d} &= \frac{\max(\Sigma_d^d)}{\tan^2\beta-1},\hspace{1em} C_{\Sigma_u^u} = \frac{-\max(\Sigma_u^u)\tan^2\beta}{\tan^2\beta-1}.
\end{align*}
The tadpole contributions $\Sigma^u_u$ and $\Sigma^d_d$ contain a sum of different contributions. All these contributions are computed individually and the maximum of these contributions is used to compute the $C_{\Sigma_u^u}$ and $C_{\Sigma_d^d}$ coefficients. 

\subsection{The Barbieri-Giudice measure}
\label{sec:bgft}
Another widely used measure is the BG measure proposed in Refs. \cite{Ellis:1986yg, Barbieri:1987fn}:
\begin{eqnarray}
\label{eq:BG}
\Delta_{\rm BG} &\equiv& {\rm max}|\Delta_p| \\
\Delta_p &\equiv& \frac{\partial \ln M_Z^2}{\partial \ln p_i},
\end{eqnarray}
where $p_i$ is one of the independent input parameters of the SUSY model. These input parameters can be defined at {\it any} scale. When the input parameters are defined at $M_{\rm SUSY}$ we will use the notation $\Delta_{\rm BG}^{\rm LS}$. On the other hand, when the input parameters are defined at $M_{\rm GUT}$, which will be the case for all of the HS models that we will be considering in this paper, we will use the notation $\Delta_{\rm BG}^{\rm HS}$.  Note that, in contrast to $\Delta_{\rm EW}$, the BG FT measure \emph{does} take dependencies between the theory parameters into account. To use the BG FT measure, the LS parameters  $m_{H_{u/d}}$ and $\mu$ that appear in Eq. (\ref{eq:mz}) need to be expressed into the fundamental parameters of the assumed SUSY model. These two quantities are related by renormalization group equations (RGEs), which can be solved numerically. The dependence of the LS parameters on the input parameters take the form of \cite{Casas:2014eca,Delgado:2014vha}
\begin{eqnarray}
m^2_{H_u}(M_{\rm SUSY}) &=& c_{M_1^2}M_1^2 + c_{M_2^2}M_2^2  + c_{M_3^2}M_3^2 + c_{M_1 M_2}M_1M_2+\dots \nonumber\\
&& +c_{A_t^2}A_t^2+c_{A_b^2}A_b^2+\dots+c_{A_tM_3}A_tM_3+\dots \nonumber\\
&& +c_{m_{H_u}^2}m_{H_u}^2+c_{m_{\tilde{Q}_3}^2}m_{\tilde{Q}_3}^2  +c_{m_{\tilde{t}_R}^2}m_{\tilde{t}_R}^2  +c_{m_{\tilde{b}_R}^2}m_{\tilde{b}_R}^2  + \dots 
\label{eq:mhu}\\
m^2_{H_d}(M_{\rm SUSY}) &=& c_{M_1^2}M_1^2 + c_{M_2^2}M_2^2  + c_{M_3^2}M_3^2 + c_{M_1 M_2}M_1M_2+\dots, \nonumber \\
&& +c_{A_t^2}A_t^2+c_{A_b^2}A_b^2+\dots+c_{A_tM_3}A_tM_3+\dots \nonumber \\
&& +c_{m_{H_d}^2}M_{m_d}^2+c_{m_{\tilde{Q}_3}^2}m_{\tilde{Q}_3}^2  +c_{m_{\tilde{t}_R}^2}m_{\tilde{t}_R}^2  +c_{m_{\tilde{b}_R}^2}m_{\tilde{b}_R}^2  + \dots 
\label{eq:mhd} \\
\mu(M_{\rm SUSY}) &=& c_{\mu}\mu,
\label{eq:mu}
\end{eqnarray}
where the dots stand for similar contributions to the LS parameter as the ones already denoted. The parameters on the right-hand sides of these equations are input parameters defined at either $M_{\rm SUSY}$ (in the case of the pMSSM) or at a HS. The numerical value of the coefficients $c_i$ depend the values of the SM matching parameters (coupling constants and masses), the scale at which the SUSY input parameters are defined, $\tan\beta$ and the SUSY breaking scale.

\subsubsection{Dependence on high-scale model assumptions}
In this subsection, we will see an example of how the chosen set of fundamental parameters impact $\Delta_{\rm BG}^{\rm HS}$. To this end, consider $m_{H_u}^2(M_{\rm SUSY})$ of Eq. (\ref{eq:mhu}), where we now explicitly show the value of the numerical coefficients using $M_{\rm SUSY}=1$ TeV, $\tan\beta$=10, a high-scale value of $10^{16}$ GeV and the usual values for the SM input parameters \cite{Casas:2014eca,Delgado:2014vha}
\begin{eqnarray}
m_{H_u}^2(1 {\text{ TeV}}) &=& -1.603 M_3^2 + 0.203 M_2^2 + 0.006 M_1^2 - 0.005 M_1 M_2 - 0.02 M_1 M_3 - 0.134 M_2 M_3 \nonumber\\
&&- 0.109 A_t^2 + 0.012 A_t M_1 + 0.068 A_t M_2+ 0.285 A_t M_3 + 0.001 A_b^2 - 0.002 A_b M_3 \nonumber\\
&& + 0.631 m_{H_u}^2 - 0.367 m_{\tilde{Q}_3}^2 - 0.025 m_{\tilde{Q}_2}^2 - 0.025 m_{\tilde{Q}_1}^2  - 0.290 m_{\tilde{t}_R}^2 \nonumber\\
&& + 0.054 m_{\tilde{c}_R}^2 + 0.054 m_{\tilde{u}_R}^2  - 0.024 m_{\tilde{b}_R}^2 - 0.025 m_{\tilde{s}_R}^2  - 0.025 m_{\tilde{d}_R}^2  + 0.026 m_{H_d}^2 \nonumber\\
&& - 0.026 m_{\tilde{\tau}_R}^2 - 0.026 m_{\tilde{\mu}_R}^2 - 0.026 m_{\tilde{e}_R}^2 + 0.025 m_{\tilde{L}_3}^2 + 0.025 m_{\tilde{L}_2}^2 + 0.025 m_{\tilde{L}_1}^2 \nonumber\\
&& +\dots,
\end{eqnarray}
where the dots indicate less important contributions. Here, one can observe that the gluino mass parameter $M_3$ and the soft SUSY breaking mass parameters $m_{\tilde{Q}_3}$, $m_{\tilde{t}_R}$ and $m_{H_u}$ will greatly contribute to the value of  $\Delta_{\rm BG}^{\rm HS}$, as their coefficients are relatively large. To demonstrate the dependence of $\Delta_{\rm BG}^{\rm HS}$ on the input parameters, consider the contribution of $m_{\tilde{Q}_3}$ to the FT
\begin{eqnarray}
\Delta_{m_{\tilde{Q}_3}} &=& \frac{m_{\tilde{Q}_3}}{M_Z^2}\frac{\partial M_Z^2}{\partial m_{\tilde{Q}_3}} \\
&\simeq& 2\frac{m_{\tilde{Q}_3}}{M_Z^2}\frac{\partial M_{H_u}^2}{\partial m_{\tilde{Q}_3}} = 4c_{m_{\tilde{Q}_3}^2}\frac{m_{\tilde{Q}_3}^2}{M_Z^2} = -1.468\frac{m_{\tilde{Q}_3}^2}{M_Z^2}. \nonumber
\end{eqnarray}
Here, one sees that a large value for $m_{\tilde{Q}_3}$ at the GUT scale will automatically lead to a large value for $\Delta_{m_{\tilde{Q}_3}}$. The same holds true for $M_3$, $m_{\tilde{t}_R}$ and $m_{H_u}$, hence the wide-spread assumption that gluinos and stops should be light to avoid large fine-tunings. Note that here one can already observe the strong dependence of $\Delta_{\rm BG}^{\rm HS}$ on the model assumptions, as there is a factor of $2$ difference in taking $m_{\tilde{Q}_3}$ or $m_{\tilde{Q}_3}^2$ as a fundamental (input) parameter
\begin{eqnarray}
\Delta_{m_{\tilde{Q}_3}^2} = \frac{m_{\tilde{Q}_3}^2}{M_Z^2}\frac{\partial M_Z^2}{\partial m_{\tilde{Q}_3}^2} \simeq  -0.734\frac{m_{\tilde{Q}_3}^2}{M_Z^2}. 
\end{eqnarray}
This shows that the overall amount of FT is reduced by a factor of $2$ if the input parameters squared are taken as fundamental parameters. In the foregoing, we have assumed that all pMSSM parameters are {\it independent} at the high scale. To explicitly see how parameter dependencies impact the value for $\Delta_{\rm BG}^{\rm HS}$, we can impose a relation between several input parameters
\begin{eqnarray*}
m_{\tilde{Q}_3} = m_{H_u} = m_{H_d} = m_{\tilde{t}_R} = \dots &\equiv& m_0, \\
M_1 = M_2 = M_3 &\equiv&M_{1/2}, \\
A_t = A_b &\equiv& A_0.
\end{eqnarray*}
This reduces the numerical values of the RGE coefficients. Using the code of Ref. \cite{Casas:2014eca}, we now obtain
\begin{eqnarray}
\label{eq:RGEs}
m_{H_u}^2(1 {\rm TeV}) &=& -2.125 M_{1/2}^2 - 0.099 A_0^2 + 0.382 M_{1/2}A_0 -0.087 m_0^2.
\end{eqnarray}
The contribution stemming from the scalar soft SUSY breaking masses to $\Delta_{\rm BG}^{\rm HS}$ is now greatly reduced compared to the previous example where all parameters were independent, as the highest value for all scalar coefficients (previously $c_{m_{H_u}}$) drops with a factor of $\sim 7$ in the new coefficient $c_{m_0}$. Of course, this unification is well known and exactly defines the unification assumed in the mSUGRA model as described in the previous section \cite{Chamseddine:1982jx,Barbieri:1982eh,Ohta:1982wn,Hall:1983iz}.\\
The goal of the foregoing was to show that  $\Delta_{\rm BG}^{\rm HS}$ will depend greatly on the chosen parameter dependencies {\it and} the chosen set of fundamental parameters. This means that two different high-scale SUSY models with {\it exactly} the same mass spectra can lead to radically {\it different} values of $\Delta_{\rm BG}^{\rm HS}$, depending on the HS model assumptions\cite{PhysRevD.89.115019,Baer:2013gva,Mustafayev:2014lqa}. Furthermore, in ref. \cite{Buckley:2016tbs} it was shown that conclusions drawn from $\Delta_{\rm BG}^{\rm HS}$ (or  $\Delta_{\rm BG}^{\rm LS}$) are very sensitive to the order of accuracy of the RGE equations.  \\
The EW FT measure $\Delta_{\rm EW}$ as given in Eq.~(\ref{eq:FT}) is not affected by these assumptions and should therefore be seen as a more conservative measure, although it is strictly only applicable to an LS SUSY model such as the pMSSM. It is evaluated from weak scale parameters containing no information on a possible HS theory. Therefore it gives only an indicative FT value for a given EW spectrum interpreted in the pMSSM. We can interpret a small value of $\Delta_{\rm EW}$ for a given SUSY spectrum as a minimal necessary condition of a natural SUSY model, but it is not sufficient. The real value of FT will depend on the exact parameter conditions that are present for the HS SUSY model and we will see in Sec. \ref{sec:pMSSMGUT} that this can be either higher {\it or lower} for some spectra. Note that the two measures can also agree for specific HS model assumptions and values for the input parameters. This may happen for example when both the EW and BG FT measures are dominated by the value of $\mu$. As the value for $c_{\mu}$ is close to one, in this case $\Delta_{\rm EW} \simeq \Delta_{\rm BG}^{\rm HS}$ if $\mu^2$ is chosen as a fundamental parameter.

\subsection{Concluding remarks}
In this section we have shown the differences between two popular FT measures used to quantify FT in SUSY models: $\Delta_{\rm EW}$ (Eq.~(\ref{eq:FT})) and  $\Delta_{\rm BG}^{\rm HS}$ (Eq.~(\ref{eq:BG})). The FT measure $\Delta_{\rm BG}^{\rm HS}$ suffers from model dependence more than  $\Delta_{\rm EW}$, as the former is extremely sensitive to which HS model and set of fundamental parameters is chosen. This leads to the confusing notion that different models with exactly the same LS spectra can give rise to different values for $\Delta_{\rm BG}^{\rm HS}$. As we do not know the exact SUSY breaking mechanism or HS model assumptions, we should be careful in using  $\Delta_{\rm BG}^{\rm HS}$ to construct natural mass ranges for sparticles or conclude anything about the exclusion of a natural realization of the MSSM. On the other hand,  $\Delta_{\rm EW}$ suffers from the fact that it can only indicate a conservative value for the FT in a given LS spectrum. To compute the actual amount of FT, one needs to construct a HS theory that can give rise to the same LS spectrum and recompute the FT taking into account all parameter dependencies. To clearly highlight the different conclusions one can draw using both FT measures, in what follows we will minimize these measures in the four different SUSY GUT scenarios that have been described in Sec. \ref{sec:susymodels}, taking into account all current constraints and future experiments, which will be described in the following section. 

\section{Analysis setup}
\label{sec:analysis}
Already a minimal model such as the pMSSM has a very rich phenomenology. Therefore, it is necessary to intelligently scan the parameter space, probing it for interesting regions, which could be missed if one adopts a random scan using, for instance, flat priors, which is often done in the fine-tuning literature. We use the Gaussian particle filter \cite{1232326} to tackle this problem. This scanning algorithm starts off by collecting an initial seed of randomly generated points. Then, in an iterative procedure, the best-fit points of the foregoing iteration are used as seeds to sample new model points, where a multi-dimensional Gaussian distribution is used around each seed parameter. The width of the Gaussian distribution in a specific dimension is chosen to be a variable fraction times the value of the seed point in that dimension. The fraction depends on the stage of the iteration. In the beginning, the fraction is chosen to be large ($1-2$) in order to be sensitive to a wide range of values. However, if one finds an interesting and possibly narrow region in the parameter space, the fraction needs to be reduced in order to efficiently probe it. In total, we have generated around $\mathcal{O}(100)$ million spectra for each GUT scale model. \\ 
To create the SUSY spectra we use \textsc{SoftSUSY} 4.0 \cite{Allanach:2001kg}, while the Higgs mass is calculated using FeynHiggs 2.14.2 \cite{Bahl:2016brp, Hahn:2013ria,Frank:2006yh,Degrassi:2002fi, Heinemeyer:1998yj}. We only select models that have the lightest neutralino $\tilde{\chi}^0_1$ as lightest supersymmetric particle (LSP) and discard spectra with tachyons or that do not satisfy the EWSB conditions. \textsc{SUSYHIT} \cite{Djouadi:2006bz} is used to calculate the decay of the SUSY and Higgs particles. \\
\textsc{MicrOMEGAs} 4.3.4 \cite{Barducci:2016pcb} is used to compute several flavor variables, the muon anomalous magnetic moment, the dark matter relic density\footnote{The computed values for $\Omega h^2_{\rm DM}$ were cross-checked with \textsc{MicrOMEGAs} 5.0.9 \cite{Belanger:2018mqt}. }($\Omega_{\rm DM} h^2$), the present-day velocity-weighted annihilation cross section ($\langle \sigma v \rangle$) and the spin-dependent and spin-independent WIMP-nucleon scattering cross sections ($\sigma_{\rm SD}$ and $\sigma_{\rm SI}$). The constraints on the WIMP-nucleon scattering cross sections stemming from various dark matter direct detection (DMDD) experiments are computed using \textsc{DDCalc} 2.0.0 \cite{ddcalc:2017lvb} where the 2018 and 2019 results from XENON1T \cite{Aprile:2018dbl,Aprile:2019dbj}, the 2017 and 2019 limits from PICO \cite{Amole:2017dex,Amole:2016pye,Amole:2019fdf} and the 2018 limits from PandaX \cite{Tan:2016zwf,Xia:2018qgs} are implemented. For DM indirect detection we only consider the limit on $\langle \sigma v \rangle$ stemming from the observation of gamma rays originating from dwarf galaxies, which we implement as a hard cut on each of the channels reported on the last page of Ref. \cite{Ackermann:2015zua}. Other DM indirect detection experiments are not taken into account as they are found not to constrain the DM properties any further. We use the central values of \textsc{MicrOMEGAs} 4.3.4 for the nuclear form factors, the DM local density and velocity. We allow for a multi-component DM, therefore the DM direct detection limits are rescaled by $f=\frac{\Omega h^2_{\rm DM}}{\Omega h^2_{\rm Planck}}$ (or $f^2$ in the case of indirect detection) if the dark matter relic abundance is less than the observed value $\Omega h^2_{\rm Planck}=0.120\pm0.001$ \cite{Aghanim:2018eyx}.\\
We use \textsc{SUSY-AI} to determine the exclusion of a model point in the pMSSM parameter space based on the ATLAS 13 TeV results \cite{Caron:2016hib, Barr:2016sho}. To cross check the \textsc{SUSY-AI} results, we have used \textsc{SModelS} \cite{Ambrogi:2018ujg,Heisig:2018kfq,Dutta:2018ioj,Ambrogi:2017neo,Kraml:2013mwa} for a selection of the models that have the lowest FT.
\textsc{HiggsBounds} 5.1.1 is used to determine whether the SUSY models satisfy the LEP, Tevatron and LHC Higgs constraints \cite{Bechtle:2015pma,Bechtle:2013wla,Bechtle:2013gu, Bechtle:2011sb,Bechtle:2008jh,Stal:2013hwa,Bechtle:2014ewa,Bechtle:2013xfa}.  Vevacious \cite{Camargo-Molina:2013qva,Lee2008,Wainwright:2011kj} is used to check that the models do not have a color/charge breaking minimum and have at least a meta-stable minimum that has a lifetime that exceeds that of our Universe. \\

\noindent We apply the following cuts on the values for certain masses and flavor observables:
\begin{itemize}
\item LEP limits on the masses of the chargino ($m_{\tilde{\chi}^{\pm}_1}~>~103.5$~GeV) and light sleptons ($m_{\tilde{l}}~>~90$~GeV) \cite{LEP:working}. For the staus we use a limit of $m_{\tilde{\tau}}~>~85$~GeV.
\item Constraints on the invisible and total width of the $Z$-boson  ($\Gamma_{Z, {\rm inv}}~=~499.0~\pm~1.5$~MeV and $\Gamma_{Z}~=~2.4952~\pm~0.0023$~GeV) \cite{Carena:2003aj}.
\item The lightest Higgs boson is required to be in the mass range of $122$ GeV $\leq m_{h_0} \leq 128$ GeV. 
\item An upper bound on the muon anomalous magnetic dipole moment $\Delta(g-2)_{\mu} < 40\times 10^{-10}$ is required, taking into account the fact that the SM prediction lies well outside the experimentally obtained value: $(24.9\pm 6.3)\times 10^{-10}$ \cite{Roberts:2010cj}.
\item Measurements of the $B/D$-meson branching fractions Br$(B_{s}^0~\rightarrow~\mu^+\mu^-)$ \cite{Aaij:2013aka},\\ Br$(\bar{B}~\rightarrow~X_s~\gamma)$ \cite{Misiak:2015xwa, Czakon:2015exa}, Br($B^+~\rightarrow~\tau^+~\nu_{\tau}$) \cite{Kronenbitter:2015kls}, Br$(D_s^+~\rightarrow~\mu^+~\nu_{\mu})$ \cite{Widhalm:2007ws} and Br($D^+_s~\rightarrow~\tau^+~\nu_{\tau})$ \cite{Onyisi:2009th}.
\end{itemize}
For all observables (except for $m_{h_0}$ and $\Delta(g-2)_{\mu}$) we require the value to lie within a $3 \sigma$ interval from the observed value. \\

\noindent The EW FT measure $\Delta_{\rm EW}$ is calculated by computing the effective potential terms and determining the maximal contribution via Eq. (\ref{eq:FT}), using the code from Ref. \cite{vanBeekveld:2016hug}. Note that this code differs from the built-in function from \textsc{SoftSUSY} 4.0 in how it handles the tadpole terms. While the latter code sums up all contributions in the tadpole terms and then computes the FT, we take the maximum value of each term in the tadpole. This ensures that we don't have a large cancellation in e.g. the stop sector, which could result in a very low value for $\Delta_{\rm EW}$. An explicit example of this mechanism can be seen in Sec.~\ref{sec:pMSSMGUT}. The BG measure of Eq. (\ref{eq:BG}) is calculated via the procedure implemented in SoftSUSY 4.0\footnote{An error was found in the calculation of the tadpole contributions to the fine-tuning, so we corrected the error in our version of SoftSUSY. This was communicated with the authors and updated in newer versions of SoftSUSY. }.

\subsection{Implementation of the limits imposed by future experiments}
\begin{table}[t]
    \centering
    \begin{tabular}{|c|c|c|c|}
    \hline
    Particle & Mass cut (HL-LHC) & Mass cut (HE-LHC) & Mass cut (CLIC) \\
    \hline
    $\tilde{g}$ & 3.2 TeV  & 5.7 TeV & - \\
    $\tilde{t}_1$ &1.7 TeV & 3.6 TeV & - \\
    $\tilde{\chi}^{\pm}_1$ (higgsino) & 350 GeV & 550 GeV & 1.5 TeV\\
    $\tilde{\tau}_1$ & 730 GeV & 1.15 TeV & - \\
    \hline
    \end{tabular}
    \caption{Exclusion potential of the HL-LHC, HE-LHC and CLIC on various SUSY particles as implemented in this analysis. The values for the HL-LHC and HE-LHC are taken from Ref. \cite{CidVidal:2018eel}. For CLIC \cite{Charles:2018vfv} we simply divided the optimal energy reach of 3 TeV by 2 as a baseline. }
    \label{tab:masscuts}
\end{table}
In this study we will consider two kinds of future experiments: DMDD experiments and colliders, as these two classes of experiments will have the biggest impact on the pMSSM parameter space. \\
Future DMDD experiments that are considered are the LZ experiment \cite{Akerib:2015cja}, DARWIN \cite{Aalbers:2016jon,Schumann:2015cpa}, Darkside-50k \cite{Aalseth:2017fik} and the PICO-500 experiment \cite{pico500}.
The first three experiments are most useful for constraining the SI WIMP-neutron scattering cross sections, while PICO-500 is more sensitive to the SD WIMP-proton scattering cross section. 
These detectors and analyses are implemented in the code \textsc{DDCalc} \cite{Workgroup:2017lvb}, which is used to calculate the exclusion limits. The limits depend on $\sigma_{\rm SD}$ and $\sigma_{\rm SI}$, the dark matter local density and velocity and the nuclear form factors assumed. We do not consider future DM indirect detection experiments as their exclusion power is not as powerful as the DMDD experiments'. \\
The predicted exclusion reach for collider experiments is implemented simply as a mass cut on the relevant mass parameter. To set the exclusions reaches we follow Ref. \cite{CidVidal:2018eel}, where the sensitivity of the High-Luminosity (HL) and High-Energy (HE) phase of the LHC on SUSY particles is discussed. The mass limits that we will use are given in Table \ref{tab:masscuts}. Given the sensitivity of the higgsino mass parameter on the amount of FT, we also add the reach of the Compact Linear Collider (CLIC)\cite{Charles:2018vfv} as its maximal energy reach ($3$ TeV) divided by 2. Note that this is a simplified approach: in reality, the limits on the masses may be lower depending on the exact SUSY spectrum and the complexity of the decays of the SUSY particles. We stress that these simple mass limits are only an indication of how far the future HL and HE-LHC can {\it maximally} reach. For the electroweak sparticles, we only implement the higgsino mass limits in this analysis, as the value of the wino mass parameter has little impact on the amount of FT.

\section{Results}
\label{S:3}
In this section we report on the resulting minimum allowed amount of FT, using either the BG FT measure $\Delta_{\rm BG}^{\rm HS}$  (defined in Eq.~(\ref{eq:BG})) or the EW FT measure $\Delta_{\rm EW}$ (defined in Eq.~(\ref{eq:mz})). As explained in Sec.~\ref{S:1}, we will report our results for four different HS SUSY models, whose spectra all are embedded in the pMSSM. We will start with the mSUGRA model, then move on to mSUGRA-var, then consider NUHGM and finally look at the pMSSM-GUT model. We will also consider some phenomenology of the spectra that have the lowest FT for both measures, and their prospects to be probed at future experiments. Every sub-section will be structured in the following way: first, we will discuss the current status of $\Delta_{\rm BG}^{\rm HS}$. Then we move on to $\Delta_{\rm EW}$, combined with a discussion on the impact of current and future DMDD experiments. Every sub-section will end with a discussion on the future collider prospects. 
\subsection{mSUGRA}
\begin{figure}[t]
  \centering
  \includegraphics[trim=0 0 0 0,clip,width=\textwidth]{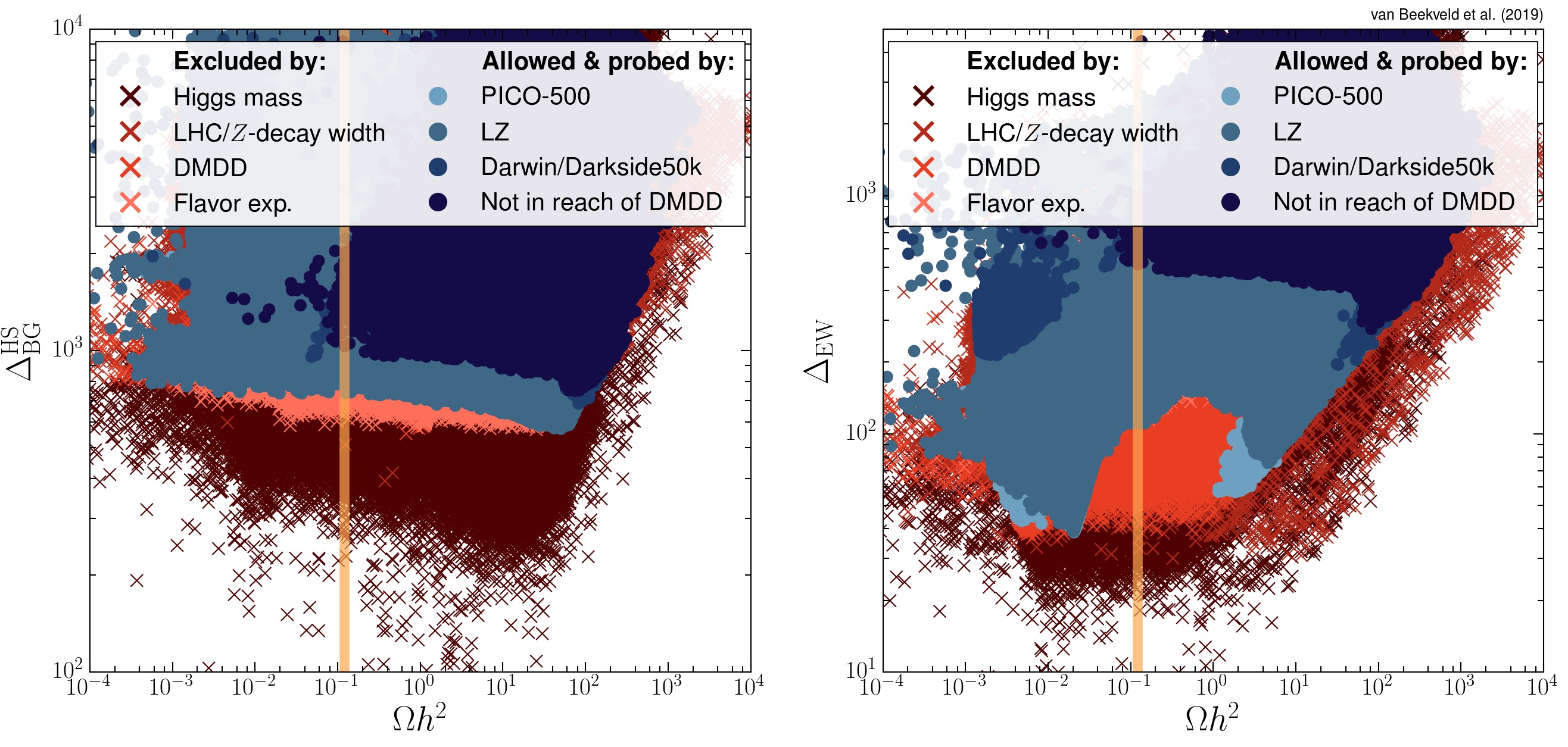}%
\caption{Generated spectra for the mSUGRA model. The left figure shows $\Delta_{\rm BG}^{\rm HS}$  as a function of $\Omega h^2$, the right figure shows $\Delta_{\rm EW}$  as a function of $\Omega h^2$. Red crosses indicate that the spectrum is excluded due to (in plotting order): the Higgs mass requirement (darkest red); limits placed on SUSY masses; by DMDD experiments; by limits placed on several flavor physics observables (lightest red) respectively. The constraints are also checked in this order. Circles in different shades of blue indicate the sensitivity of DMDD experiments, where in order of increasing brightness we indicate: PICO-500, LZ, Darwin/Darkside50k or unconstrained by DMDD experiments. Unconstrained spectra lie on top of more constrained and excluded spectra. The orange band indicates $\Omega h^2 = 0.12$. 
}
\label{fig:sugraomega}
\end{figure}

\begin{figure}[t]
  \centering
  \includegraphics[trim=0 0 0 0,clip,width=\textwidth]{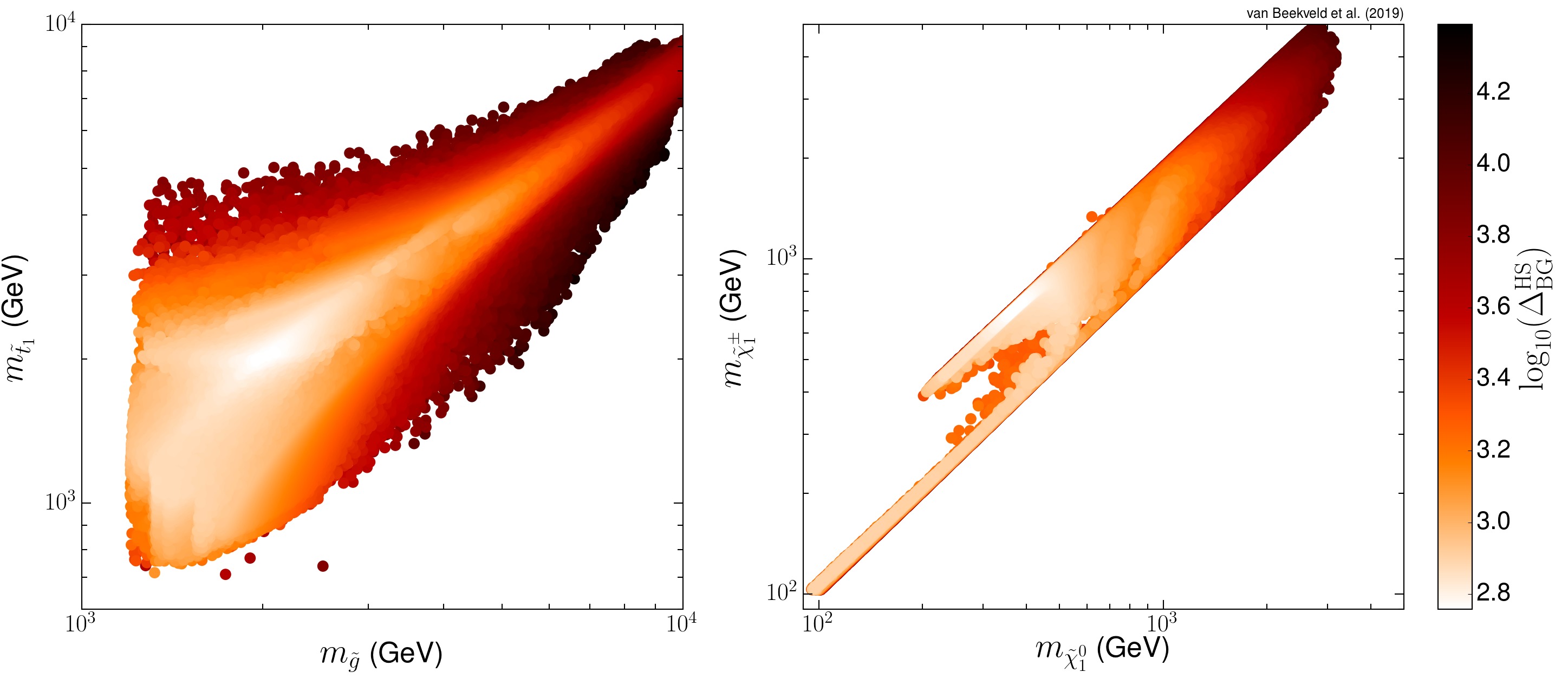}%
\caption{Left: the gluino mass ($m_{\tilde{g}}$) against the lightest stop mass ($m_{\tilde{t}_1}$) for mSUGRA, showing only the allowed spectra. Right: the lightest chargino mass ($m_{\tilde{\chi}^{\pm}_1}$) against the lightest neutralino mass ($m_{\tilde{\chi}^0_1}$) for the allowed spectra. All the masses are shown in units of GeV. The color code indicates the value for $\log_{10}(\Delta_{\rm BG}^{\rm HS})$. Spectra with lower values for $\log_{10}(\Delta_{\rm BG}^{\rm HS})$ lie on top of spectra with higher values.}
\label{fig:sugragluino}
\end{figure}

The resulting values for $\Delta_{\rm BG}^{\rm HS}$ and  $\Delta_{\rm EW}$ for all generated mSUGRA spectra are shown in Figure~\ref{fig:sugraomega} as a function of the dark matter relic density $\Omega h^2$. The lowest value for $\Delta_{\rm BG}^{\rm HS}$ is 571. The minimal value for $\Delta_{\rm BG}^{\rm HS}$ is constrained mainly by the Higgs mass requirement, and to a lesser extent by the limit placed on ${\rm Br}\left(B_{s} \rightarrow \mu^+ \mu^-\right)$. Dropping the Higgs mass requirement, while keeping all other constraints, would result in a value for $\Delta_{\rm BG}^{\rm HS}$ of about $240$. The region where $\Delta_{\rm BG}^{\rm HS}$ is minimized corresponds to values of $M_{1/2} \simeq 800$ GeV, $A_0 \simeq -3$ TeV and $m_0 \simeq 2.5$ TeV. The value for $\tan\beta$ is less constrained and lies between 10 and 50 in this region. There is a clear reason why these specific values for $m_0$, $A_0$ and $M_{1/2}$ are preferred. If the value for $m_0$ is lowered, $|\mu|$ needs to increase to still satisfy the EWSB requirement, so the amount of fine-tuning is increased because of $|\mu|$. If $m_0$ is increased, $\Delta_{\rm BG}^{\rm HS}$ increases due to $m_0$. The Higgs mass requirement prevents $M_{1/2}$ and $A_0$ to get lower, although lowering the absolute value of these two parameters would result in a lower $\Delta_{\rm BG}^{\rm HS}$. Another reason why low values of $M_{1/2}$ are not allowed is the limit placed on ${\rm Br}\left(B_{s} \rightarrow \mu^+ \mu^-\right)$, as it leads to a too light pseudo-scalar Higgs boson. Hence we find a special region where $\Delta_{\rm BG}^{\rm HS}$ is minimized, driven mainly by the observed value of the Higgs mass.  \\
This region can also be observed in Figure \ref{fig:sugragluino}, where on the left-hand side the gluino mass ($m_{\tilde{g}}$) is plotted against the lightest stop mass ($m_{\tilde{t}_1}$). The value for $\Delta_{\rm BG}^{\rm HS}$ is shown as a color code, whose minimum is reached for gluino and stop masses of $\mathcal{O}(2\text{ TeV})$. One observes that $\Delta_{\rm BG}^{\rm HS}$ increases for higher stop and gluino masses. Both $m_{H_u}$ and $m_{H_d}$ depend on the gaugino mass parameter $M_{1/2}$ with a large RGE coefficient. Therefore, $M_{1/2}$ needs to be as low as experimentally allowed in order to keep the value for $\Delta_{\rm BG}^{\rm HS}$ as small as possible. A higher value for $M_{1/2}$ will result in a higher value for both $m_{\tilde{g}}$ and $m_{\tilde{t}_1}$. Therefore, by increasing $m_{\tilde{g}}$ and $m_{\tilde{t}_1}$, one also sees that the value for $\Delta_{\rm BG}^{\rm HS}$ is increased. \\
In the neutralino-chargino sector, there is not much freedom for the allowed masses due to the unification of the gaugino masses, as can be seen on the right-hand side of Figure \ref{fig:sugragluino}. The gaugino mass parameter $M_{1/2}$ needs to be large at the GUT scale for the allowed spectra to satisfy the observed value for the Higgs boson mass, evade the gluino mass limits and the limit on ${\rm Br}\left(B_{s} \rightarrow \mu^+ \mu^-\right)$. The ratio of the gaugino masses at the SUSY scale is roughly $M_3 \simeq 2.7 M_2 \simeq 5 M_1$ due to the unification of the gaugino masses at the GUT scale, where the exact ratio depends on the numerical value of the GUT and SUSY scale. Due to this relation, $M_1$ is prevented to get lower than about $200$ GeV at the SUSY scale, otherwise, the gluino mass would also get too. LSPs with a mass around 100 GeV can then only be higgsino-like and are necessarily accompanied by a higgsino-like chargino with a similar mass. For slightly higher LSP masses, the LSP is also allowed to be bino-like. For these models, the chargino will be wino-like with a higher mass than the LSP. This explains the presence of the two hard lines in Figure \ref{fig:sugravaromega}. Spectra with mixed LSP compositions lie in-between these two hard lines. The value for $\Delta_{\rm BG}^{\rm HS}$ is minimized in the second region where the LSP is bino-like.

\subsubsection*{$\Delta_{\rm EW}$ and the impact of DMDD experiments}
\noindent While the observed value of the Higgs boson mass constrains the minimal value for $\Delta_{\rm BG}^{\rm HS}$, limits placed by current DMDD experiments constrain the minimal value for $\Delta_{\rm EW}$, which is 38. This can be seen on the right-hand side of Figure~\ref{fig:sugraomega}. The spectra that minimize $\Delta_{\rm EW}$ all feature a higgsino-dominated LSP with a negligible wino component, a small bino component ($<10\%$) and a mass of around $100-400$ GeV. These spectra result in values for $\Omega h^2$ around $10^{-3}-10^{-2}$, where $\Omega h^2$ increases with higher values for $m_{\tilde{\chi}^0_1}$ and/or a larger bino component of $m_{\tilde{\chi}^0_1}$. The size of the SD cross section is proportional to the $\tilde{\chi}^0_{1}\tilde{\chi}^0_{1}Z$ coupling, which is proportional to the difference between the two higgsino components of the LSP ($|N_{13}|^2 - |N_{14}|^2$). There is a higgsino asymmetry ($|N_{13}|\neq |N_{14}|$) for these higgsino-dominated LSPs, therefore the $\tilde{\chi}^0_{1}\tilde{\chi}^0_{1}Z$ coupling is generally high. For this reason, these spectra will be fully probed by future DMDD experiments, despite the suppression factor that they receive due to the fact that $\Omega h^2 < 0.12$. \\
Starting at $\Omega h^2 \sim 0.1$, the LSPs become bino dominated with a small higgsino component. These models correspond to the models in-between the lower and upper band in Figure \ref{fig:sugragluino}, where it can be seen that the LSPs have masses around $300-700$ GeV. In this regime, the DMDD experiments that are sensitive to either the SI or SD cross-sections constrain $\Delta_{\rm EW}$. At even higher values of $\Omega h^2$ the LSP will be a pure bino, so the DMDD experiments lose their sensitivity in this regime. The higgsino component keeps decreasing for higher $\Omega h^2$, which is the cause for the increase of $\Delta_{\rm EW}$ for $\Omega h^2 > 5$. The impact of future DMDD experiments is sizable, as these increase the minimal value for $\Delta_{\rm EW}$ to $275$. For models that have $0.09 < \Omega h^2 < 0.15$, the minimal value for $\Delta_{\rm EW}$ is around $515$. On the other hand, the minimal value of $\Delta_{\rm BG}^{\rm HS}$ can be increased to 750 by the reach of future DMDD experiments. 
 \\

\begin{figure}[t]
  \centering
  \includegraphics[trim=0 0 0 0,clip,width=\textwidth]{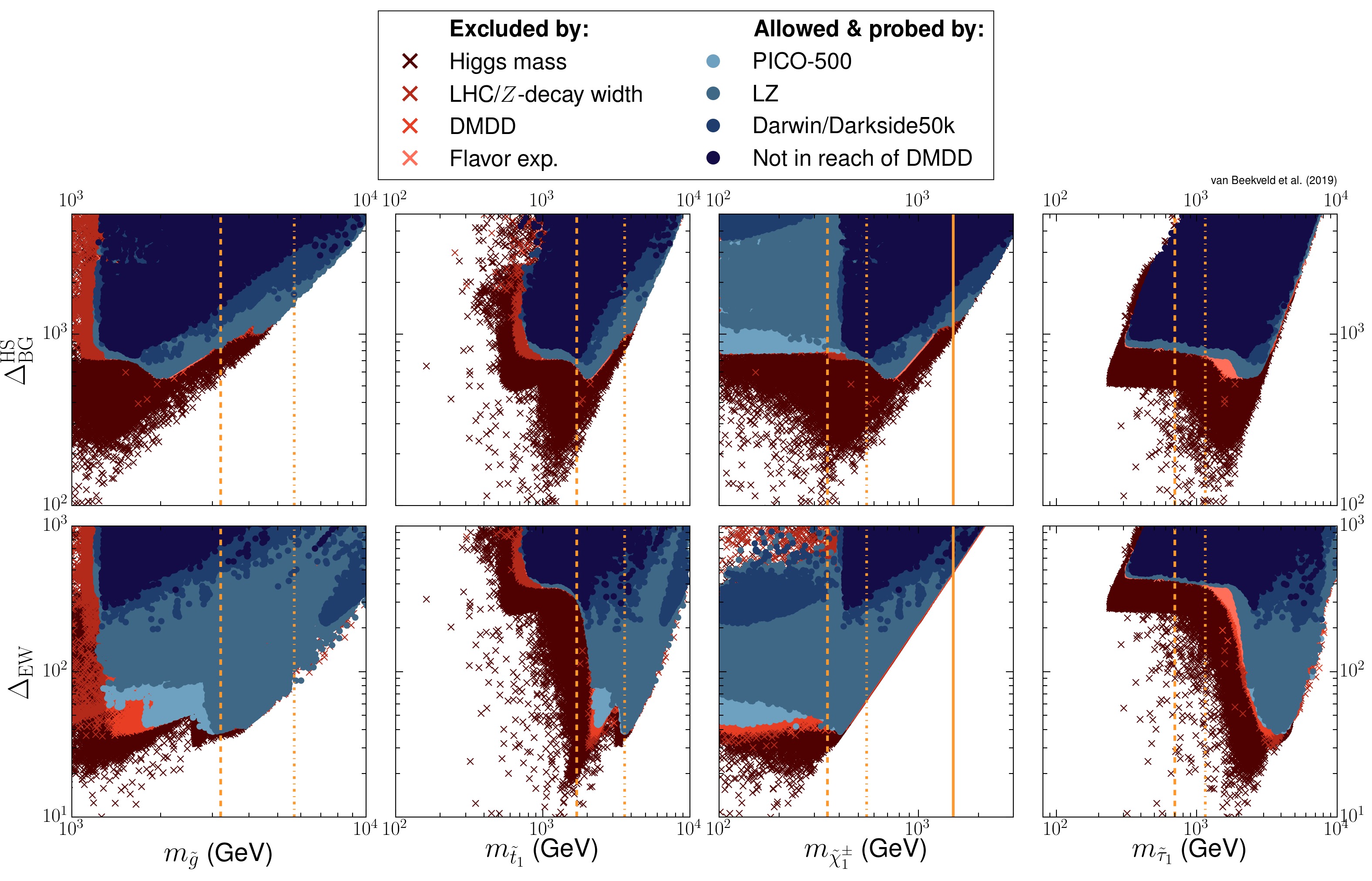}%
\caption{The values for $\Delta_{\rm BG}^{\rm HS}$ (top) and $\Delta_{\rm EW}$ (bottom) against (from left to right) the gluino mass ($m_{\tilde{g}}$), stop mass ($m_{\tilde{t}_1}$), chargino mass ($m_{\tilde{\chi}^{\pm}_1}$) and stau mass ($m_{\tilde{\tau}_1}$) for all generated mSUGRA spectra. The masses are given in units of GeV. The color code and plotting order is the same as in Figure \ref{fig:sugraomega}. The dashed, dash-dotted and solid orange line shows the exclusion potential of the HL-, HE-LHC and CLIC on the masses of various SUSY particles (see Table~\ref{tab:masscuts}). The solid orange line in the chargino mass plot shows the exclusion potential of CLIC.}
\label{fig:sugraFTvsPole}
\end{figure}
\subsubsection*{The impact of future collider experiments} 
\noindent The dependence of $\Delta_{\rm BG}^{\rm HS}$ and $\Delta_{\rm EW}$ on $m_{\tilde{g}}$, $m_{\tilde{t}_1}$, lightest chargino mass ($m_{\tilde{\chi}^{\pm}_1}$) and lightest stau mass ($m_{\tilde{\tau}_1}$)  is shown in Figure \ref{fig:sugraFTvsPole}. We also show the reach of the future HL-LHC, HE-LHC and CLIC experiments as a dashed, dash-dotted and solid orange line respectively. One observes that the HL-HLC can bring the minimal value for $\Delta_{\rm BG}^{\rm HS}$ from 571 to about 848 by its power to constrain $m_{\tilde{g}}$. Due to its increased energy reach, the HE-LHC can bring the minimal value for $\Delta_{\rm BG}^{\rm HS}$ to about 1700. The HE-LHC machine will be the most constraining for $\Delta_{\rm BG}^{\rm HS}$, as CLIC can constrain $\Delta_{\rm BG}^{\rm HS}$  to about 1240 in the case of a non-observation. The impact on $\Delta_{\rm EW}$ is less sizeable and has two origins, namely the exclusion reach of the HL and HE-LHC on both $m_{\tilde{g}}$ and $m_{\tilde{\chi}^{\pm}_1}$. The HL-LHC increases the minimal value for $\Delta_{\rm EW}$ to about $38$, while the HE-LHC increases the value to about $86$. The impact of CLIC on $\Delta_{\rm EW} $ is significant, as it can increase the current limit of $\Delta_{\rm EW}$ to about 530. 

\subsection{mSUGRA-var}

\begin{figure}[t]
  \centering
  \includegraphics[trim=0 0 0 0,clip,width=\textwidth]{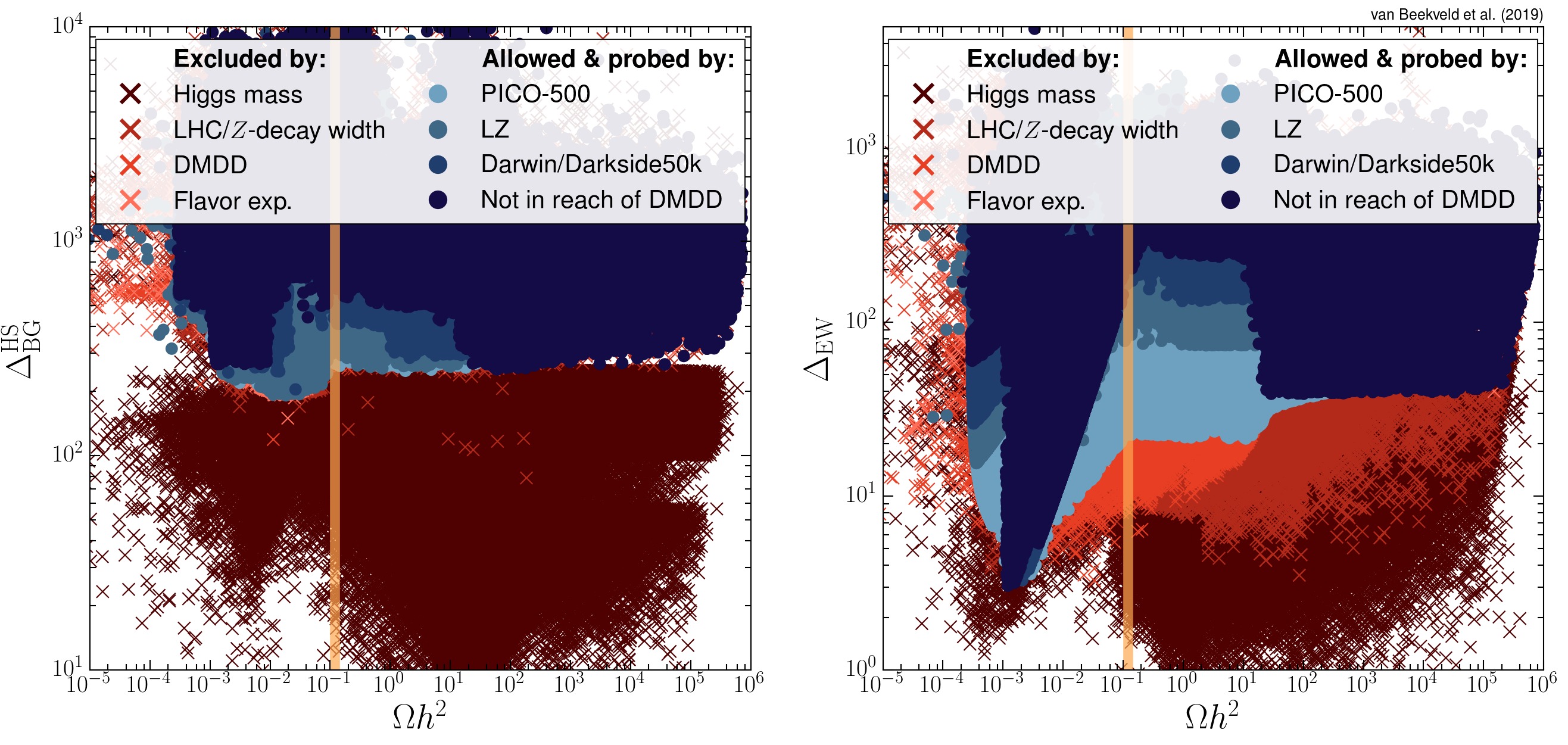}%
\caption{Generated spectra for the mSUGRA-var model. The left figure shows $\Delta_{\rm BG}^{\rm HS}$ as a function of $\Omega h^2$, the right figure shows $\Delta_{\rm EW}$ as a function of $\Omega h^2$. The color code and plotting order is the same as in Figure \ref{fig:sugraomega}. The orange band indicates $\Omega h^2 = 0.12$.}
\label{fig:sugravaromega}
\end{figure}

\begin{figure}[t]
  \centering
  \includegraphics[width=0.7\textwidth]{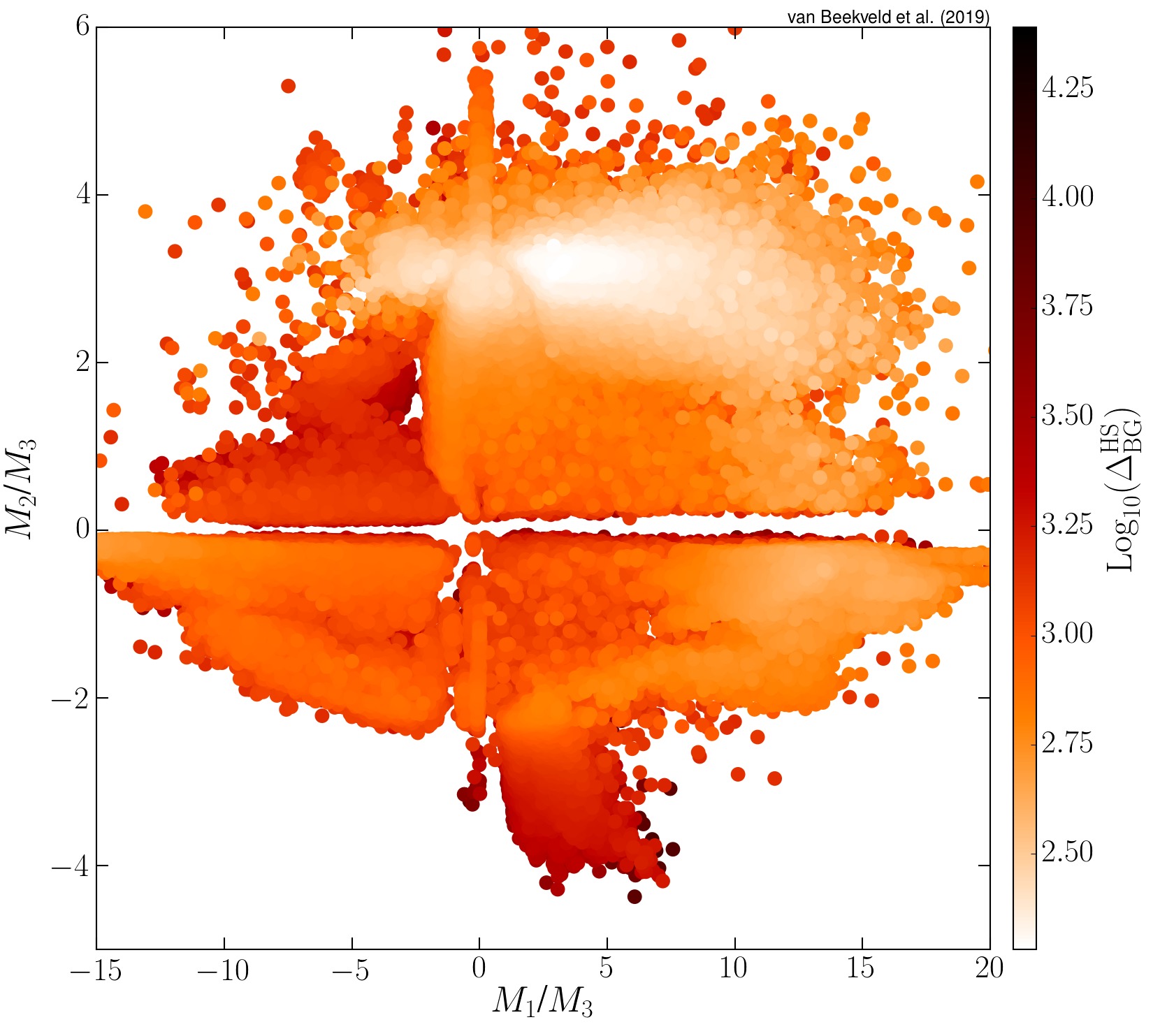}%
\caption{Ratios of $M_1/M_3$ and $M_2/M_3$ for the allowed mSUGRA-var spectra. The color code indicates the value for $\log_{10}(\Delta_{\rm BG}^{\rm HS})$. Spectra with lower values for $\log_{10}(\Delta_{\rm BG}^{\rm HS})$ lie on top of spectra with higher values.}
\label{fig:sugraf1f2}
\end{figure}

\begin{figure}[t]
  \centering
  \includegraphics[trim=0 0 0 0,clip,width=\textwidth]{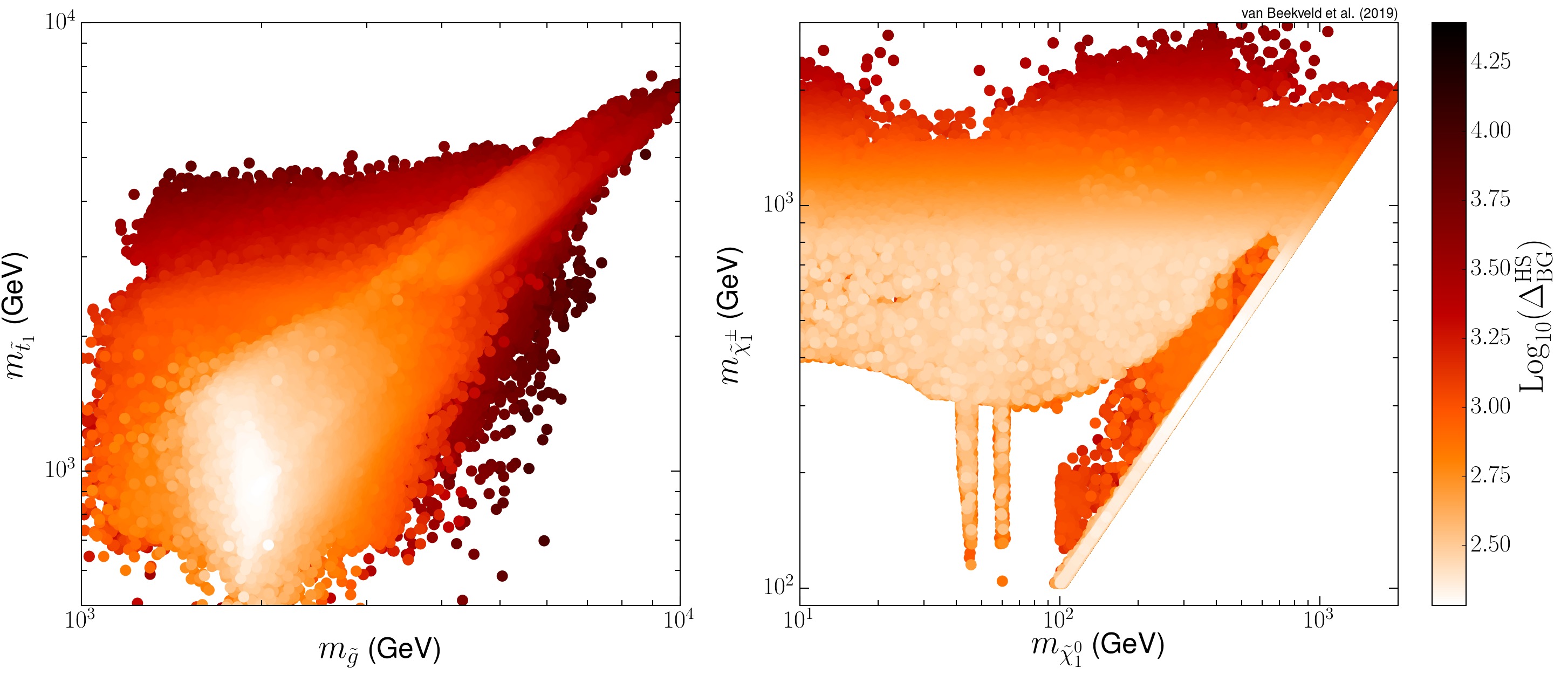}%
\caption{Left: $m_{\tilde{g}}$ against $m_{\tilde{t}_1}$ for mSUGRA-var, showing only the allowed spectra. Right: $m_{\tilde{\chi}^{\pm}_1}$ against $m_{\rm DM}$  for the allowed spectra. The masses are given in units of GeV. The color code indicates the value for $\log_{10}(\Delta_{\rm BG}^{\rm HS})$. Spectra with lower values for $\log_{10}(\Delta_{\rm BG}^{\rm HS})$ lie on top of spectra with higher values.}
\label{fig:sugravargluino}
\end{figure}

This model is closely related to the mSUGRA model that was considered in the previous section, but differs in the fact that it has more freedom in the gaugino sector. This feature is directly reflected in the minimal values for both of the FT measures. We will again first discuss  $\Delta_{\rm BG}^{\rm HS}$, whose value as a function of $\Omega h^2$ can be seen on the left-hand side of Figure~\ref{fig:sugravaromega}. The lowest allowed value for $\Delta_{\rm BG}^{\rm HS}$ is $191$, which is a decrease of around $400$ when compared to the mSUGRA model. The cause of this big decrease in $\Delta_{\rm BG}^{\rm HS}$ is that we treat $M_{1/2}$ as the only independent parameter in FT the computation, {\it and} at the same time let the ratios of $M_1$ and $M_2$ to $M_3$ be unconstrained. This shows an explicit example of the aforementioned dependence of $\Delta_{\rm BG}^{\rm HS}$ on the assumed dependencies that are present in the model. The Higgs mass requirement is again the strongest constraint for the minimum value of $\Delta_{\rm BG}^{\rm HS}$. Dropping this requirement, while keeping all of the other current collider and DM constraints, shows a decrease of $\Delta_{\rm BG}^{\rm HS}$ to a value of about $100$. Note that this effect cannot directly be seen in the left panel of Figure \ref{fig:sugravaromega}, as the spectra that are excluded because of the Higgs mass requirement can also be excluded by other constraints.\\
The optimal value for the ratio of $M_2$ to $M_3$ at the GUT scale is around 3, as can be seen in Figure \ref{fig:sugraf1f2}. The ratio of $M_1$ to $M_3$ is less constrained. This result can be understood by inspection of Eq.~(\ref{eq:RGEs}). For moderate values of $\tan\beta$, the biggest contribution to $\Delta_{\rm BG}^{\rm HS}$ comes from the sensitivity of $m_{H_u}$ on the input parameters. As can be seen in Eq.~(\ref{eq:RGEs}), the dependence of $m_{H_u}$ on the unified gaugino mass $M_{1/2}$ is minimal for $M_2 = f_2 M_3 \simeq 2.7 M_3$ for a GUT scale value of $10^{16}$ GeV, SUSY scale of $1$ TeV and a $\tan\beta$ of 10. This explains why we find $M_2/M_3 \simeq 3$ in our scan. The bino mass parameter $M_1$ has a very small RGE coefficient, hence we don't expect the ratio of $M_1$ to $M_3$ to influence the value for $\Delta_{\rm BG}^{\rm HS}$ by a big amount, which is indeed what is observed in the figure. \\
On the left-hand side of Figure \ref{fig:sugravargluino} it can be seen that $\Delta_{\rm BG}^{\rm HS}$ is again minimized in the region where $m_{\tilde{g}} \simeq 2$ TeV and $m_{\tilde{t}_1} \simeq 1$ TeV, which is happening for the same reason as in the mSUGRA model. On the right-hand side of Figure \ref{fig:sugravargluino}, one can observe that the LSP can drop below 100 GeV. This can happen due to the increased freedom in the gaugino sector compared to the mSUGRA model. These low-mass LSPs are necessarily all bino-like, otherwise,  $\tilde{\chi}^{\pm}_1$ would be excluded by LEP. One might wonder about the appearance of the two funnels around $m_{\tilde{\chi}^0_1}\simeq 45$ GeV and $m_{\tilde{\chi}^0_1}\simeq 65$ GeV. All spectra surrounding these funnels are excluded by DMDD experiments, since for these spectra the higgsino or wino component is generally too high. However, in the funnel regions, these components are allowed to increase as these spectra typically have values for $\Omega h^2$ that are less than $0.12$. This means that $m_{\tilde{\chi}^{\pm}_1}$ is allowed to be lower as well, which creates the funnels. The models that have $m_{\tilde{\chi}^{\pm}_1} \simeq m_{\tilde{\chi}^0_1}$ have an LSP that is nearly $100\%$ higgsino-like or wino-like. However, only $\tilde{\chi}^{\pm}_1$ and $\tilde{\chi}^{0}_1$ with small wino components feature in spectra that result in a low value of $\Delta_{\rm BG}^{\rm HS}$ have is $<5\%$, which is due to the preferred GUT scale ratio of $M_2 \simeq 2.7 M_3$. The spectra that minimize $\Delta_{\rm BG}^{\rm HS}$ all have a nearly-pure higgsino LSP with a mass around $200-500$ GeV.  


\subsubsection*{$\Delta_{\rm EW}$ and the impact of DMDD experiments}

\noindent The minimal value of  $\Delta_{\rm EW}$ is 3 (right panel of Figure \ref{fig:sugravaromega}), which is a factor of $10$ smaller than in the mSUGRA model. The origin of this decrease is again caused by the increase in freedom of the gaugino sector. The radiative corrections of $m_{H_u}$ in both mSUGRA and mSUGRA-var are mainly driven by $M_{1/2}$ through $M_3$. Therefore, the value for $M_{1/2}$ at the GUT scale is constrained by the EWSB conditions. By no longer demanding the unifying condition of $M_1 = M_2 = M_3$ at the GUT-scale, one allows $M_2$ and $M_1$ to decouple from $M_3$. The decoupling creates a region in parameter space where the LSP is a pure higgsino LSP that is much larger than the region in mSUGRA. The purity of the LSP depletes the $\tilde{\chi}^0_1 \tilde{\chi}^0_1 Z$ coupling, therefore these spectra are not excluded by the DMDD experiments. Precisely these spectra minimize the value of $\Delta_{\rm EW}$. \\
Allowed spectra that result in a very low value for $\Omega h^2 < 10^{-3}$ have a wino-like LSP. These models were absent in mSUGRA since the wino mass parameter $M_2$ will never drop below $M_1$ at the SUSY scale due to the HS unification constraint. The LSPs of the allowed spectra with $\Omega h^2 \simeq 10^{-4}$ and $\Delta_{\rm EW}\simeq 500$ have a mass around $100$ GeV. The mass of the wino LSP increases for higher values of $\Omega h^2$. For slightly higher values of  $10^{-3} < \Omega h^2 < 10^{-2}$, spectra appear that feature a pure higgsino LSP. The minimum of $\Delta_{\rm EW}$ is reached for spectra that have $\Omega h^2 \simeq 10^{-3}$, corresponding to a pure higgsino LSP with a mass of $\mathcal{O}(100)$ GeV. As the LEP limits prevent $m_{\chi^{\pm}_1}$ to get smaller than $103.5$ GeV, it is impossible to further minimize the value for $\Delta_{\rm EW}$. Future DMDD experiments are not able to constrain $\Delta_{\rm EW}$ in this regime.
We do see a large sensitivity of the current and future DMDD experiments on $\Delta_{\rm EW}$ for spectra that saturate the dark matter relic density exactly. The minimal still allowed value of $\Delta_{\rm EW}$ for these spectra is $20$, while future DMDD experiments increase this value to $126$. The spectra in this regime all have a bino-like LSP with a small higgsino component. 
The sensitivity of the future DMDD experiments decreases rapidly for $\Omega h^2 \gtrsim 10$. This is explained due to the fact that in this regime spectra appear that have a bino-like LSP with a very small mass $(<10 $ GeV). In this mass regime, the DMDD experiments loose their sensitivity. To escape the limits on the size of the invisible $Z$-decay width, these light bino LSPs must have a negligible higgsino component. This prevents $|\mu|$ to get too low, which puts a lower limit on the value for $\Delta_{\rm EW}$ of around $40$. Note that also collider experiments constrain the wino component for these spectra. To escape detection at the LHC, a wino-like $\tilde{\chi}^0_2$ or $\tilde{\chi}^{\pm}_1$ needs to be heavier than about $600$ GeV \cite{Aaboud:2018sua,Sirunyan:2018lul}.  Future DMDD experiments constrain the minimal value of $\Delta_{\rm BG}^{\rm HS}$ to $252$. If we only consider the spectra that result in $0.09 < \Omega h^2 < 0.15$, the future DMDD constrain $\Delta_{\rm BG}^{\rm HS}$ to $545$. 
\begin{figure}[t]
  \centering
  \includegraphics[width=\textwidth]{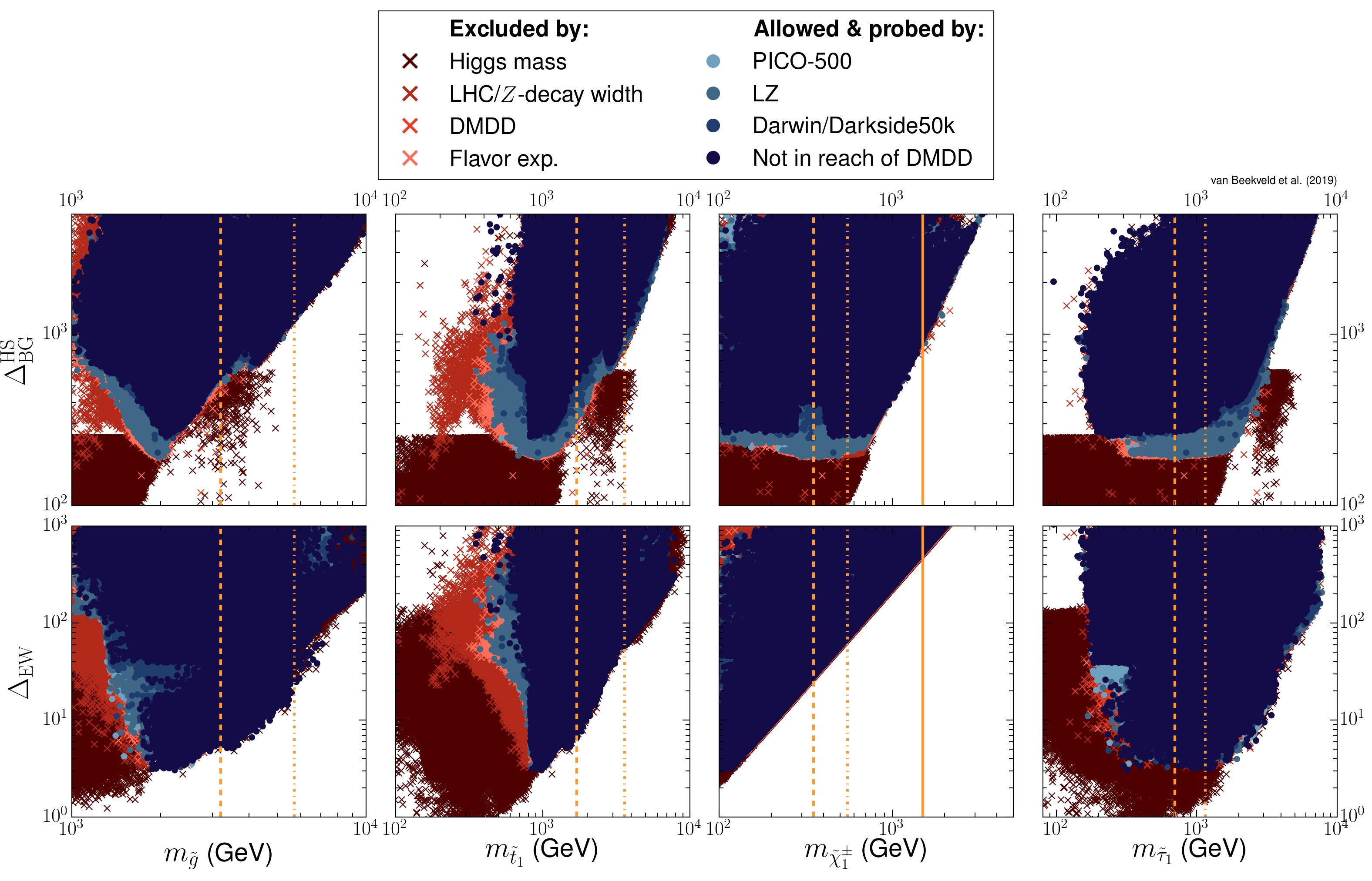}%
\caption{The values for $\Delta_{\rm BG}^{\rm HS}$ (top) and $\Delta_{\rm EW}$ (bottom) against (from left to right) $m_{\tilde{g}}$, $m_{\tilde{t}_1}$, $m_{\tilde{\chi}^{\pm}_1}$ and $m_{\tilde{\tau}_1}$ for all generated mSUGRA-var spectra. The masses are given in units of GeV. The color code and plotting order is the same as in Figure \ref{fig:sugraomega}. The dashed, dash-dotted and solid orange line shows the exclusion potential of the HL-, HE-LHC and CLIC on the masses of various SUSY particles  (see Table~\ref{tab:masscuts}). The solid orange line in the chargino mass plot shows the exclusion potential of CLIC. }
\label{fig:sugravarimpact}
\end{figure}

\subsubsection*{The impact of future collider experiments}
\noindent The impact of the HL-LHC, HE-LHC and CLIC on the two FT measures can be seen in Figure~\ref{fig:sugravarimpact}. Like in the mSUGRA model we observe that the impact on $\Delta_{\rm BG}^{\rm HS}$ is driven by the exclusion reach of $m_{\tilde{g}}$. The HL-LHC can increase the minimal value of $\Delta_{\rm BG}^{\rm HS}$ to about $530$, and the HE-LHC can constrain it to about $1220$. CLIC can constrain it to about 900 due to the sensitivity on $|\mu|$. The minimal value of $\Delta_{\rm EW}$ is not constrained by the exclusion reach of $m_{\tilde{g}}$, but only by the exclusion reach on $m_{\tilde{\chi}^{\pm}_1}$. As the HL-LHC is able to probe $m_{\tilde{\chi}^{\pm}_1}$ up to about $350$ GeV, $\Delta_{\rm EW}$ can be constrained to about $28$ in the event of a non-observation, whereas the HE-LHC can constrain $\Delta_{\rm EW}$ to about $70$. The exclusion reach of the HE-LHC on the stop mass increases the minimal value of $\Delta_{\rm EW}$ to about $60$. CLIC can constrain the value of $\Delta_{\rm EW}$ to about $530$.\\ 

\subsection{NUHGM}

\begin{figure}[t]
  \centering
  \includegraphics[width=\textwidth]{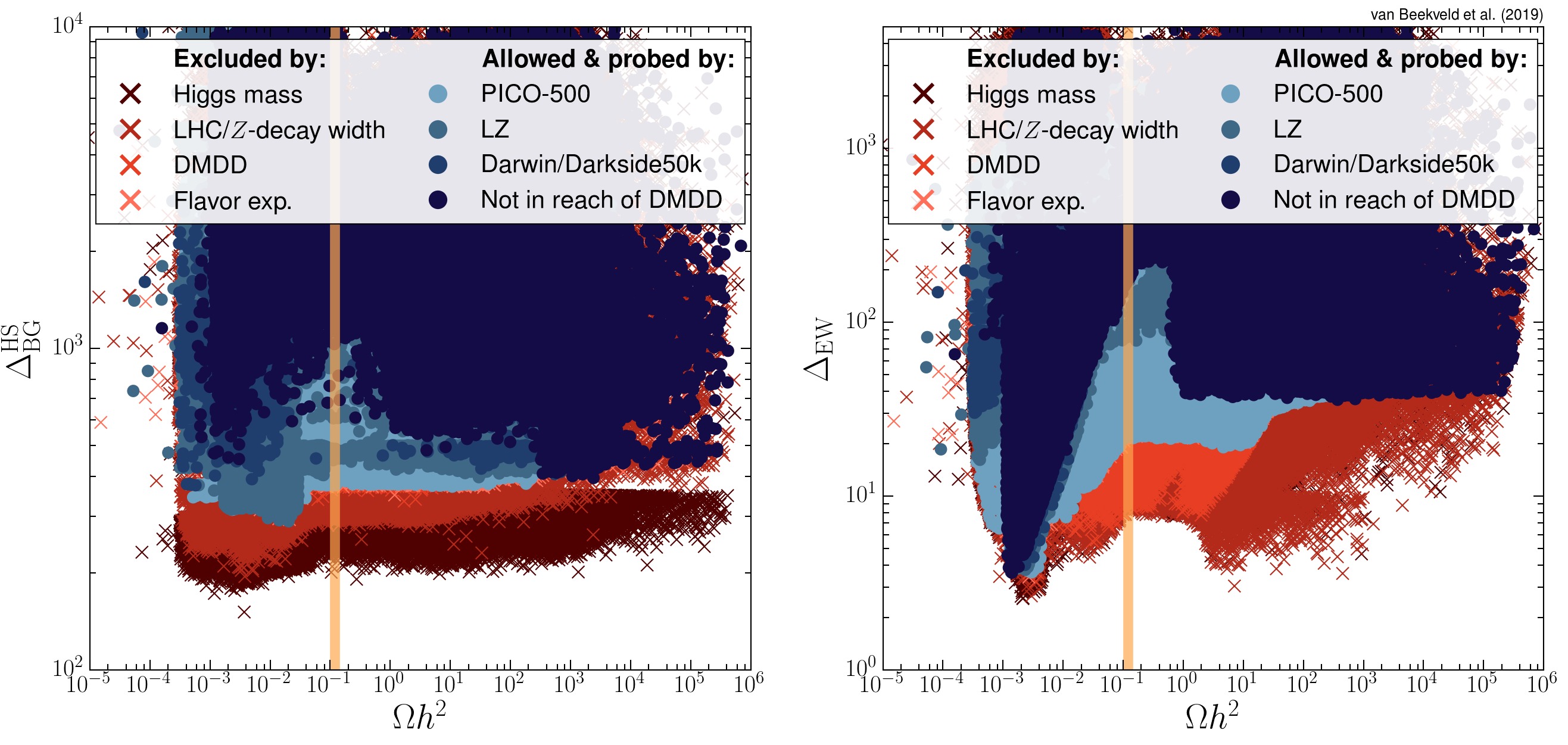}%
\caption{Generated spectra for the NUHGM model. The left figure shows $\Delta_{\rm BG}^{\rm HS}$ as a function of $\Omega h^2$, the right figure shows $\Delta_{\rm EW}$ as a function of $\Omega h^2$. The color code and plotting order is the same as in Figure \ref{fig:sugraomega}. The orange band indicates $\Omega h^2 = 0.12$.}
\label{fig:nuhm8omega}
\end{figure}
\begin{figure}[t]
  \centering
  \includegraphics[trim=0 0 0 0,clip,width=\textwidth]{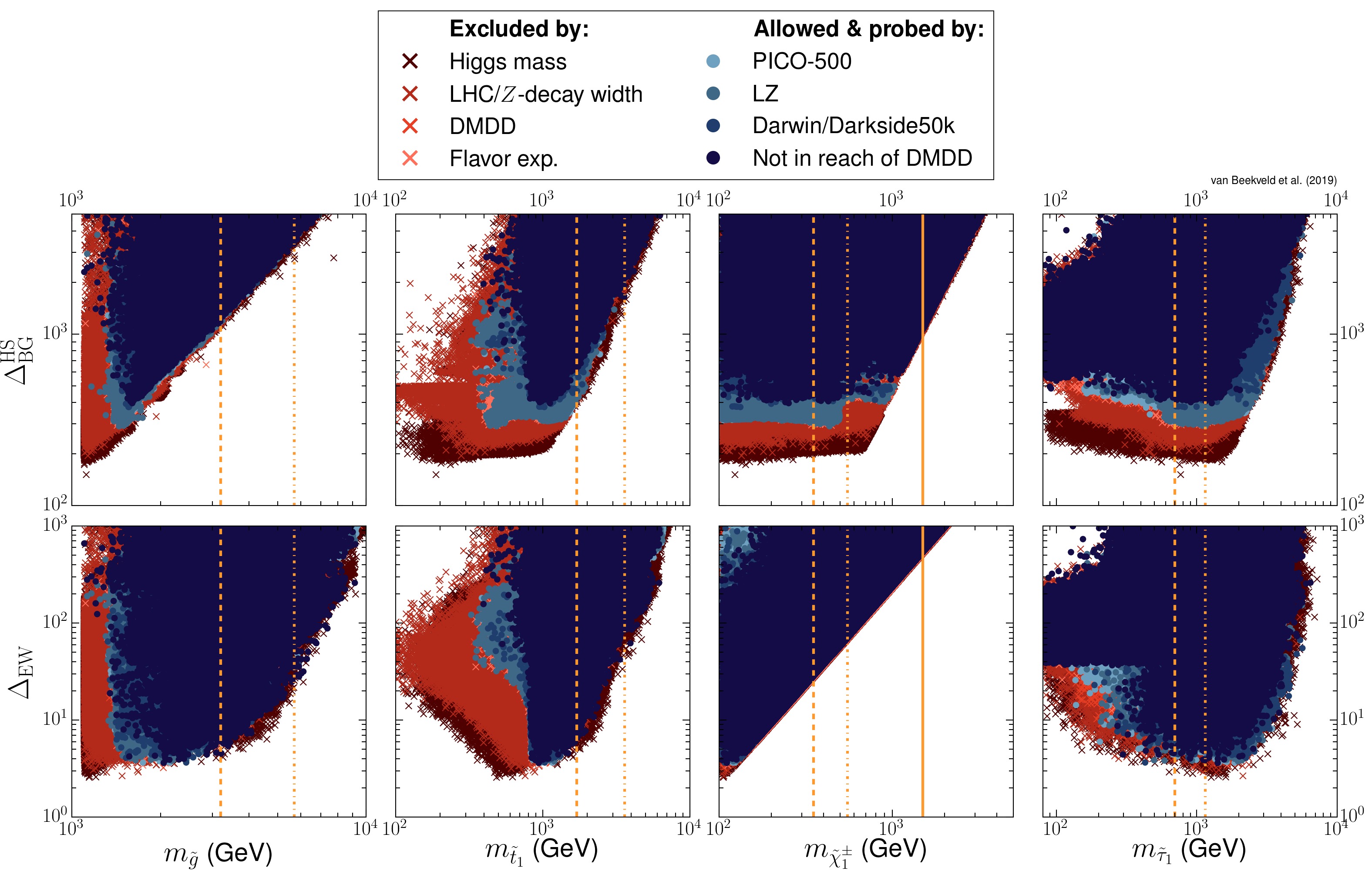}\caption{The values for $\Delta_{\rm BG}^{\rm HS}$ (top) and $\Delta_{\rm EW}$ (bottom) against (from left to right) $m_{\tilde{g}}$, $m_{\tilde{t}_1}$, $m_{\tilde{\chi}^{\pm}_1}$ and $m_{\tilde{\tau}_1}$ for all generated NUHGM spectra. The masses are given in units of GeV. The color code and plotting order is the same as in Figure \ref{fig:sugraomega}. The dashed, dash-dotted and solid orange line shows the exclusion potential of the HL-, HE-LHC and CLIC on the masses of various SUSY particles  (see Table~\ref{tab:masscuts}). The solid orange line in the chargino mass plot shows the exclusion potential of CLIC. }
\label{fig:nuhmFTvsPole}
\end{figure}

In this HS model we don't assume a relation between  $M_1$, $M_2$ and $M_3$, but instead treat them as free parameters. We furthermore increase the freedom of this model by having separate left and right-handed mass parameters for the sfermions and include $M_A$ and $\mu$ as input parameters. Although this model has more free parameters than the mSUGRA-var model, the resulting minimum for $\Delta_{\rm BG}^{\rm HS}$ is higher. The minimum value that we obtain for $\Delta_{\rm BG}^{\rm HS}$ is 290. The increase with respect to the mSUGRA-var model, where the minimum of $\Delta_{\rm BG}^{\rm HS}$ was found at 191, is due to the fact that now the ratios of $M_1$ and $M_2$ to $M_3$ are assumed to be independent at the HS. This indeed shows the dependence of $\Delta_{\rm BG}^{\rm HS}$ on the assumed HS model dependencies very clearly, as dropping the requirement of having a common parameter that generates mass for the entire gaugino sector increases the minimal value of $\Delta_{\rm BG}^{\rm HS}$ by 100. \\
Different from the earlier discussed HS models, for NUHGM the minimal value for $\Delta_{\rm BG}^{\rm HS}$ is not only constrained by the Higgs mass requirement, but also by limits placed on SUSY particles. Dropping these two requirements decreases $\Delta_{\rm BG}^{\rm HS}$ with about $100$. Another difference with respect to the previous two models is that now the minimal value for $\Delta_{\rm BG}^{\rm HS}$ is reached for the lowest still allowed value for $m_{\tilde{g}}$ and $m_{\tilde{t}_1}$ (Figure~\ref{fig:nuhmFTvsPole}). This is due to the fact that in this GUT model, $m_0$ and $M_{3}$ are not depending on $\mu$. This was not the case in the mSUGRA or mSUGRA-var model, where the value for $\mu$ is set by the value for $m_0$ and $M_{3}$ via the EWSB requirement.\\

\subsubsection*{$\Delta_{\rm EW}$ and the impact of DMDD experiments}
\noindent The minimal value obtained for $\Delta_{\rm EW}$ is again $3$. As expected, the increase in freedom for the HS parameters in this model did not result in a lower value for $\Delta_{\rm EW}$. The same is true for spectra that result in $0.09 < \Omega h^2 < 0.15$, where the minimal value for $\Delta_{\rm EW}$ is again found around $20$. Future DMDD experiments are able to constrain $\Delta_{\rm EW}$ for these spectra to $146$. The spectra that escape detection by future DMDD experiments have pure higgsino LSPs with masses around $800$~GeV. For values of $\Omega h^2 \gtrsim 1$, the future DMDD experiments quickly lose their sensitivity, which is again caused by the presence of light ($\lesssim 10$ GeV) bino-like LSPs. However, the value of $\Omega h^2$ where the DMDD experiments lose their constraining power is a factor of $10$ lower than for the mSUGRA-var model. One can observe in the right panel of Figure~\ref{fig:nuhmFTvsPole} that $m_{\tilde{\tau}_1}$ is allowed to drop below $\sim 200$ GeV. In the mSUGRA-var model, this is not allowed, as then the Higgs mass requirement cannot be satisfied. In the present case, the small value for $m_{\tilde{\tau}_1}$ allows for a more efficient annihilation of LSPs into tau leptons via a t-channel $\tilde{\tau}_1$ exchange. This decreases the value of $\Omega h^2$, but at the same time does not give rise to a higher value for the SI or SD cross sections, as $\tilde{\tau}_1$ does not couple to nucleons directly. \\

\subsubsection*{The impact of future collider experiments} 
\noindent The impact of the HL-LHC, HE-LHC and CLIC on the two FT measures can be seen in Figure~\ref{fig:nuhmFTvsPole}. As before, we observe that the impact on $\Delta_{\rm BG}^{\rm HS}$ is driven by a higher reach on $m_{\tilde{g}}$. Surprisingly, the impact of the gluino mass exclusion on the minimal value of $\Delta_{\rm BG}^{\rm HS}$ is roughly a factor of 3-4 higher in this model than for mSUGRA-var. The HL-LHC can increase the minimal value of $\Delta_{\rm BG}^{\rm HS}$ to about $1195$, while the HE-LHC constrains it about $3567$. CLIC can constrain $\Delta_{\rm BG}^{\rm HS}$ to about $1070$. The impact of the future colliders on the minimal value for $\Delta_{\rm EW}$ is similar here as to mSUGRA-var: the HL-LHC can increase $\Delta_{\rm EW}$ to about $28$, the HE-LHC can increase it to about $70$, and CLIC can increase it to about $540$, which is similar as in the mSUGRA-var model.\\

\subsection{pMSSM-GUT}
\label{sec:pMSSMGUT}
The last model we analyze is the pMSSM-GUT model. As explained in Sec.~\ref{sec:susymodels}, it has the same number of free parameters as the pMSSM model, but the input parameters are given at $M_{\rm GUT}$ (defined as the scale where the coupling constants $g_1$, $g_2$ and $g_3$ unify). Having essentially the same parameters at $M_{\rm SUSY}$ and at $M_{\rm GUT}$ allows us to study the influence of the RGE running on the obtained FT of a particular spectrum. To this end, we will use {\it three} different FT measures in this section. The first two are the same as before: the low scale FT measure $\Delta_{\rm EW}$ defined according to Eq.~(\ref{eq:FT}) and the high scale FT measure $\Delta_{\rm BG}^{\rm HS}$ defined according to Eq.~(\ref{eq:BG}). For the third FT measure we will use Eq.~(\ref{eq:BG}), but set the matching conditions for the input parameters at  $M_{\rm SUSY} = \sqrt{m_{\tilde{t}_1}m_{\tilde{t}_2}}$ instead of at $M_{\rm GUT}$, like in the case of the pMSSM. This FT measure will be indicated by $\Delta_{\rm BG}^{\rm LS}$. This section will be structured differently compared to the previous three sections, as here we will first compare the three FT measures, and subsequently move on to the discussion of some phenomenology of the spectra with the lowest FT values. 

\begin{figure}[t]
  \centering
  \includegraphics[trim=0 0 0 0,clip,width=\textwidth]{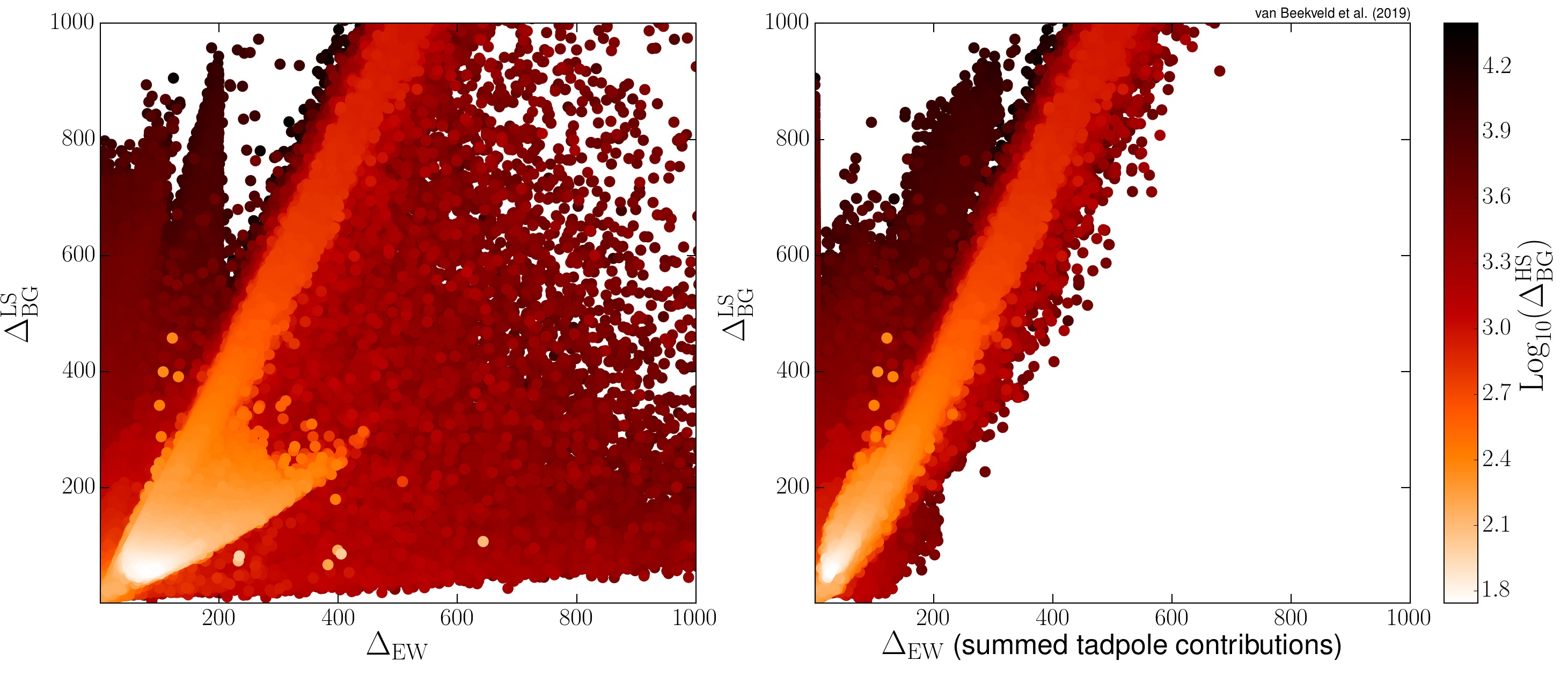}%
\caption{Left: $\Delta_{\rm EW}$ against $\Delta_{\rm BG}^{\rm LS}$ with $\log_{10}(\Delta_{\rm BG}^{\rm HS})$ as color code. Here, $\Delta_{\rm EW}$ is computed considering each tadpole independently. Right: $\Delta_{\rm EW}$ against $\Delta_{\rm BG}^{\rm LS}$ where $\Delta_{\rm EW}$ is computed by summing up the tadpole contributions before evaluating their size. All computed pMSSM-GUT spectra are shown. Spectra with lower values for $\log_{10}(\Delta_{\rm BG}^{\rm HS})$ lie on top of spectra with higher values. }
\label{fig:ewbg}
\end{figure}

\subsubsection*{Comparison of the fine-tuning measures}
We begin by comparing the two LS FT measures: $\Delta_{\rm EW}$ and $\Delta_{\rm BG}^{\rm LS}$, which are plotted against each other in the left panel of Figure \ref{fig:ewbg}. One observes that the two LS FT measures mostly agree for low values of $\Delta_{\rm BG}^{\rm HS}$, taken into account the fact that $\Delta_{\rm BG}^{\rm LS}$ is in general a factor of 2 higher than  $\Delta_{\rm EW}$. This difference can indeed be traced back to the FT definitions in Eq.~(\ref{eq:FT}) and Eq.~(\ref{eq:BG}): $\mu$ is used as a fundamental parameter in  Eq.~(\ref{eq:BG}), which creates an extra factor of 2 as explained in Sec.~\ref{sec:bgft}. The discrepancy between the measures generally grows for higher values of $\Delta_{\rm BG}^{\rm HS}$. We observe that $\Delta_{\rm EW}$ can either underestimate or \emph{overestimate} $\Delta_{\rm BG}^{\rm LS}$. This overestimation happens because in the computation of $\Delta_{\rm EW}$, all tadpole contributions are assumed to be independent (see Eq.~(\ref{eq:FT})). When instead the total tadpole contributions are summed up, the spectra where $\Delta_{\rm EW}$ overestimates $\Delta_{\rm BG}^{\rm LS}$ mostly disappear, as can be seen on the right-hand side of Figure~\ref{fig:ewbg}. This phenomenon is observed for spectra for which the tadpole corrections belonging to the same particle type are large individually, but carry opposite signs. This may happen for example when the stop masses are degenerate (see appendix A of Ref. \cite{Baer:2012cf}), and shows that it is {\it not true} that $\Delta_{\rm EW}$ is always the most conservative measure. \\
On the other hand, one can see that for some spectra the value for $\Delta_{\rm EW}$ greatly underestimates the value for $\Delta_{\rm BG}^{\rm LS}$. The value for $\Delta_{\rm EW}$ for these spectra is determined by the size of $m_{H_u}$, while the value for $\Delta_{\rm BG}^{\rm LS}$ is mainly determined by variations in $m_{\tilde{Q}_3}$, $m_{\tilde{t}_R}$ and $M_3$. When the one-loop corrections of $m_{H_{u}}$ actually {\it determine} the size of $m_{H_{u}}$ (for example when the stop and/or gluino masses are large), the value of $\Delta_{\rm BG}^{\rm LS}$ will be driven by these parameters. In this case, when varying either one of these parameters, one induces a large change in $m_{H_{u}}$, which gives rise to a large value for $\Delta_{\rm BG}^{\rm LS}$. However, in $\Delta_{\rm EW}$, merely the size of $m_{H_{u}}$ and its tadpole terms are taken into account and these are not necessarily big for these spectra. It is precisely in this case that $\Delta_{\rm EW}$ can lead to an underestimation of $\Delta_{\rm BG}^{\rm LS}$. A second (subdominant) effect originates from the value of the SUSY scale. In general, a higher value for $M_{\rm SUSY}$ increases the dependence of $m_{H_u}$ on $M_3$, $m_{\tilde{Q}_3}$ and $m_{\tilde{t}_R}$, which can lead to $\Delta_{\rm EW}$ underestimating $\Delta_{\rm BG}^{\rm LS}$.\\
We now move on to the comparison between $\Delta_{{\rm BG}}^{{\rm HS}}$ and $\Delta_{{\rm BG}}^{{\rm LS}}$.  In general, $\Delta_{{\rm BG}}^{{\rm HS}}$ will be larger than $\Delta_{{\rm BG}}^{{\rm LS}}$ due to the RGE running. Since the influence of the RGE running is almost absent in $\Delta_{{\rm BG}}^{{\rm LS}}$ for modest values ($\mathcal{O}(1)$ TeV) of the SUSY breaking scale, this leads to a reduced sensitivity of $m_{H_u}$ on e.g. $M_3$, $m_{\tilde{Q}_3}$ and $m_{\tilde{t}_R}$, and therefore a smaller value for $\Delta_{{\rm BG}}^{{\rm LS}}$ if these terms dominate its size. For $\Delta_{{\rm BG}}^{{\rm HS}}$ however, the effect of the RGE running cannot be neglected and in general the value of $\Delta_{{\rm BG}}^{{\rm HS}}$ will grow large for large values for the GUT values of $M_3$, $m_{\tilde{Q}_3}$ or $m_{\tilde{t}_R}$. Therefore, the discrepancy between $\Delta_{{\rm BG}}^{{\rm LS}}$ and $\Delta_{{\rm BG}}^{{\rm HS}}$ generally grows bigger for higher values of either $M_3$, $m_{\tilde{Q}_3}$ or $m_{\tilde{t}_R}$. Interestingly, we also find spectra where $\Delta_{{\rm BG}}^{{\rm LS}}\simeq \Delta_{{\rm BG}}^{{\rm HS}}\simeq 2 \Delta_{{\rm EW}}$. For these spectra, the value of $|\mu|$ determines the value of the FT for all measures. As $c_{\mu}$ in Eq.~(\ref{eq:mu}) is close to 1, for these spectra indeed only the size of $|\mu|$ matters in the computation of the FT and therefore the three FT measures reduce to approximately the same value.

\subsubsection*{The phenomenology of low-$\Delta_{\rm BG}^{\rm HS}$ spectra}
\begin{figure}[t]
  \centering
  \includegraphics[width=\textwidth]{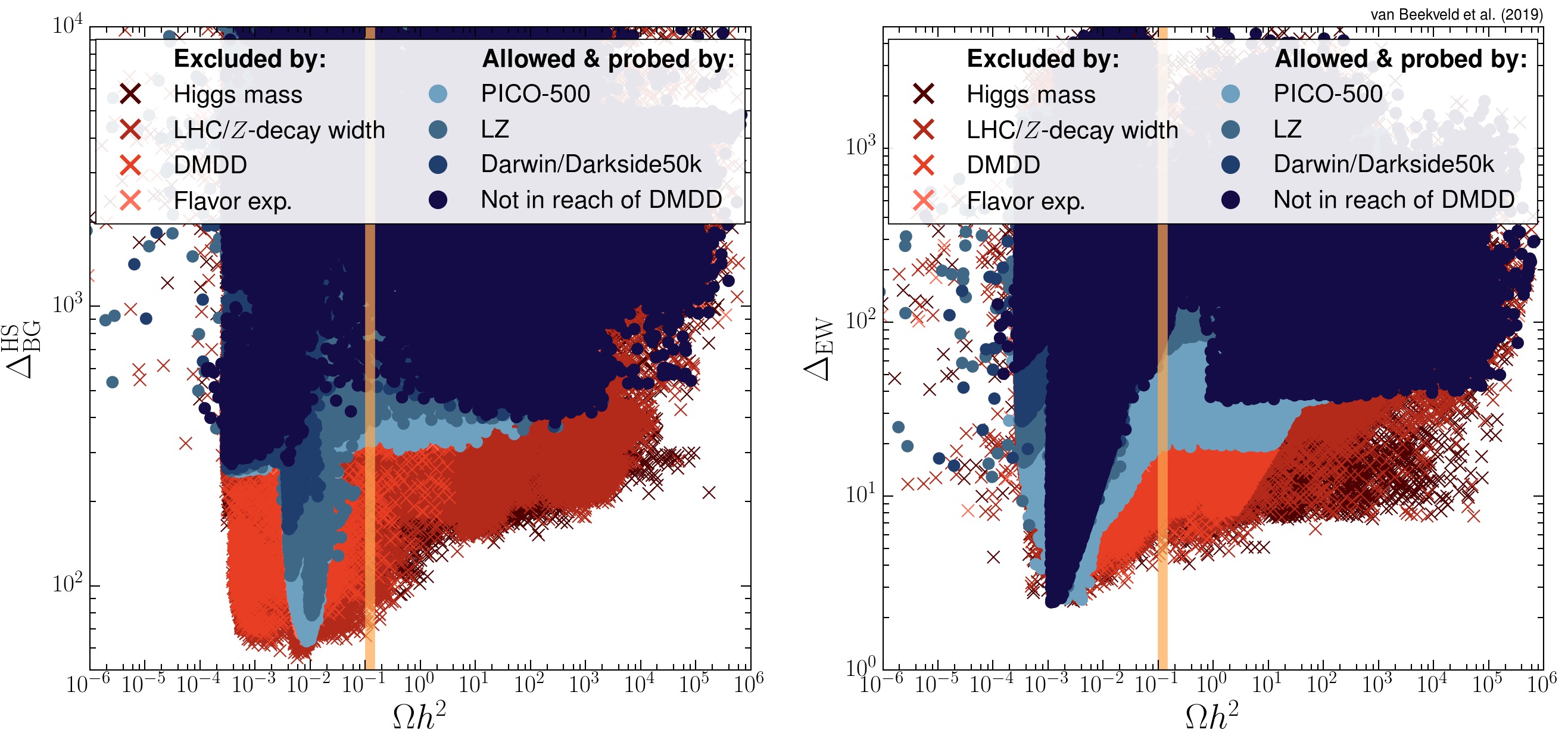}%
\caption{Generated spectra for the pMSSM-GUT model. The left figure shows $\Delta_{\rm BG}^{\rm HS}$ as a function of $\Omega h^2$, the right figure shows $\Delta_{\rm EW}$ as a function of $\Omega h^2$. The color code and plotting order is the same as in Figure \ref{fig:sugraomega}. The orange band indicates $\Omega h^2 = 0.12$.}
\label{fig:pmssmomega}
\end{figure}
\begin{figure}[b]
  \centering
  \includegraphics[width=1.1\textwidth]{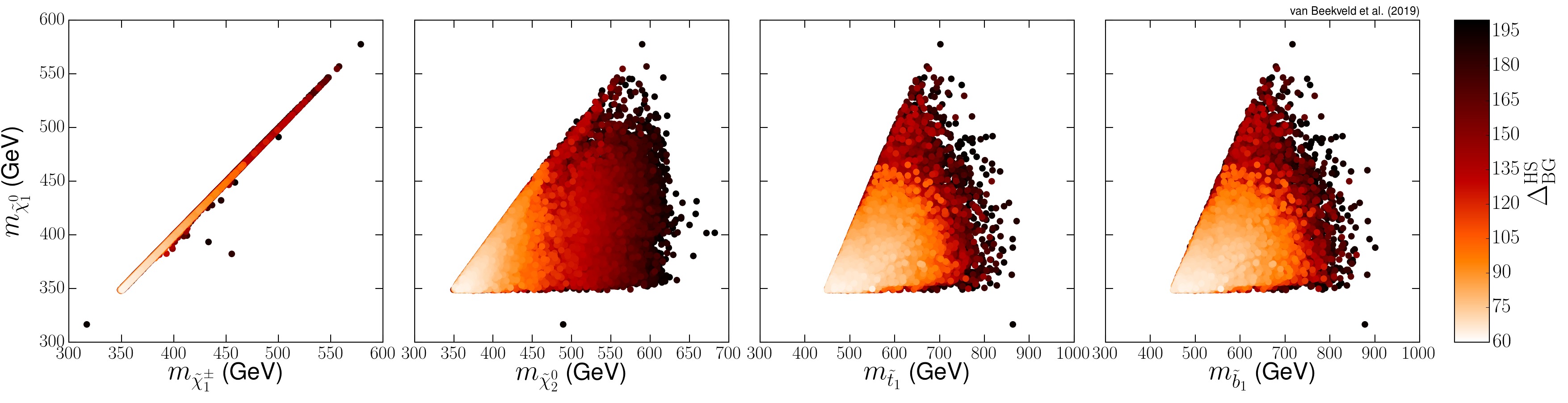}
\caption{The masses of $\tilde{\chi}^{\pm}_1$, $\tilde{\chi}^{0}_2$, $\tilde{t}_1$ and $\tilde{b}_1$ against $m_{\tilde{\chi}^{0}_1}$ in GeV with $\Delta_{\rm BG}^{\rm HS}$ in color for the allowed pMSSM-GUT spectra and have $\Delta_{\rm BG}^{\rm HS}<200$. The spectra with lower values for $\Delta_{\rm BG}^{\rm HS}$ lie on top of those with higher values.   }
\label{fig:pMSSMstopsbot}
\end{figure}
\begin{figure}[t]
  \centering
  \includegraphics[width=\textwidth]{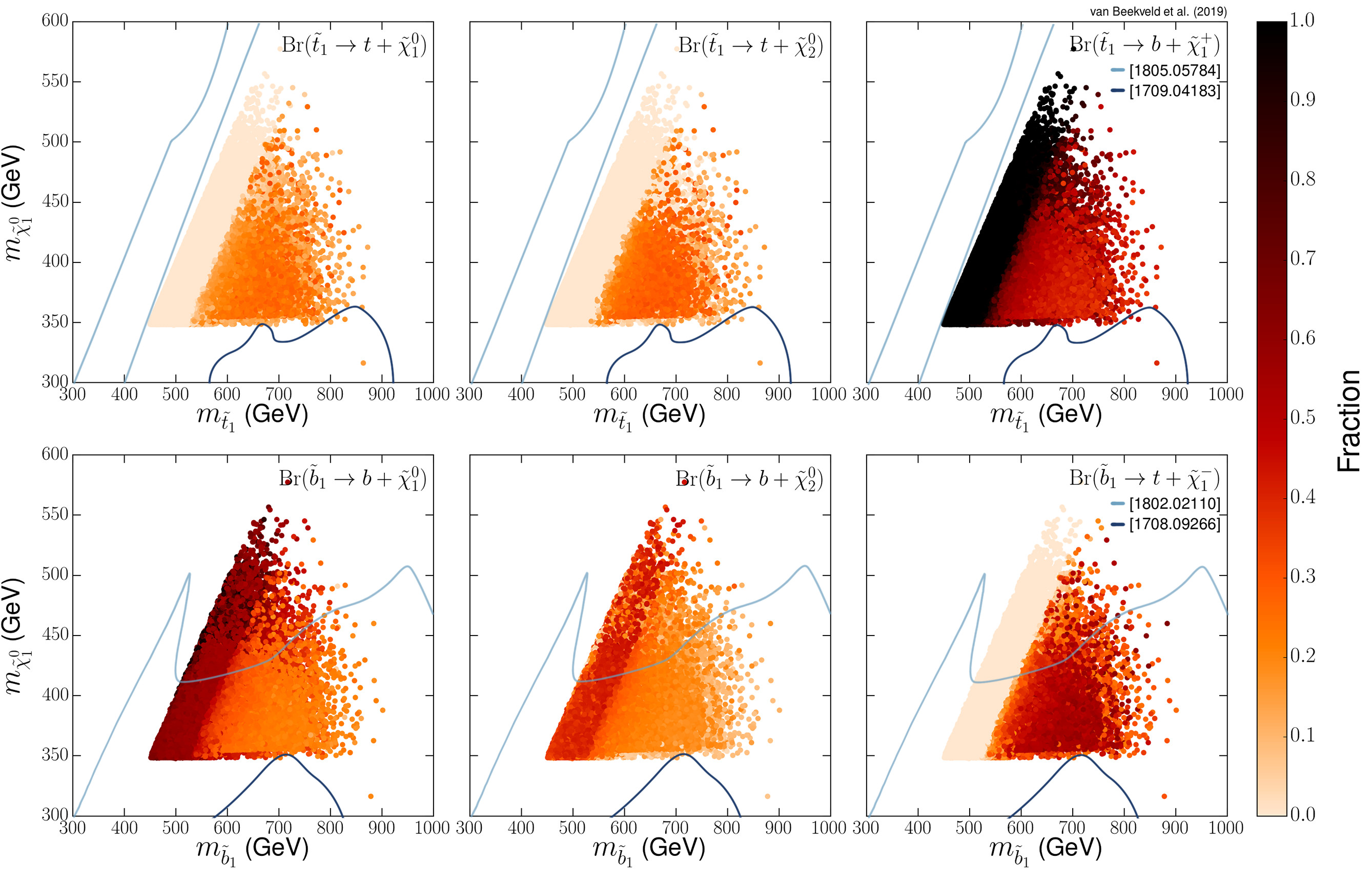}
\caption{Branching ratios of $\tilde{t}_1$ and $\tilde{b}_1$ for the allowed pMSSM-GUT spectra with $\Delta_{\rm BG} < 200$, where spectra with lower values for $\Delta_{\rm BG}^{\rm HS}$ lie on top of those with higher values. The top (bottom) row shows $m_{\tilde{t}_1}$ ( $m_{\tilde{b}_1}$) against $m_{\tilde{\chi}^0_1}$ in GeV, and the relevant decay process is indicated in the upper right corner of each plot. The two lines in the plots on the top panel show the exclusion limits of Ref. \cite{Sirunyan:2018omt} (light blue), where ${\rm Br}(\tilde{t}_1 \rightarrow t \tilde{\chi}^0_1) = 100\%$ is assumed, and Ref. \cite{Aaboud:2017ayj} (dark blue), where a mixed decay scenario is assumed. The two lines in the plots of the bottom panel show the exclusion limits of Ref. \cite{Sirunyan:2018vjp} (light blue), where ${\rm Br}(\tilde{b}_1 \rightarrow b \tilde{\chi}^0_1) = 100\%$ is assumed, and Ref. \cite{Aaboud:2017wqg} (dark blue), where mixed decay scenarios are considered. }
\label{fig:pMSSMstopsbotBR}
\end{figure}

The resulting values for $\Delta_{\rm BG}^{\rm HS}$ (and $\Delta_{\rm EW}$) as a function of $\Omega h^2$ are shown in Figure \ref{fig:pmssmomega}. The minimal value for $\Delta_{\rm BG}^{\rm HS}$ is 63, which is lower than for all other GUT models previously considered. The spectra that minimize $\Delta_{\rm BG}^{\rm HS}$ all feature higgsino LSPs, and some of the LSPs also have a sizable wino component. Due to the presence of a higgsino assymmetry, and a non-zero wino component, the future DMDD experiments are sensitive to these spectra despite the fact that the relic density is not saturated. In Figure \ref{fig:pMSSMstopsbot} we show $m_{\tilde{\chi}^{0}_1}$ against $m_{\tilde{\chi}^{\pm}_1}$, $m_{\tilde{\chi}^{0}_2}$, $m_{\tilde{t}_1}$ and $m_{\tilde{b}_1}$ of the allowed spectra with  $\Delta_{\rm BG}^{\rm HS} < 200$. These spectra are characterized by low SUSY breaking scales of $300-700$ GeV, accompanied by low values of $m_{\tilde{t}_1}\simeq 400-800$~GeV and $m_{\tilde{b}_1} \simeq 450-800$~GeV. These masses are driven by $m_{\tilde{Q}_3}$, whose value at the SUSY scale lies around $100$~GeV. The masses of $\tilde{t}_2$ and $\tilde{b}_2$ are less constrained and have values ranging from $600$~GeV to $2$~TeV. Surprisingly, the gluino mass is also unconstrained in this region. The lightest chargino is ultracompressed with the LSP, but their masses are too high to be discovered by the analysis of Ref. \cite{Sirunyan:2018iwl,Aaboud:2017leg}. In the stop case, the analyses performed in Ref. \cite{Sirunyan:2018omt, Aaboud:2017ayj,Aaboud:2017aeu} are most relevant when $\Delta(m_{\tilde{\chi}^0_1},m_{\tilde{t}_1}) < m_{t}$. In tihs region, $\tilde{t}_1$ decays to $\tilde{\chi}^{+}_1$ in association with a bottom quark with a branching ratio (BR) of $100\%$ (see top panel of Figure~\ref{fig:pMSSMstopsbotBR}). The lightest chargino subsequently decays into a fermion-anti-fermion pair via an off-shell $W$-boson, where the fermion are ultra soft due to the mass compression between $\tilde{\chi}^{\pm}_1$ and $\tilde{\chi}^0_1$. One observes that the spectra where ${\rm Br}(\tilde{t}_1 \rightarrow \tilde{\chi}^{+}_1 b) \simeq 100\%$ and $\Delta(m_{\tilde{\chi}^0_1}, m_{\tilde{t}_1}) < 100$~GeV are under pressure, therefore we have introduced an explicit hard cut to exclude these spectra from our final results. In the case that $\Delta(m_{\tilde{\chi}^0_1},m_{\tilde{t}_1}) > m_{t}$, the analyses in Ref. \cite{Sirunyan:2017leh, Aaboud:2017ayj, Aaboud:2017aeu} are the most relevant, as we observe a mixed decay scenario for this region where ${\rm Br}(\tilde{t}_1 \rightarrow \tilde{\chi}^0_1 t) \simeq 25\%$, ${\rm Br}(\tilde{t}_1 \rightarrow \tilde{\chi}^0_2 t) \simeq 25\%$ and ${\rm Br}(\tilde{t}_1 \rightarrow \tilde{\chi}^{+}_1 b) \simeq 50\%$. \\

\noindent In the sbottom case (lower panel of Figure \ref{fig:pMSSMstopsbotBR}), the searches presented in Ref. \cite{Sirunyan:2018vjp}, where ${\rm Br}(\tilde{b}_1 \rightarrow \tilde{\chi}^0_1 + b)= 100\%$ is assumed, and in particular in Ref. \cite{Aaboud:2017wqg} are relevant, where the latter analysis considers a mixed sbottom decay scenario. The spectra with ${\rm Br}(\tilde{b}_1 \rightarrow \tilde{\chi}^0_1 + b)= 100\%$ are under pressure, and we exclude the spectra within the exclusion limit of Ref. \cite{Sirunyan:2018vjp} that have ${\rm Br}(\tilde{b}_1 \rightarrow \tilde{\chi}^0_1 + b)> 75\%$. The spectra with the lowest FT values escape the limits set by the experiments, showing that a dedicated search for low mass sbottom and/or stop sparticles assuming \emph{mixed} decay scenarios is needed to cover this region. 

\subsubsection*{$\Delta_{\rm EW}$ and the impact of DMDD experiments}

\begin{figure}[t]
  \centering
  \includegraphics[width=\textwidth]{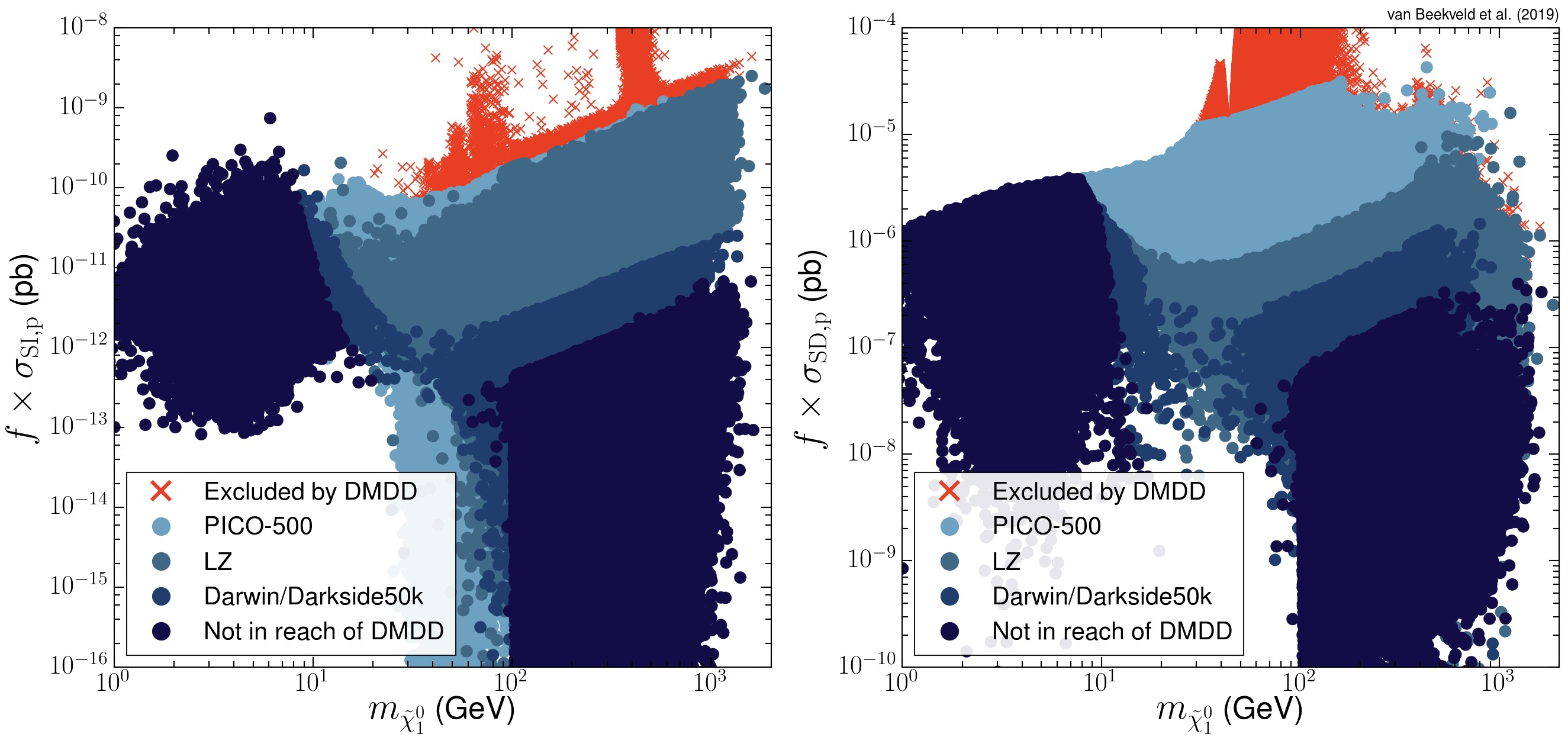}
\caption{The mass of $\tilde{\chi}^0_1$ in GeV against $f \sigma_{\rm SI,p}$ in pb (right) and $f \sigma_{\rm SD,p}$ in pb (left), where $f$ is the rescaling factor when $\Omega h^2 < 0.12$. The color code and plotting order is the same as in Figure \ref{fig:sugraomega}. }
\label{fig:pMSSMsigmaSD}
\end{figure}
\noindent The minimal value for $\Delta_{\rm EW}$ is again around 3 (Figure \ref{fig:pmssmomega}). The spectra that result in $\Omega h^2 \simeq 0.12$ give a minimum value of $\Delta_{\rm EW} \simeq 20$. The PICO 2019 limit on $\sigma_{\rm SD,p}$\cite{Amole:2019fdf} is particularly constraining the minimally fine-tuned spectra that have the correct relic density. The LSPs that feature in these spectra are bino-higgsino mixtures, and have masses around $M_Z/2$, $m_h/2$ or 100 GeV. The lightest charginos have masses ranging from 100 to 300 GeV for these spectra. The spectra that escape detection by future DMDD experiments are pure higgsino models, just like was the case for mSUGRA-var and NUHGM. In Figure \ref{fig:pMSSMsigmaSD} we show $m_{\tilde{\chi}^0_1}$ against $f \,\sigma_{\rm SI,p}$ and  $f \,\sigma_{\rm SD,p}$, where $f$ is the rescaling factor defined in Sec. \ref{sec:analysis}.  One observes a sharp increase in the number of spectra that evade the limits of future DMDD experiments completely around $m_{\tilde{\chi}^0_1} = 100$~GeV. These are the models with a pure higgsino LSP that deplete the SD coupling due to the higgsino symmetry, and in addition have a rescaling factor of around $10^{-2}$. On the other hand, one observes that also low mass LSPs with $m_{\tilde{\chi}^0_1}$ are out of reach of future DMDD experiments, as these simply are not sensitive to these mass ranges.

\begin{figure}[t]
  \centering
  \includegraphics[width=\textwidth]{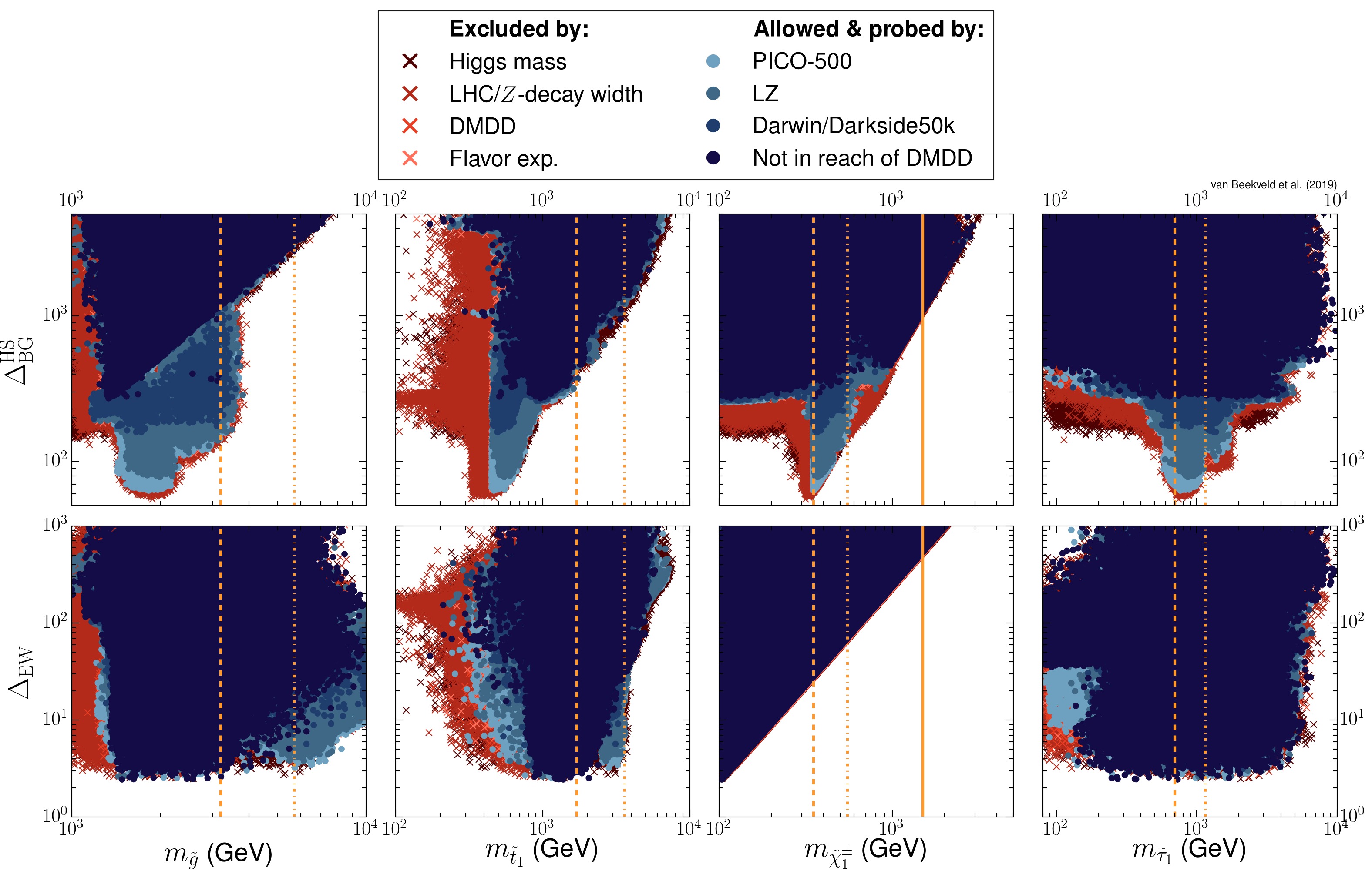}
\caption{ The values for $\Delta_{\rm BG}^{\rm HS}$ (top) and $\Delta_{\rm EW}$ (bottom) against (from left to right) $m_{\tilde{g}}$, $m_{\tilde{t}_1}$, $m_{\tilde{\chi}^{\pm}_1}$ and $m_{\tilde{\tau}_1}$ for all generated pMSSM-GUT spectra. The masses are given in units of GeV. The color code and plotting order is the same as in Figure \ref{fig:sugraomega}. The dashed, dash-dotted and solid orange line shows the exclusion potential of the HL-, HE-LHC and CLIC on the masses of various SUSY particles  (see Table~\ref{tab:masscuts}). The solid orange line in the chargino mass plot shows the exclusion potential of CLIC.}
\label{fig:pMSSMFTvsPole}
\end{figure}

\subsubsection*{The impact of future collider experiments}
The HL(HE)-LHC increases $\Delta_{\rm EW}$ to $\sim$ 70 (28) by constraining the higgsino mass (Figure \ref{fig:pMSSMFTvsPole}), while CLIC constrains it to 540. These numbers are similar than those that are found in the previous models. The HS FT measure $\Delta_{\rm BG}^{\rm HS}$ is again more constrained than $\Delta_{\rm EW}$ due to the increased gluino and stop mass reach of the HL- and HE-LHC. The HL-LHC can raise $\Delta_{\rm BG}^{\rm HS}$ to about $328$ (driven by the limit that is placed on $m_{\tilde{t}_1}$), while the HE-LHC can raise $\Delta_{\rm BG}^{\rm HS}$ to about $3030$ (due to the gluino mass reach). One observes a sharp increase in $\Delta_{\rm BG}^{\rm HS}$ for gluino masses around $3.5$~TeV. There is no physical reason why this happens, it simply takes a large number of computer resources to find these spectra. One can expect to find spectra with low values for $\Delta_{\rm BG}^{\rm HS}$ that have $m_{\tilde{g}} \lesssim 3.5$~TeV. CLIC can constrain the minimal value of $\Delta_{\rm BG}^{\rm HS}$ to $1100$.\\
Remarkable is that for all high-scale models and both FT measures result in roughly the same number in case of a non-observation at CLIC (taking into account the factor of $2$ difference between $\Delta_{\rm BG}^{\rm HS}$ and  $\Delta_{\rm EW}$). This shows that $|\mu|$ is the most model-independent parameter to determine the minimal possible amount of FT in the pMSSM.   \\
\noindent One observes that $m_{\tilde{g}}$ is allowed to get as low as 1 TeV for some spectra. These gluinos have a very complicated decay chain and they decay in at least three different neutralinos and one chargino. The mass differences between the neutralinos/charginos and the gluino are less than $50$ GeV, and the chargino and heavier neutralino states decay into off-shell $W$/$Z$-bosons. 
Pair production of gluinos in the LHC thus mimics the QCD background, since the missing transverse energy (MET) is typically not large enough to discriminate the events. This allows them to escape detection at the LHC. Interestingly, these low mass gluinos do not appear in the spectra that create the minimum of $\Delta_{\rm BG}^{\rm HS}$ or $\Delta_{\rm EW}$. This is due to the fact that for both measures $|\mu|$ needs to be as low as possible. A low mass gluino is therefore excluded in spectra with low values for $\Delta_{\rm BG}^{\rm HS}$ or $\Delta_{\rm EW}$, as a low $|\mu|$-value reduces the mass compression of the NLSPs with the gluinos. This will allow pair production of the low mass gluinos to stand out of the QCD background due to a higher MET in these events. \\

\section{Conclusion}

\begin{table}[b]
    \centering
    \begin{tabular}{|c|c|c|c|c|c|}
    \hline
    \textbf{Model} & \textbf{Current} & \textbf{Future DD} & \textbf{HL-LHC} & \textbf{HE-LHC} & \textbf{CLIC}\\
    \hline
    mSUGRA $\Delta_{\rm BG}^{\rm HS}$     &571(737) & 750(1032) & 848 & 1669 & 1237 \\
    mSUGRA $\Delta_{\rm EW}$    &38(110) & 275(515) & 39 & 86 & 529 \\
    \hline 
    mSUGRA-var  $\Delta_{\rm BG}^{\rm HS}$     &191(262) & 252(545) & 529 & 1222 & 888  \\
    mSUGRA-var $\Delta_{\rm EW}$   &3(20) & 3(126) & 28 & 70 & 529 \\
    \hline
    NUHGM  $\Delta_{\rm BG}^{\rm HS}$    &290(375) & 395(691) & 1195 & 3567 &1070  \\
    NUHGM $\Delta_{\rm EW}$    &3(20) & 3(146) & 28 & 70 &537 \\
    \hline 
    pMSSM-GUT $\Delta_{\rm BG}^{\rm HS}$    &63(272) & 328(517) & 372 & 3258& 1108  \\
    pMSSM-GUT $\Delta_{\rm EW}$    &3(19) & 3(63) & 28 & 70& 544  \\
    \hline
    \end{tabular}
    \caption{Summary of the high-scale models and their minimal amount of FT (rounded to integers). The first column indicates the high-scale model (definitions can be found in Sec.~\ref{sec:susymodels}). The second column shows the minimal FT both for $\Delta_{\rm BG}^{\rm HS}$ and $\Delta_{\rm EW}$ after applying only the current constraints. Between brackets we quote the result for the spectra that result in $0.09 < \Omega h^2 < 0.15$. The third column gives the minimum of $\Delta_{\rm BG}^{\rm HS}$ and $\Delta_{\rm EW}$ after all the future DMDD experiments. The remaining columns give the maximal constraining power on $\Delta_{\rm BG}^{\rm HS}$ and $\Delta_{\rm EW}$ using the reach of the HL-LHC, HE-LHC and CLIC.  }
    \label{tab:summary}
\end{table}

\begin{figure}[t]
  \centering
  \includegraphics[width=\textwidth]{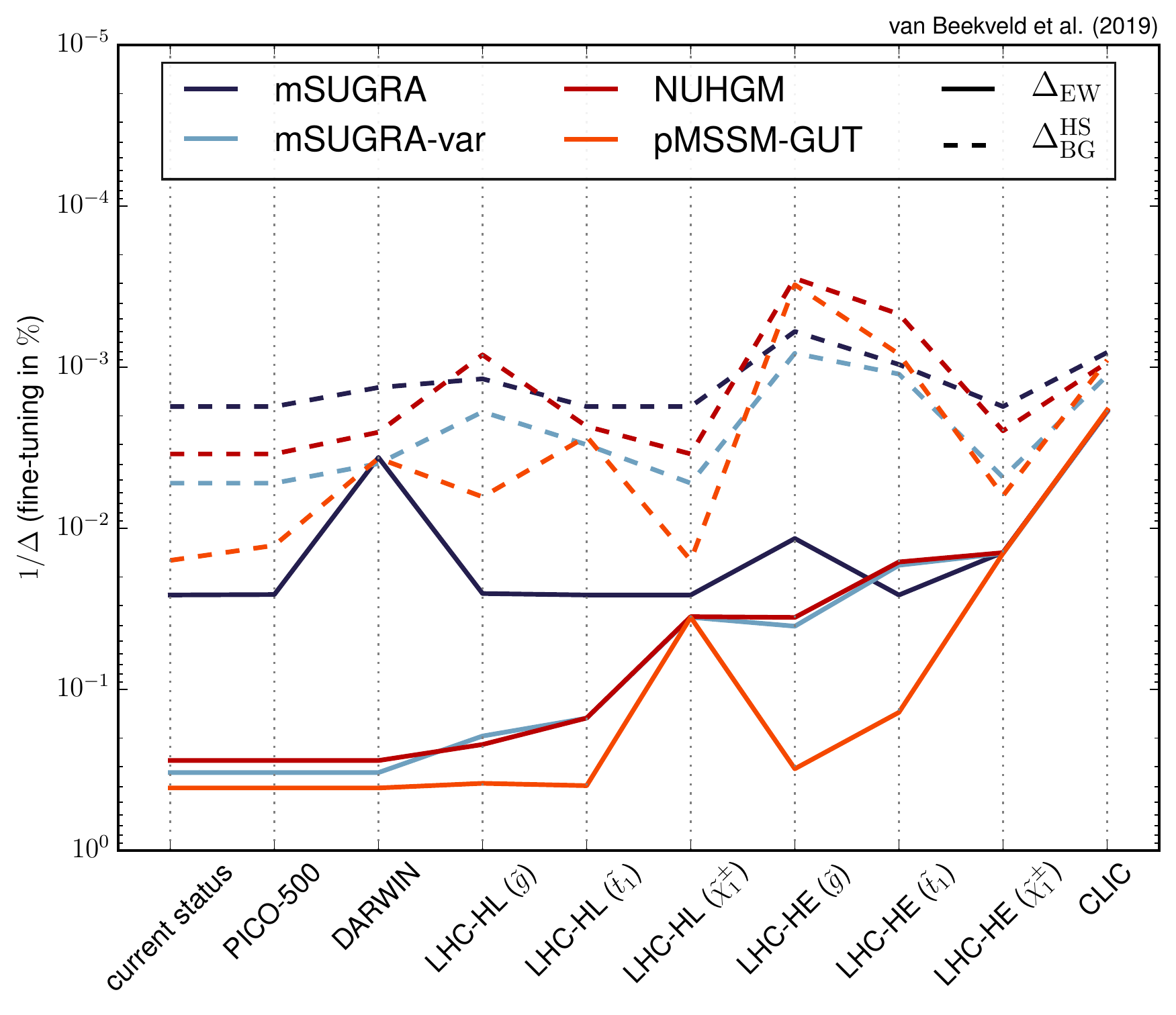}
\caption{Summary of the high-scale models and minimal amount of FT still allowed after various future experiments. The solid (dashed) line shows the result for $\Delta_{\rm EW}$ ($\Delta_{\rm BG}^{\rm HS})$. The colors indicate various high-scale models, where dark blue indicates mSUGRA, light blue indicates mSUGRA-var, red indicated NUGHM and orange indicates pMSSM-GUT. }
\label{fig:superfine}
\end{figure}

\label{sec:conclusion}
In this paper, we have considered four different high-scale models, all with a low-scale realization in the pMSSM. We have minimized two different FT measures, $\Delta_{\rm EW}$  (Eq. \ref{eq:FT}) and  $\Delta_{\rm BG}^{\rm HS}$  (Eq. \ref{eq:BG}), for these four scenarios, and computed the impact of current/future collider experiments and current/future dark matter experiments on the minimal allowed FT. The results are summarized in Table \ref{tab:summary} and Figure \ref{fig:superfine}. One observes that the obtained minimal FT values for $\Delta_{\rm EW}$ and $\Delta_{\rm BG}^{\rm HS}$ differ by at least an order of magnitude for each GUT model. For the mSUGRA and mSUGRA-var, the minimum of $\Delta_{\rm BG}^{\rm HS}$ is determined by the observed value for the Higgs boson mass, whereas for the NUHGM, the stop/gluino mass exclusion limits determine the minimum value of $\Delta_{\rm BG}^{\rm HS}$. The stop and sbottom searches constrain the minimal value of $\Delta_{\rm BG}^{\rm HS}$ for the pMSSM-GUT model. Interestingly, this value is one order of magnitude lower than for the other GUT models, whose minimal $\Delta_{\rm BG}^{\rm HS}$ values are of $\mathcal{O}(100)$. The minimum value for $\Delta_{\rm BG}^{\rm HS}$ in the pMSSM-GUT model is found at $63$, which corresponds to no more than $2\%$ fine-tuning. This region could be further constrained by sbottom and stop searches at the LHC, where we stress that mixed decay scenarios should be further explored. \\
At the same time, the minimum of $\Delta_{\rm EW}$ is determined solely by $|\mu|$. The minimum value of this measure is less affected by the high-scale model. The minimal value that we have found for $\Delta_{\rm EW}$ is 3 (corresponding to $33\%$) and is determined by the LEP chargino limits. The minimally fine-tuned spectra that predict the observed dark matter relic density are constrained by the dark matter direct detection experiments, and in particular by the PICO experiment \cite{Amole:2019fdf}. The LSPs that feature in these spectra are bino-higgsino mixtures and have masses around $M_Z/2$, $m_h/2$ or 100 GeV. The lightest charginos have masses ranging from 100 to 300 GeV for these spectra, which are unconstrained by the current LHC experiments as the production cross section is generally too small \cite{Sirunyan:2018iwl,vanBeekveld:2016hbo}.  \\
\noindent For all GUT models, we found spectra where $2 \Delta_{\rm EW}$ is roughly equal to $\Delta_{\rm BG}^{\rm HS}$. These spectra have in common that $|\mu|$ determines the value of $\Delta_{\rm BG}^{\rm HS}$. Therefore, to constrain the minimal FT in the most measure- and model-independent way, one must constrain $|\mu|$\footnote{There are ways to reduce the FT induced by $\mu$ that go beyond the MSSM (see Ref. \cite{Bae:2019dgg} for a recent overview), for example by including an explicit soft-SUSY breaking Higgsino mass term  \cite{Ross:2016pml,Ross:2017kjc}. }. Future experiments that can constrain the value of $|\mu|$ are therefore favorable to settle the naturalness question of the pMSSM. Since a pure low-mass (around 100 GeV) higgsino LSP is produced under-abundantly in the early universe, the near-future DM direct detection experiments will not be sensitive to these spectra. Note that this statement changes if one considers only LSPs that satisfy the dark matter relic density constraint. In case of a non-observation, the future dark matter direct detection experiments can increase the limit on fine-tuning for spectra with LSPs that satisfy the dark matter relic density constraint from $19$ ($5\%$) to $63$ ($2\%$). Interestingly, we have found a correlation between the relic density and $\Delta_{\rm EW}$, as spectra that result in a low value for $\Delta_{\rm EW}$ give rise to $10^{-3} < \Omega h^2 < 1$. \\
On the other hand, dedicated higgsino searches at proton-proton colliders are particularly challenging due to small cross-sections (compared to QCD-induced processes or wino-induced production of the charginos and second-lightest neutralino) and low missing transverse energy (due to the mass compression). This makes the $|\mu|$ coverage relatively poor (see e.g. Ref. \cite{Athron:2017yua,Athron:2018vxy}). The ATLAS and CMS experiments both have targeted soft lepton searches to observe these particles, which seem to see small deviations from the SM background hypothesis in these regions \cite{Aaboud:2018jiw,Aaboud:2018sua,Aaboud:2018zeb,Sirunyan:2017qaj}, although this deviation was not seen in the soft lepton analysis of Ref. \cite{Sirunyan:2018ubx}. As can be seen in Figure \ref{fig:superfine}, the low- and high-scale fine-tuning measures reduce to the same value in case of a non-observation of higgsinos at CLIC. Therefore, CLIC may be the preferred collider to constrain the FT value in the most model-independent way. Our study shows that it is too early to conclude on the fate of the MSSM, and more measurements with less simplifying assumptions are needed to cover the parameter space of low fine-tuned MSSM spectra.

\acknowledgments
We wish to thank Wim Beenakker, Alberto Casas, Rohini Godbole and Ruud Peeters for useful discussions. R. RdA acknowledges partial funding/support from the Elusives ITN (Marie Sk\l{}odowska-Curie grant agreement No 674896), the  
``SOM Sabor y origen de la Materia" (FPA 2017-85985-P) and the Spanish MINECO Centro de Excelencia Severo Ochoa del IFIC 
program under grant SEV-2014-0398. MvB acknowledges support from the Dutch NWO-I program 156, "Higgs as Probe and Portal". 

\bibliographystyle{JHEP}
\bibliography{bibliography}

\providecommand{\href}[2]{#2}\begingroup\raggedright\begin{thebibliography}{100}

\bibitem{Susskind:1978ms}
L.~Susskind, {\it {Dynamics of Spontaneous Symmetry Breaking in the Weinberg-
  Salam Theory}},  {\em Phys. Rev.} {\bf D20} (1979) 2619--2625.

\bibitem{Veltman:1980mj}
M.~J.~G. Veltman, {\it {The Infrared - Ultraviolet Connection}},  {\em Acta
  Phys. Polon.} {\bf B12} (1981) 437.

\bibitem{tHooft:1979rat}
G.~'t~Hooft, {\it {Naturalness, chiral symmetry, and spontaneous chiral
  symmetry breaking}},  {\em NATO Sci. Ser. B} {\bf 59} (1980) 135--157.

\bibitem{Aad:2012tfa}
{{\bf ATLAS} Collaboration}, {\it {Observation of a new particle in the search
  for the Standard Model Higgs boson with the ATLAS detector at the LHC}},
  {\em Phys. Rev.} {\bf B716} (2012) 1--29,
  [\href{http://arxiv.org/abs/1207.7214}{{\tt arXiv:1207.7214}}].

\bibitem{Chatrchyan:2012ufa}
{{\bf CMS} Collaboration}, {\it {Observation of a new boson at a mass of 125
  GeV with the CMS experiment at the LHC}},  {\em Phys. Rev.} {\bf B716} (2012)
  30--61, [\href{http://arxiv.org/abs/1207.7235}{{\tt arXiv:1207.7235}}].

\bibitem{Martin:1997ns}
S.~P. Martin, {\it {A supersymmetry primer}},
  \href{http://arxiv.org/abs/9709356}{{\tt 9709356}}.

\bibitem{WITTEN1981513}
E.~Witten, {\it Dynamical breaking of supersymmetry},  {\em Nucl. Phys.} {\bf
  B188} (1981), no.~3 513 -- 554.

\bibitem{KAUL198219}
R.~K. Kaul, {\it Gauge hierarchy in a supersymmetric model},  {\em Phys. Lett.}
  {\bf B109} (1982), no.~1 19 -- 24.

\bibitem{Sakai:1981gr}
N.~Sakai, {\it {Naturalness in supersymmetric GUTs}},  {\em Zeit. Phys.} {\bf
  C11} (1981) 153.

\bibitem{PhysRevLett.109.161802}
H.~Baer, V.~Barger, P.~Huang, A.~Mustafayev, and X.~Tata, {\it {Radiative
  natural supersymmetry with a 125 GeV Higgs boson }},  {\em Phys. Rev. Lett.}
  {\bf 109} (Oct, 2012) 161802, [\href{http://arxiv.org/abs/1207.3343}{{\tt
  arXiv:1207.3343}}].

\bibitem{PhysRevD.73.095004}
R.~Kitano and Y.~Nomura, {\it {Supersymmetry, naturalness, and signatures at
  the CERN LHC}},  {\em Phys. Rev.} {\bf D73} (May, 2006) 095004,
  [\href{http://arxiv.org/abs/0602096}{{\tt 0602096}}].

\bibitem{Hall2012}
L.~J. Hall, D.~Pinner, and J.~T. Ruderman, {\it {{A natural SUSY Higgs near 125
  GeV}}},  {\em JHEP} {\bf 1212} (2012), no.~4 131,
  [\href{http://arxiv.org/abs/1112.2703}{{\tt arXiv:1112.2703}}].

\bibitem{Baer:2012cf}
H.~Baer, V.~Barger, P.~Huang, D.~Mickelson, A.~Mustafayev, and X.~Tata, {\it
  {Radiative natural supersymmetry: Reconciling electroweak fine-tuning and the
  Higgs boson mass}},  {\em Phys. Rev.} {\bf D87} (2013), no.~11 115028,
  [\href{http://arxiv.org/abs/1212.2655}{{\tt arXiv:1212.2655}}].

\bibitem{Barbieri:1987fn}
R.~Barbieri and G.~F. Giudice, {\it {Upper Bounds on Supersymmetric Particle
  Masses}},  {\em Nucl. Phys.} {\bf B306} (1988) 63--76.

\bibitem{Feng:1999zg}
J.~L. Feng, K.~T. Matchev, and T.~Moroi, {\it {Focus points and naturalness in
  supersymmetry}},  {\em Phys. Rev.} {\bf D61} (2000) 075005,
  [\href{http://arxiv.org/abs/hep-ph/9909334}{{\tt hep-ph/9909334}}].

\bibitem{Kitano:2005wc}
R.~Kitano and Y.~Nomura, {\it {A solution to the supersymmetric fine-tuning
  problem within the MSSM}},  {\em Phys. Lett.} {\bf B631} (2005) 58--67,
  [\href{http://arxiv.org/abs/hep-ph/0509039}{{\tt hep-ph/0509039}}].

\bibitem{Cabrera:2012vu}
M.~E. Cabrera, J.~A. Casas, and R.~Ruiz~de Austri, {\it {The health of SUSY
  after the Higgs discovery and the XENON100 data}},  {\em JHEP} {\bf 1307}
  (2013) 182, [\href{http://arxiv.org/abs/1212.4821}{{\tt arXiv:1212.4821}}].

\bibitem{Fichet:2012sn}
S.~Fichet, {\it {Quantified naturalness from Bayesian statistics}},  {\em Phys.
  Rev.} {\bf D86} (2012) 125029, [\href{http://arxiv.org/abs/1204.4940}{{\tt
  arXiv:1204.4940}}].

\bibitem{Baer:2012mv}
H.~Baer, V.~Barger, P.~Huang, D.~Mickelson, A.~Mustafayev, and X.~Tata, {\it
  {Post-LHC7 fine-tuning in the minimal supergravity/CMSSM model with a 125 GeV
  Higgs boson}},  {\em Phys. Rev.} {\bf D87} (2013), no.~3 035017,
  [\href{http://arxiv.org/abs/1210.3019}{{\tt arXiv:1210.3019}}].

\bibitem{Boehm:2013gst}
C.~Boehm, P.~S.~B. Dev, A.~Mazumdar, and E.~Pukartas, {\it {Naturalness of
  light neutralino dark matter in pMSSM after LHC, XENON100 and Planck data}},
  {\em JHEP} {\bf 1306} (2013) 113, [\href{http://arxiv.org/abs/1303.5386}{{\tt
  arXiv:1303.5386}}].

\bibitem{Balazs:2013qva}
C.~Balazs, A.~Buckley, D.~Carter, B.~Farmer, and M.~White, {\it {Should we
  still believe in constrained supersymmetry?}},  {\em Eur. Phys. J.} {\bf C73}
  (2013) 2563, [\href{http://arxiv.org/abs/1205.1568}{{\tt arXiv:1205.1568}}].

\bibitem{Kim:2013uxa}
D.~Kim, P.~Athron, C.~Bal\'{a}zs, B.~Farmer, and E.~Hutchison, {\it {Bayesian
  naturalness of the CMSSM and CNMSSM}},  {\em Phys. Rev.} {\bf D90} (2014),
  no.~5 055008, [\href{http://arxiv.org/abs/1312.4150}{{\tt arXiv:1312.4150}}].

\bibitem{Casas:2014eca}
J.~A. Casas, J.~M. Moreno, S.~Robles, K.~Rolbiecki, and B.~Zald\'{i}var, {\it
  {What is a Natural SUSY scenario?}},  {\em JHEP} {\bf 1506} (2015) 070,
  [\href{http://arxiv.org/abs/1407.6966}{{\tt arXiv:1407.6966}}].

\bibitem{PhysRevD.91.075005}
H.~Baer, V.~Barger, P.~Huang, D.~Mickelson, M.~Padeffke-Kirkland, and X.~Tata,
  {\it {Natural SUSY with a bino- or wino-like LSP}},  {\em Phys. Rev.} {\bf
  D91} (Apr, 2015) 075005, [\href{http://arxiv.org/abs/1501.06357}{{\tt
  arXiv:1501.06357}}].

\bibitem{Drees:2015aeo}
M.~Drees and J.~S. Kim, {\it {Minimal natural supersymmetry after the LHC8}},
  {\em Phys. Rev.} {\bf D93} (2016), no.~9 095005,
  [\href{http://arxiv.org/abs/1511.04461}{{\tt arXiv:1511.04461}}].

\bibitem{Baer:2015rja}
H.~Baer, V.~Barger, and M.~Savoy, {\it {Upper bounds on sparticle masses from
  naturalness or how to disprove weak scale supersymmetry}},  {\em Phys. Rev.}
  {\bf D93} (2016), no.~3 035016, [\href{http://arxiv.org/abs/1509.02929}{{\tt
  arXiv:1509.02929}}].

\bibitem{Kim:2016rsd}
J.~S. Kim, K.~Rolbiecki, R.~Ruiz, J.~Tattersall, and T.~Weber, {\it {Prospects
  for natural SUSY}},  {\em Phys. Rev.} {\bf D94} (2016), no.~9 095013,
  [\href{http://arxiv.org/abs/1606.06738}{{\tt arXiv:1606.06738}}].

\bibitem{Casas:2016xnl}
J.~A. Casas, J.~M. Moreno, S.~Robles, and K.~Rolbiecki, {\it {Reducing the
  fine-tuning of gauge-mediated SUSY breaking}},  {\em Eur. Phys. J.} {\bf C76}
  (2016), no.~8 450, [\href{http://arxiv.org/abs/1602.06892}{{\tt
  arXiv:1602.06892}}].

\bibitem{vanBeekveld:2016hug}
M.~van Beekveld, W.~Beenakker, S.~Caron, R.~Peeters, and R.~Ruiz~de Austri,
  {\it {Supersymmetry with dark matter is still natural}},  {\em Phys. Rev.}
  {\bf D96} (2017), no.~3 035015, [\href{http://arxiv.org/abs/1612.06333}{{\tt
  arXiv:1612.06333}}].

\bibitem{Cici:2016oqr}
A.~\c{C}i\c{c}i, Z.~Kirca, and C.~S. Un, {\it {Light stops and fine-tuning in
  MSSM}},  {\em Eur. Phys. J.} {\bf C78} (2018), no.~1 60,
  [\href{http://arxiv.org/abs/1611.05270}{{\tt arXiv:1611.05270}}].

\bibitem{Cabrera:2016wwr}
M.~E. Cabrera, J.~A. Casas, A.~Delgado, S.~Robles, and R.~Ruiz~de Austri, {\it
  {Naturalness of MSSM dark matter}},  {\em JHEP} {\bf 1608} (2016) 058,
  [\href{http://arxiv.org/abs/1604.02102}{{\tt arXiv:1604.02102}}].

\bibitem{Buckley:2016kvr}
M.~R. Buckley, D.~Feld, S.~Macaluso, A.~Monteux, and D.~Shih, {\it {Cornering
  Natural SUSY at LHC Run II and Beyond}},  {\em JHEP} {\bf 1708} (2017) 115,
  [\href{http://arxiv.org/abs/1610.08059}{{\tt arXiv:1610.08059}}].

\bibitem{Baer:2017pba}
H.~Baer, V.~Barger, J.~S. Gainer, H.~Serce, and X.~Tata, {\it {Reach of the
  high-energy LHC for gluinos and top squarks in SUSY models with light
  Higgsinos}},  {\em Phys. Rev.} {\bf D96} (2017), no.~11 115008,
  [\href{http://arxiv.org/abs/1708.09054}{{\tt arXiv:1708.09054}}].

\bibitem{Abdughani:2017dqs}
M.~Abdughani, L.~Wu, and J.~M. Yang, {\it {Status and prospects of light bino
  higgsino dark matter in natural SUSY}},  {\em Eur. Phys. J.} {\bf C78}
  (2018), no.~1 4, [\href{http://arxiv.org/abs/1705.09164}{{\tt
  arXiv:1705.09164}}].

\bibitem{Fundira:2017vip}
P.~Fundira and A.~Purves, {\it {Bayesian naturalness, simplicity, and
  testability applied to the $B−L$ MSSM GUT}},  {\em Int. J. Mod. Phys.} {\bf
  A33} (2018), no.~11 1841004, [\href{http://arxiv.org/abs/1708.07835}{{\tt
  arXiv:1708.07835}}].

\bibitem{Baer:2018rhs}
H.~Baer, V.~Barger, D.~Sengupta, and X.~Tata, {\it {Is natural higgsino-only
  dark matter excluded?}},  {\em Eur. Phys. J.} {\bf C78} (2018), no.~10 838,
  [\href{http://arxiv.org/abs/1803.11210}{{\tt arXiv:1803.11210}}].

\bibitem{Ellis:1986yg}
J.~R. Ellis, K.~Enqvist, D.~V. Nanopoulos, and F.~Zwirner, {\it {Observables in
  Low-Energy Superstring Models}},  {\em Mod. Phys. Lett.} {\bf A1} (1986) 57.

\bibitem{Kane:1993td}
G.~L. Kane, C.~F. Kolda, L.~Roszkowski, and J.~D. Wells, {\it {Study of
  constrained minimal supersymmetry}},  {\em Phys. Rev.} {\bf D49} (1994)
  6173--6210, [\href{http://arxiv.org/abs/hep-ph/9312272}{{\tt
  hep-ph/9312272}}].

\bibitem{Anderson:1994dz}
G.~W. Anderson and D.~J. Castano, {\it {Measures of fine-tuning}},  {\em Phys.
  Lett.} {\bf B347} (1995) 300--308,
  [\href{http://arxiv.org/abs/hep-ph/9409419}{{\tt hep-ph/9409419}}].

\bibitem{Anderson:1994tr}
G.~W. Anderson and D.~J. Castano, {\it {Naturalness and superpartner masses or
  when to give up on weak scale supersymmetry}},  {\em Phys. Rev.} {\bf D52}
  (1995) 1693--1700, [\href{http://arxiv.org/abs/hep-ph/9412322}{{\tt
  hep-ph/9412322}}].

\bibitem{Dimopoulos:1995mi}
S.~Dimopoulos and G.~F. Giudice, {\it {Naturalness constraints in
  supersymmetric theories with nonuniversal soft terms}},  {\em Phys. Lett.}
  {\bf B357} (1995) 573--578, [\href{http://arxiv.org/abs/hep-ph/9507282}{{\tt
  hep-ph/9507282}}].

\bibitem{Chankowski:1997zh}
P.~H. Chankowski, J.~R. Ellis, and S.~Pokorski, {\it {The fine-tuning price of
  LEP}},  {\em Phys. Lett.} {\bf B423} (1998) 327--336,
  [\href{http://arxiv.org/abs/hep-ph/9712234}{{\tt hep-ph/9712234}}].

\bibitem{Chankowski:1998xv}
P.~H. Chankowski, J.~R. Ellis, M.~Olechowski, and S.~Pokorski, {\it {Haggling
  over the fine-tuning price of LEP}},  {\em Nucl. Phys.} {\bf B544} (1999)
  39--63, [\href{http://arxiv.org/abs/hep-ph/9808275}{{\tt hep-ph/9808275}}].

\bibitem{Casas:2003jx}
J.~A. Casas, J.~R. Espinosa, and I.~Hidalgo, {\it {The MSSM fine-tuning
  problem: a way out}},  {\em JHEP} {\bf 0401} (2004) 008,
  [\href{http://arxiv.org/abs/hep-ph/0310137}{{\tt hep-ph/0310137}}].

\bibitem{Papucci:2011wy}
M.~Papucci, J.~T. Ruderman, and A.~Weiler, {\it {Natural SUSY Endures}},  {\em
  JHEP} {\bf 1209} (2012) 035, [\href{http://arxiv.org/abs/1110.6926}{{\tt
  arXiv:1110.6926}}].

\bibitem{Liu:2013ula}
M.~Liu and P.~Nath, {\it {Higgs boson mass, proton decay, naturalness, and
  constraints of the LHC and Planck data}},  {\em Phys. Rev.} {\bf D87} (2013),
  no.~9 095012, [\href{http://arxiv.org/abs/1303.7472}{{\tt arXiv:1303.7472}}].

\bibitem{Kowalska:2013ica}
K.~Kowalska and E.~M. Sessolo, {\it {Natural MSSM after the LHC 8 TeV run}},
  {\em Phys. Rev.} {\bf D88} (2013), no.~7 075001,
  [\href{http://arxiv.org/abs/1307.5790}{{\tt arXiv:1307.5790}}].

\bibitem{Arvanitaki:2013yja}
A.~Arvanitaki, M.~Baryakhtar, X.~Huang, K.~van Tilburg, and G.~Villadoro, {\it
  {The last vestiges of naturalness}},  {\em JHEP} {\bf 1403} (2014) 022,
  [\href{http://arxiv.org/abs/1309.3568}{{\tt arXiv:1309.3568}}].

\bibitem{Lykken:2014bca}
J.~Lykken and M.~Spiropulu, {\it {Supersymmetry and the crisis in physics}},
  {\em Sci. Am.} {\bf 310N5} (2014) 36--39.

\bibitem{Cassel:2009ps}
S.~Cassel, D.~M. Ghilencea, and G.~G. Ross, {\it {Fine-tuning as an indication
  of physics beyond the MSSM}},  {\em Nucl. Phys.} {\bf B825} (2010) 203--221,
  [\href{http://arxiv.org/abs/0903.1115}{{\tt arXiv:0903.1115}}].

\bibitem{Ross:2011xv}
G.~G. Ross and K.~Schmidt-Hoberg, {\it {The fine-tuning of the generalised
  NMSSM}},  {\em Nucl. Phys.} {\bf B862} (2012) 710--719,
  [\href{http://arxiv.org/abs/1108.1284}{{\tt arXiv:1108.1284}}].

\bibitem{Gogoladze:2012yf}
I.~Gogoladze, F.~Nasir, and Q.~Shafi, {\it {Non-universal gaugino masses and
  natural supersymmetry}},  {\em Int. J. Mod. Phys.} {\bf A28} (2013) 1350046,
  [\href{http://arxiv.org/abs/1212.2593}{{\tt arXiv:1212.2593}}].

\bibitem{Gogoladze:2013wva}
I.~Gogoladze, F.~Nasir, and Q.~Shafi, {\it {SO(10) as a framework for natural
  supersymmetry}},  {\em JHEP} {\bf 1311} (2013) 173,
  [\href{http://arxiv.org/abs/1306.5699}{{\tt arXiv:1306.5699}}].

\bibitem{Kaminska:2013mya}
A.~Kaminska, G.~G. Ross, and K.~Schmidt-Hoberg, {\it {Non-universal gaugino
  masses and fine-tuning implications for SUSY searches in the MSSM and the
  GNMSSM}},  {\em JHEP} {\bf 1311} (2013) 209,
  [\href{http://arxiv.org/abs/1308.4168}{{\tt arXiv:1308.4168}}].

\bibitem{Cao:2016nix}
J.~Cao, Y.~He, L.~Shang, W.~Su, and Y.~Zhang, {\it {Natural NMSSM after LHC Run
  I and the higgsino dominated dark matter scenario}},  {\em JHEP} {\bf 1608}
  (2016) 037, [\href{http://arxiv.org/abs/1606.04416}{{\tt arXiv:1606.04416}}].

\bibitem{Cao:2016cnv}
J.~Cao, Y.~He, L.~Shang, W.~Su, P.~Wu, and Y.~Zhang, {\it {Strong constraints
  of LUX-2016 results on the natural NMSSM}},  {\em JHEP} {\bf 1610} (2016)
  136, [\href{http://arxiv.org/abs/1609.00204}{{\tt arXiv:1609.00204}}].

\bibitem{Li:2017fbg}
C.~Li, B.~Zhu, and T.~Li, {\it {Naturalness, dark matter, and the muon
  anomalous magnetic moment in supersymmetric extensions of the standard model
  with a pseudo-Dirac gluino}},  {\em Nucl. Phys.} {\bf B927} (2018) 255--273,
  [\href{http://arxiv.org/abs/1704.05584}{{\tt arXiv:1704.05584}}].

\bibitem{Zhu:2017moa}
B.~Zhu, F.~Staub, and R.~Ding, {\it {Naturalness and a light $Z'$}},  {\em
  Phys. Rev.} {\bf D96} (2017), no.~3 035038,
  [\href{http://arxiv.org/abs/1707.03101}{{\tt arXiv:1707.03101}}].

\bibitem{Athron:2017fxj}
P.~Athron, C.~Balazs, B.~Farmer, A.~Fowlie, D.~Harries, and D.~Kim, {\it
  {Bayesian analysis and naturalness of (next-to-)minimal supersymmetric
  models}},  {\em JHEP} {\bf 1710} (2017) 160,
  [\href{http://arxiv.org/abs/1709.07895}{{\tt arXiv:1709.07895}}].

\bibitem{Alvarado:2018rfl}
C.~Alvarado, A.~Delgado, and A.~Martin, {\it {Constraining the $R$-symmetric
  chargino NLSP at the LHC}},  {\em Phys. Rev.} {\bf D97} (2018), no.~11
  115044, [\href{http://arxiv.org/abs/1803.00624}{{\tt arXiv:1803.00624}}].

\bibitem{Du:2018pko}
X.~K. Du, G.-L. Liu, F.~Wang, W.~Wang, J.~M. Yang, and Y.~Zhang, {\it {NMSSM
  with generalized deflected mirage mediation}},  {\em Eur. Phys. J.} {\bf C79}
  (2019), no.~5 397, [\href{http://arxiv.org/abs/1804.07335}{{\tt
  arXiv:1804.07335}}].

\bibitem{Badziak:2018nnf}
M.~Badziak and K.~Harigaya, {\it {Impact of an extra gauge interaction on
  naturalness of supersymmetry}},  {\em JHEP} {\bf 1808} (2018) 080,
  [\href{http://arxiv.org/abs/1806.07900}{{\tt arXiv:1806.07900}}].

\bibitem{Kobakhidze:2018vuy}
A.~Kobakhidze and M.~Talia, {\it {Supersymmetric naturalness beyond MSSM}},
  \href{http://arxiv.org/abs/1806.08502}{{\tt arXiv:1806.08502}}.

\bibitem{Yanagida:2018arr}
T.~T. Yanagida and N.~Yokozaki, {\it {Focus point gauge mediation without a
  severe fine-tuning}},  {\em JHEP} {\bf 1810} (2018) 149,
  [\href{http://arxiv.org/abs/1809.00787}{{\tt arXiv:1809.00787}}].

\bibitem{Badziak:2018ijy}
M.~Badziak, N.~Desai, C.~Hugonie, and R.~Ziegler, {\it {Extended gauge
  mediation in the NMSSM with displaced LHC signals}},  {\em Eur. Phys. J.}
  {\bf C79} (2019), no.~1 67, [\href{http://arxiv.org/abs/1810.05618}{{\tt
  arXiv:1810.05618}}].

\bibitem{Cao:2018rix}
J.~Cao, Y.~He, L.~Shang, Y.~Zhang, and P.~Zhu, {\it {Current status of a
  natural NMSSM in light of LHC 13 TeV data and XENON-1T results}},  {\em Phys.
  Rev.} {\bf D99} (2019), no.~7 075020,
  [\href{http://arxiv.org/abs/1810.09143}{{\tt arXiv:1810.09143}}].

\bibitem{Wang:2018vxp}
K.~Wang, F.~Wang, J.~Zhu, and Q.~Jie, {\it {The semi-constrained NMSSM in light
  of muon g-2, LHC, and dark matter constraints}},  {\em Chin. Phys.} {\bf C42}
  (2018), no.~10 103109--103109, [\href{http://arxiv.org/abs/1811.04435}{{\tt
  arXiv:1811.04435}}].

\bibitem{Baer:2013gva}
H.~Baer, V.~Barger, and D.~Mickelson, {\it {How conventional measures
  overestimate electroweak fine-tuning in supersymmetric theory}},  {\em Phys.
  Rev.} {\bf D88} (2013), no.~9 095013,
  [\href{http://arxiv.org/abs/1309.2984}{{\tt arXiv:1309.2984}}].

\bibitem{Cabrera:2008tj}
M.~E. Cabrera, J.~A. Casas, and R.~Ruiz~de Austri, {\it {Bayesian approach and
  naturalness in MSSM analyses for the LHC}},  {\em JHEP} {\bf 0903} (2009)
  075, [\href{http://arxiv.org/abs/0812.0536}{{\tt arXiv:0812.0536}}].

\bibitem{Djouadi:1998di}
{\textbf{MSSM Working Group}}, {\it {The minimal supersymmetric standard model:
  group summary report}},  in {\em {GDR (Groupement De Recherche) -
  Supersymetrie Montpellier, France, April 15-17, 1998}}, 1998.
\newblock \href{http://arxiv.org/abs/9901246}{{\tt 9901246}}.

\bibitem{Coleman:1973jx}
S.~R. Coleman and E.~J. Weinberg, {\it {Radiative corrections as the origin of
  spontaneous symmetry breaking}},  {\em Phys. Rev.} {\bf D7} (1973)
  1888--1910.

\bibitem{Chamseddine:1982jx}
A.~H. Chamseddine, R.~L. Arnowitt, and P.~Nath, {\it {Locally supersymmetric
  grand unification}},  {\em Phys. Rev. Lett.} {\bf 49} (1982) 970.

\bibitem{Barbieri:1982eh}
R.~Barbieri, S.~Ferrara, and C.~A. Savoy, {\it {Gauge models with spontaneously
  broken local supersymmetry}},  {\em Phys. Lett.} {\bf B119} (1982) 343.

\bibitem{Ohta:1982wn}
N.~Ohta, {\it Grand unified theories based on local supersymmetry},  {\em Prog.
  Theor. Phys.} {\bf 70} (1983) 542.

\bibitem{Hall:1983iz}
L.~J. Hall, J.~D. Lykken, and S.~Weinberg, {\it {Supergravity as the messenger
  of supersymmetry breaking}},  {\em Phys. Rev.} {\bf D27} (1983) 2359--2378.

\bibitem{Cabrera:2013jya}
M.~E. Cabrera, A.~Casas, R.~Ruiz~de Austri, and G.~Bertone, {\it {LHC and dark
  matter phenomenology of the NUGHM}},  {\em JHEP} {\bf 1412} (2014) 114,
  [\href{http://arxiv.org/abs/1311.7152}{{\tt arXiv:1311.7152}}].

\bibitem{Peiro:2016ykr}
M.~Peiro and S.~Robles, {\it {Low-mass neutralino dark matter in supergravity
  scenarios: phenomenology and naturalness}},  {\em JCAP} {\bf 1705} (2017),
  no.~05 010, [\href{http://arxiv.org/abs/1612.00460}{{\tt arXiv:1612.00460}}].

\bibitem{Delgado:2014vha}
A.~Delgado, M.~Quiros, and C.~Wagner, {\it {General focus point in the MSSM}},
  {\em JHEP} {\bf 1404} (2014) 093, [\href{http://arxiv.org/abs/1402.1735}{{\tt
  arXiv:1402.1735}}].

\bibitem{PhysRevD.89.115019}
H.~Baer, V.~Barger, D.~Mickelson, and M.~Padeffke-Kirkland, {\it {SUSY models
  under siege: LHC constraints and electroweak fine-tuning}},  {\em Phys. Rev.}
  {\bf D89} (Jun, 2014) 115019.

\bibitem{Mustafayev:2014lqa}
A.~Mustafayev and X.~Tata, {\it {Supersymmetry, naturalness, and light
  higgsinos}},  {\em Indian J. Phys.} {\bf 88} (2014) 991--1004,
  [\href{http://arxiv.org/abs/1404.1386}{{\tt arXiv:1404.1386}}].

\bibitem{Buckley:2016tbs}
M.~R. Buckley, A.~Monteux, and D.~Shih, {\it {Precision corrections to
  fine-tuning in SUSY}},  {\em JHEP} {\bf 1706} (2017) 103,
  [\href{http://arxiv.org/abs/1611.05873}{{\tt arXiv:1611.05873}}].

\bibitem{1232326}
J.~H. Kotecha and P.~M. Djuric, {\it Gaussian particle filtering},  {\em IEEE
  Transactions on Signal Processing} {\bf 51} (Oct, 2003) 2592--2601.

\bibitem{Allanach:2001kg}
B.~C. Allanach, {\it {SOFTSUSY: a program for calculating supersymmetric
  spectra}},  {\em Comput. Phys. Commun.} {\bf 143} (2002) 305--331,
  [\href{http://arxiv.org/abs/hep-ph/0104145}{{\tt hep-ph/0104145}}].

\bibitem{Bahl:2016brp}
H.~Bahl and W.~Hollik, {\it {Precise prediction for the light MSSM Higgs boson
  mass combining effective field theory and fixed-order calculations}},  {\em
  Eur. Phys. J.} {\bf C76} (2016), no.~9 499,
  [\href{http://arxiv.org/abs/1608.01880}{{\tt arXiv:1608.01880}}].

\bibitem{Hahn:2013ria}
T.~Hahn, S.~Heinemeyer, W.~Hollik, H.~Rzehak, and G.~Weiglein, {\it
  {High-precision predictions for the Light CP-even Higgs boson mass of the
  Minimal Supersymmetric Standard Model}},  {\em Phys. Rev. Lett.} {\bf 112}
  (2014), no.~14 141801, [\href{http://arxiv.org/abs/1312.4937}{{\tt
  arXiv:1312.4937}}].

\bibitem{Frank:2006yh}
M.~Frank, T.~Hahn, S.~Heinemeyer, W.~Hollik, H.~Rzehak, and G.~Weiglein, {\it
  {The Higgs boson masses and mixings of the complex MSSM in the
  Feynman-diagrammatic approach}},  {\em JHEP} {\bf 0702} (2007) 047,
  [\href{http://arxiv.org/abs/0611326}{{\tt 0611326}}].

\bibitem{Degrassi:2002fi}
G.~Degrassi, S.~Heinemeyer, W.~Hollik, P.~Slavich, and G.~Weiglein, {\it
  {Towards high precision predictions for the MSSM Higgs sector}},  {\em Eur.
  Phys. J.} {\bf C28} (2003) 133--143,
  [\href{http://arxiv.org/abs/0212020}{{\tt 0212020}}].

\bibitem{Heinemeyer:1998yj}
S.~Heinemeyer, W.~Hollik, and G.~Weiglein, {\it {FeynHiggs: A program for the
  calculation of the masses of the neutral CP even Higgs bosons in the MSSM}},
  {\em Comput. Phys. Commun.} {\bf 124} (2000) 76--89,
  [\href{http://arxiv.org/abs/9812320}{{\tt 9812320}}].

\bibitem{Djouadi:2006bz}
A.~Djouadi, M.~M. Muhlleitner, and M.~Spira, {\it {Decays of supersymmetric
  particles: the program SUSY-HIT (SUspect-SdecaY-Hdecay-InTerface)}},  {\em
  Acta Phys. Polon.} {\bf B38} (2007) 635--644,
  [\href{http://arxiv.org/abs/hep-ph/0609292}{{\tt hep-ph/0609292}}].

\bibitem{Barducci:2016pcb}
D.~Barducci, G.~Belanger, J.~Bernon, F.~Boudjema, J.~Da~Silva, S.~Kraml,
  U.~Laa, and A.~Pukhov, {\it {Collider limits on new physics within
  MicrOMEGAs-4.3}},  {\em Comput. Phys. Commun.} {\bf 222} (2018) 327--338,
  [\href{http://arxiv.org/abs/1606.03834}{{\tt arXiv:1606.03834}}].

\bibitem{Belanger:2018mqt}
G.~B\'{e}langer, F.~Boudjema, A.~Goudelis, A.~Pukhov, and B.~Zaldivar, {\it
  {MicrOMEGAs-5.0 : freeze-in}},  {\em Comput. Phys. Commun.} {\bf 231} (2018)
  173--186, [\href{http://arxiv.org/abs/1801.03509}{{\tt arXiv:1801.03509}}].

\bibitem{ddcalc:2017lvb}
{\textbf{GAMBIT Dark Matter Workgroup}}, {\it {DarkBit: A GAMBIT module for
  computing dark matter observables and likelihoods}},
  \href{http://arxiv.org/abs/1705.07920}{{\tt arXiv:1705.07920}}.

\bibitem{Aprile:2018dbl}
{\bf XENON} Collaboration, E.~Aprile et~al., {\it {Dark matter search results
  from a one ton-year exposure of XENON1T}},  {\em Phys. Rev. Lett.} {\bf 121}
  (2018), no.~11 111302, [\href{http://arxiv.org/abs/1805.12562}{{\tt
  arXiv:1805.12562}}].

\bibitem{Aprile:2019dbj}
{\bf XENON} Collaboration, E.~Aprile et~al., {\it {Constraining the
  spin-dependent WIMP-nucleon cross sections with XENON1T}},  {\em Phys. Rev.
  Lett.} {\bf 122} (2019), no.~14 141301,
  [\href{http://arxiv.org/abs/1902.03234}{{\tt arXiv:1902.03234}}].

\bibitem{Amole:2017dex}
{\bf PICO} Collaboration, C.~Amole et~al., {\it {Dark Matter Search Results
  from the PICO-60 C$_3$F$_8$ Bubble Chamber}},  {\em Phys. Rev. Lett.} {\bf
  118} (2017), no.~25 251301, [\href{http://arxiv.org/abs/1702.07666}{{\tt
  arXiv:1702.07666}}].

\bibitem{Amole:2016pye}
{\bf PICO} Collaboration, C.~Amole et~al., {\it {Improved dark matter search
  results from PICO-2L Run 2}},  {\em Phys. Rev.} {\bf D93} (2016), no.~6
  061101, [\href{http://arxiv.org/abs/1601.03729}{{\tt arXiv:1601.03729}}].

\bibitem{Amole:2019fdf}
{\bf PICO} Collaboration, C.~Amole et~al., {\it {Dark Matter Search Results
  from the Complete Exposure of the PICO-60 C$_3$F$_8$ Bubble Chamber}},
  \href{http://arxiv.org/abs/1902.04031}{{\tt arXiv:1902.04031}}.

\bibitem{Tan:2016zwf}
{\bf PandaX-II} Collaboration, A.~Tan et~al., {\it {Dark matter results from
  first 98.7 days of data from the PandaX-II experiment}},  {\em Phys. Rev.
  Lett.} {\bf 117} (2016), no.~12 121303,
  [\href{http://arxiv.org/abs/1607.07400}{{\tt arXiv:1607.07400}}].

\bibitem{Xia:2018qgs}
{\bf PandaX-II} Collaboration, J.~Xia et~al., {\it {PandaX-II constraints on
  spin-dependent WIMP-nucleon effective interactions}},  {\em Phys. Lett.} {\bf
  B792} (2019) 193--198, [\href{http://arxiv.org/abs/1807.01936}{{\tt
  arXiv:1807.01936}}].

\bibitem{Ackermann:2015zua}
{\bf Fermi-LAT} Collaboration, M.~Ackermann et~al., {\it {Searching for Dark
  Matter Annihilation from Milky Way Dwarf Spheroidal Galaxies with Six Years
  of Fermi Large Area Telescope Data}},  {\em Phys. Rev. Lett.} {\bf 115}
  (2015), no.~23 231301, [\href{http://arxiv.org/abs/1503.02641}{{\tt
  arXiv:1503.02641}}].

\bibitem{Aghanim:2018eyx}
{\bf Planck} Collaboration, N.~Aghanim et~al., {\it {Planck 2018 results. VI.
  Cosmological parameters}},  \href{http://arxiv.org/abs/1807.06209}{{\tt
  arXiv:1807.06209}}.

\bibitem{Caron:2016hib}
S.~Caron, J.~S. Kim, K.~Rolbiecki, R.~Ruiz~de Austri, and B.~Stienen, {\it {The
  BSM-AI project: SUSY-AI–generalizing LHC limits on supersymmetry with
  machine learning}},  {\em Eur. Phys. J.} {\bf C77} (2017), no.~4 257,
  [\href{http://arxiv.org/abs/1605.02797}{{\tt arXiv:1605.02797}}].

\bibitem{Barr:2016sho}
A.~Barr and J.~Liu, {\it {Analysing parameter space correlations of recent 13
  TeV gluino and squark searches in the pMSSM}},  {\em Eur. Phys. J.} {\bf C77}
  (2017), no.~3 202, [\href{http://arxiv.org/abs/1608.05379}{{\tt
  arXiv:1608.05379}}].

\bibitem{Ambrogi:2018ujg}
F.~Ambrogi et~al., {\it {SModelS v1.2: long-lived particles, combination of
  signal regions, and other novelties}},
  \href{http://arxiv.org/abs/1811.10624}{{\tt arXiv:1811.10624}}.

\bibitem{Heisig:2018kfq}
J.~Heisig, S.~Kraml, and A.~Lessa, {\it {Constraining new physics with searches
  for long-lived particles: implementation into SModelS}},  {\em Phys. Lett.}
  {\bf B788} (2019) 87--95, [\href{http://arxiv.org/abs/1808.05229}{{\tt
  arXiv:1808.05229}}].

\bibitem{Dutta:2018ioj}
J.~Dutta, S.~Kraml, A.~Lessa, and W.~Waltenberger, {\it {SModelS extension with
  the CMS supersymmetry search results from Run 2}},  {\em LHEP} {\bf 1}
  (2018), no.~1 5--12, [\href{http://arxiv.org/abs/1803.02204}{{\tt
  arXiv:1803.02204}}].

\bibitem{Ambrogi:2017neo}
F.~Ambrogi, S.~Kraml, S.~Kulkarni, U.~Laa, A.~Lessa, V.~Magerl, J.~Sonneveld,
  M.~Traub, and W.~Waltenberger, {\it {SModelS v1.1 user manual: Improving
  simplified model constraints with efficiency maps}},  {\em Comput. Phys.
  Commun.} {\bf 227} (2018) 72--98,
  [\href{http://arxiv.org/abs/1701.06586}{{\tt arXiv:1701.06586}}].

\bibitem{Kraml:2013mwa}
S.~Kraml, S.~Kulkarni, U.~Laa, A.~Lessa, W.~Magerl, D.~Proschofsky-Spindler,
  and W.~Waltenberger, {\it {SModelS: a tool for interpreting simplified-model
  results from the LHC and its application to supersymmetry}},  {\em Eur. Phys.
  J.} {\bf C74} (2014) 2868, [\href{http://arxiv.org/abs/1312.4175}{{\tt
  arXiv:1312.4175}}].

\bibitem{Bechtle:2015pma}
P.~Bechtle, S.~Heinemeyer, O.~St\r{a}l, T.~Stefaniak, and G.~Weiglein, {\it
  {Applying exclusion likelihoods from LHC searches to extended Higgs
  sectors}},  {\em Eur. Phys. J.} {\bf C75} (2015), no.~9 421,
  [\href{http://arxiv.org/abs/1507.06706}{{\tt arXiv:1507.06706}}].

\bibitem{Bechtle:2013wla}
P.~Bechtle, O.~Brein, S.~Heinemeyer, O.~St\r{a}l, T.~Stefaniak, G.~Weiglein,
  and K.~E. Williams, {\it {$\mathsf{HiggsBounds}-4$: Improved tests of
  extended Higgs sectors against exclusion bounds from LEP, the Tevatron and
  the LHC}},  {\em Eur. Phys. J.} {\bf C74} (2014), no.~3 2693,
  [\href{http://arxiv.org/abs/1311.0055}{{\tt arXiv:1311.0055}}].

\bibitem{Bechtle:2013gu}
P.~Bechtle, O.~Brein, S.~Heinemeyer, O.~St\r{a}l, T.~Stefaniak, G.~Weiglein,
  and K.~Williams, {\it {Recent developments in HiggsBounds and a preview of
  HiggsSignals}},  {\em PoS} {\bf 2012} (2012) 024,
  [\href{http://arxiv.org/abs/1301.2345}{{\tt arXiv:1301.2345}}].

\bibitem{Bechtle:2011sb}
P.~Bechtle, O.~Brein, S.~Heinemeyer, G.~Weiglein, and K.~E. Williams, {\it
  {HiggsBounds 2.0.0: Confronting neutral and charged Higgs sector predictions
  with exclusion bounds from LEP and the Tevatron}},  {\em Comput. Phys.
  Commun.} {\bf 182} (2011) 2605--2631,
  [\href{http://arxiv.org/abs/1102.1898}{{\tt arXiv:1102.1898}}].

\bibitem{Bechtle:2008jh}
P.~Bechtle, O.~Brein, S.~Heinemeyer, G.~Weiglein, and K.~E. Williams, {\it
  {HiggsBounds: confronting arbitrary Higgs sectors with exclusion bounds from
  LEP and the Tevatron}},  {\em Comput. Phys. Commun.} {\bf 181} (2010)
  138--167, [\href{http://arxiv.org/abs/0811.4169}{{\tt arXiv:0811.4169}}].

\bibitem{Stal:2013hwa}
O.~St\r{a}l and T.~Stefaniak, {\it {Constraining extended Higgs sectors with
  HiggsSignals}},  {\em PoS} {\bf 2013} (2013) 314,
  [\href{http://arxiv.org/abs/1310.4039}{{\tt arXiv:1310.4039}}].

\bibitem{Bechtle:2014ewa}
P.~Bechtle, S.~Heinemeyer, O.~St\r{a}l, T.~Stefaniak, and G.~Weiglein, {\it
  {Probing the Standard Model with Higgs signal rates from the Tevatron, the
  LHC and a future ILC}},  {\em JHEP} {\bf 1411} (2014) 039,
  [\href{http://arxiv.org/abs/1403.1582}{{\tt arXiv:1403.1582}}].

\bibitem{Bechtle:2013xfa}
P.~Bechtle, S.~Heinemeyer, O.~St\r{a}l, T.~Stefaniak, and G.~Weiglein, {\it
  {$HiggsSignals$: Confronting arbitrary Higgs sectors with measurements at the
  Tevatron and the LHC}},  {\em Eur. Phys. J.} {\bf C74} (2014), no.~2 2711,
  [\href{http://arxiv.org/abs/1305.1933}{{\tt arXiv:1305.1933}}].

\bibitem{Camargo-Molina:2013qva}
J.~E. Camargo-Molina, B.~O'Leary, W.~Porod, and F.~Staub, {\it
  {$\mathbf{Vevacious}$: A tool for finding the global minima of one-loop
  effective potentials with many scalars}},  {\em Eur. Phys. J.} {\bf C73}
  (2013), no.~10 2588, [\href{http://arxiv.org/abs/1307.1477}{{\tt
  arXiv:1307.1477}}].

\bibitem{Lee2008}
T.~L. Lee, T.~Y. Li, and C.~H. Tsai, {\it Hom4ps-2.0: a software package for
  solving polynomial systems by the polyhedral homotopy continuation method},
  {\em Computing} {\bf 83} (2008), no.~2 109.

\bibitem{Wainwright:2011kj}
C.~L. Wainwright, {\it {CosmoTransitions: Computing cosmological phase
  transition temperatures and bubble profiles with multiple fields}},  {\em
  Comput. Phys. Commun.} {\bf 183} (2012) 2006--2013,
  [\href{http://arxiv.org/abs/1109.4189}{{\tt arXiv:1109.4189}}].

\bibitem{LEP:working}
``{LEP2 SUSY Working Group, ALEPH, DELPHI, L3 and OPAL experiments}.''

\bibitem{Carena:2003aj}
M.~Carena, A.~de~Gouvea, A.~Freitas, and M.~Schmitt, {\it {Invisible Z boson
  decays at $e^+ e^-$ colliders}},  {\em Phys. Rev.} {\bf D68} (2003) 113007,
  [\href{http://arxiv.org/abs/0308053}{{\tt 0308053}}].

\bibitem{Roberts:2010cj}
B.~L. Roberts, {\it {Status of the Fermilab muon $(g-2)$ experiment}},  {\em
  Chin. Phys.} {\bf C34} (2010) 741--744,
  [\href{http://arxiv.org/abs/1001.2898}{{\tt arXiv:1001.2898}}].

\bibitem{Aaij:2013aka}
{\textbf{LHCb} collaboration}, {\it {Measurement of the $B^0_s \to \mu^+ \mu^-$
  branching fraction and search for $B^0 \to \mu^+ \mu^-$ decays at the LHCb
  experiment}},  {\em Phys. Rev. Lett.} {\bf 111} (2013) 101805,
  [\href{http://arxiv.org/abs/1307.5024}{{\tt arXiv:1307.5024}}].

\bibitem{Misiak:2015xwa}
M.~Misiak et~al., {\it {Updated NNLO QCD predictions for the weak radiative
  $B$-meson decays}},  {\em Phys. Rev. Lett.} {\bf 114} (2015), no.~22 221801,
  [\href{http://arxiv.org/abs/1503.01789}{{\tt arXiv:1503.01789}}].

\bibitem{Czakon:2015exa}
M.~Czakon, P.~Fiedler, T.~Huber, M.~Misiak, T.~Schutzmeier, and M.~Steinhauser,
  {\it {The $(Q_{7}, Q_{1,2})$ contribution to $ \overline{B}\to {X}_s\gamma $
  at $ \mathcal{O}\left({\alpha}_{\mathrm{s}}^2\right) $}},  {\em JHEP} {\bf
  1504} (2015) 168, [\href{http://arxiv.org/abs/1503.01791}{{\tt
  arXiv:1503.01791}}].

\bibitem{Kronenbitter:2015kls}
B.~Kronenbitter et~al., {\it {Measurement of the branching fraction of $B^+
  \rightarrow \tau^+ \nu_{\tau}$ decays with the semileptonic tagging method}},
   {\em Phys. Rev.} {\bf D92} (03, 2015).

\bibitem{Widhalm:2007ws}
{\bf Belle} Collaboration, L.~Widhalm et~al., {\it {Measurement of B(D(s)+ --->
  mu(nu))}},  {\em Phys. Rev. Lett.} {\bf 100} (2008) 241801,
  [\href{http://arxiv.org/abs/0709.1340}{{\tt arXiv:0709.1340}}].

\bibitem{Onyisi:2009th}
P.~U.~E. Onyisi et~al.

\bibitem{CidVidal:2018eel}
X.~Cid~Vidal et~al., {\it {Beyond the Standard Model Physics at the HL-LHC and
  HE-LHC}},  \href{http://arxiv.org/abs/1812.07831}{{\tt arXiv:1812.07831}}.

\bibitem{Charles:2018vfv}
{\bf CLIC} Collaboration, T.~K. Charles et~al., {\it {The Compact Linear
  Collider (CLIC) - 2018 Summary Report}},  {\em CERN Yellow Rep. Monogr.} {\bf
  1802} (2018) 1--98, [\href{http://arxiv.org/abs/1812.06018}{{\tt
  arXiv:1812.06018}}].

\bibitem{Akerib:2015cja}
{\bf LZ} Collaboration, D.~S. Akerib et~al., {\it {LUX-ZEPLIN (LZ) Conceptual
  Design Report}},  \href{http://arxiv.org/abs/1509.02910}{{\tt
  arXiv:1509.02910}}.

\bibitem{Aalbers:2016jon}
{\bf DARWIN} Collaboration, J.~Aalbers et~al., {\it {DARWIN: towards the
  ultimate dark matter detector}},  {\em JCAP} {\bf 1611} (2016) 017,
  [\href{http://arxiv.org/abs/1606.07001}{{\tt arXiv:1606.07001}}].

\bibitem{Schumann:2015cpa}
M.~Schumann, L.~Baudis, L.~Butikofer, A.~Kish, and M.~Selvi, {\it {Dark matter
  sensitivity of multi-ton liquid xenon detectors}},  {\em JCAP} {\bf 1510}
  (2015), no.~10 016, [\href{http://arxiv.org/abs/1506.08309}{{\tt
  arXiv:1506.08309}}].

\bibitem{Aalseth:2017fik}
C.~E. Aalseth et~al., {\it {DarkSide-20k: A 20 tonne two-phase LAr TPC for
  direct dark matter detection at LNGS}},  {\em Eur. Phys. J. Plus} {\bf 133}
  (2018) 131, [\href{http://arxiv.org/abs/1707.08145}{{\tt arXiv:1707.08145}}].

\bibitem{pico500}
``{Toward a next-generation dark matter search with the PICO-40L bubble
  chamber}.'' \url{
  https://indico.cern.ch/event/606690/contributions/2623446/attachments/1497228/2330240/Fallows_2017_07_24__TAUP__PICO-40L_v1.2.pdf}.
\newblock Accessed: 2019-01-17.

\bibitem{Workgroup:2017lvb}
{\textbf{The GAMBIT Dark Matter Workgroup}}, {\it {DarkBit: A GAMBIT module for
  computing dark matter observables and likelihoods}},  {\em Eur. Phys. J.}
  {\bf C77} (2017), no.~12 831, [\href{http://arxiv.org/abs/1705.07920}{{\tt
  arXiv:1705.07920}}].

\bibitem{Aaboud:2018sua}
{\bf ATLAS} Collaboration, {\textbf{ATLAS} collaboration}, {\it {Search for
  chargino-neutralino production using recursive jigsaw reconstruction in final
  states with two or three charged leptons in proton-proton collisions at
  $\sqrt{s}=13$ TeV with the ATLAS detector}},  {\em Phys. Rev.} {\bf D98}
  (2018), no.~9 092012, [\href{http://arxiv.org/abs/1806.02293}{{\tt
  arXiv:1806.02293}}].

\bibitem{Sirunyan:2018lul}
{\textbf{CMS} collaboration}, {\it {Searches for pair production of charginos
  and top squarks in final states with two oppositely charged leptons in
  proton-proton collisions at $\sqrt{s}=$ 13 TeV}},  {\em JHEP} {\bf 1811}
  (2018) 079, [\href{http://arxiv.org/abs/1807.07799}{{\tt arXiv:1807.07799}}].

\bibitem{Sirunyan:2018omt}
{\textbf{CMS} collaboration}, {\it {Search for top squarks decaying via
  four-body or chargino-mediated modes in single-lepton final states in
  proton-proton collisions at $\sqrt{s} =$ 13 TeV}},  {\em JHEP} {\bf 09}
  (2018) 065, [\href{http://arxiv.org/abs/1805.05784}{{\tt arXiv:1805.05784}}].

\bibitem{Aaboud:2017ayj}
{\textbf{ATLAS} collaboration}, {\it {Search for a scalar partner of the top
  quark in the jets plus missing transverse momentum final state at
  $\sqrt{s}$=13 TeV with the ATLAS detector}},  {\em JHEP} {\bf 1712} (2017)
  085, [\href{http://arxiv.org/abs/1709.04183}{{\tt arXiv:1709.04183}}].

\bibitem{Sirunyan:2018vjp}
{\textbf{CMS} collaboration}, {\it {Search for natural and split supersymmetry
  in proton-proton collisions at $ \sqrt{s}=13 $ TeV in final states with jets
  and missing transverse momentum}},  {\em JHEP} {\bf 05} (2018) 025,
  [\href{http://arxiv.org/abs/1802.02110}{{\tt arXiv:1802.02110}}].

\bibitem{Aaboud:2017wqg}
{\textbf{ATLAS} collaboration}, {\it {Search for supersymmetry in events with
  $b$-tagged jets and missing transverse momentum in $pp$ collisions at
  $\sqrt{s}=13$ TeV with the ATLAS detector}},  {\em JHEP} {\bf 11} (2017) 195,
  [\href{http://arxiv.org/abs/1708.09266}{{\tt arXiv:1708.09266}}].

\bibitem{Sirunyan:2018iwl}
{\textbf{CMS} collaboration}, {\it {Search for new physics in events with two
  soft oppositely charged leptons and missing transverse momentum in
  proton-proton collisions at $\sqrt{s}=$ 13 TeV}},  {\em Phys. Lett.} {\bf
  B782} (2018) 440--467, [\href{http://arxiv.org/abs/1801.01846}{{\tt
  arXiv:1801.01846}}].

\bibitem{Aaboud:2017leg}
{\textbf{ATLAS} collaboration}, {\it {Search for electroweak production of
  supersymmetric states in scenarios with compressed mass spectra at
  $\sqrt{s}=13$ TeV with the ATLAS detector}},  {\em Phys. Rev.} {\bf D97}
  (2018), no.~5 052010, [\href{http://arxiv.org/abs/1712.08119}{{\tt
  arXiv:1712.08119}}].

\bibitem{Aaboud:2017aeu}
{\textbf{ATLAS} collaboration}, {\it {Search for top-squark pair production in
  final states with one lepton, jets, and missing transverse momentum using 36
  fb$^{−1}$ of $ \sqrt{s}=13 $ TeV pp collision data with the ATLAS
  detector}},  {\em JHEP} {\bf 1806} (2018) 108,
  [\href{http://arxiv.org/abs/1711.11520}{{\tt arXiv:1711.11520}}].

\bibitem{Sirunyan:2017leh}
{\textbf{CMS} collaboration}, {\it {Search for top squarks and dark matter
  particles in opposite-charge dilepton final states at $\sqrt{s}=$ 13 TeV}},
  {\em Phys. Rev.} {\bf D97} (2018), no.~3 032009,
  [\href{http://arxiv.org/abs/1711.00752}{{\tt arXiv:1711.00752}}].

\bibitem{vanBeekveld:2016hbo}
M.~van Beekveld, W.~Beenakker, S.~Caron, and R.~Ruiz~de Austri, {\it {The case
  for 100 GeV bino dark matter: A dedicated LHC tri-lepton search}},  {\em
  JHEP} {\bf 1604} (2016) 154, [\href{http://arxiv.org/abs/1602.00590}{{\tt
  arXiv:1602.00590}}].

\bibitem{Bae:2019dgg}
K.~J. Bae, H.~Baer, V.~Barger, and D.~Sengupta, {\it {Revisiting the SUSY mu
  problem and its solutions in the LHC era}},
  \href{http://arxiv.org/abs/1902.10748}{{\tt arXiv:1902.10748}}.

\bibitem{Ross:2016pml}
G.~G. Ross, K.~Schmidt-Hoberg, and F.~Staub, {\it {On the MSSM higgsino mass
  and fine-tuning}},  {\em Phys. Lett.} {\bf B759} (2016) 110--114,
  [\href{http://arxiv.org/abs/1603.09347}{{\tt arXiv:1603.09347}}].

\bibitem{Ross:2017kjc}
G.~G. Ross, K.~Schmidt-Hoberg, and F.~Staub, {\it {Revisiting fine-tuning in
  the MSSM}},  {\em JHEP} {\bf 1703} (2017) 021,
  [\href{http://arxiv.org/abs/1701.03480}{{\tt arXiv:1701.03480}}].

\bibitem{Athron:2017yua}
{\bf GAMBIT} Collaboration, {\it {A global fit of the MSSM with GAMBIT}},  {\em
  Eur. Phys. J.} {\bf C77} (2017), no.~12 879,
  [\href{http://arxiv.org/abs/1705.07917}{{\tt arXiv:1705.07917}}].

\bibitem{Athron:2018vxy}
{\bf GAMBIT} Collaboration, {\it {Combined collider constraints on neutralinos
  and charginos}},  {\em Eur. Phys. J.} {\bf C79} (2019), no.~5 395,
  [\href{http://arxiv.org/abs/1809.02097}{{\tt arXiv:1809.02097}}].

\bibitem{Aaboud:2018jiw}
{\textbf{ATLAS} collaboration}, {\it {Search for electroweak production of
  supersymmetric particles in final states with two or three leptons at
  $\sqrt{s}=13\,$TeV with the ATLAS detector}},  {\em Eur. Phys. J.} {\bf C78}
  (2018), no.~12 995, [\href{http://arxiv.org/abs/1803.02762}{{\tt
  arXiv:1803.02762}}].

\bibitem{Aaboud:2018zeb}
{\textbf{ATLAS} collaboration}, {\it {Search for supersymmetry in events with
  four or more leptons in $\sqrt{s}=13$ TeV $pp$ collisions with ATLAS}},  {\em
  Phys. Rev.} {\bf D98} (2018), no.~3 032009,
  [\href{http://arxiv.org/abs/1804.03602}{{\tt arXiv:1804.03602}}].

\bibitem{Sirunyan:2017qaj}
{\textbf{CMS} collaboration}, {\it {Search for new phenomena in final states
  with two opposite-charge, same-flavor leptons, jets, and missing transverse
  momentum in pp collisions at $ \sqrt{s}=13 $ TeV}},  {\em JHEP} {\bf 1803}
  (2018) 076, [\href{http://arxiv.org/abs/1709.08908}{{\tt arXiv:1709.08908}}].

\bibitem{Sirunyan:2018ubx}
{The \textbf{CMS} collaboration}, {\it {Combined search for electroweak
  production of charginos and neutralinos in proton-proton collisions at
  $\sqrt{s} =$ 13 TeV}},  {\em JHEP} {\bf 1803} (2018) 160,
  [\href{http://arxiv.org/abs/1801.03957}{{\tt arXiv:1801.03957}}].

\end{thebibliography}\endgroup

\end{document}